\documentclass[bibyear]{aa}  

\usepackage{graphicx}
\usepackage{txfonts}
\usepackage{pdflscape}
\usepackage{afterpage}
\usepackage{scrextend}

\usepackage{threeparttable}
\usepackage{lipsum,amsmath}
\usepackage{multirow}
 \usepackage{color}
 \usepackage{natbib,twoopt}
 \usepackage{amssymb,mathtools}

\usepackage[colorlinks, citecolor=blue]{hyperref}      

\newcommand\Tstrut{\rule{0pt}{2.6ex}}         
\newcommand\Bstrut{\rule[-0.9ex]{0pt}{0pt}}   
\newcommand\BBstrut{\rule[-1.3ex]{0pt}{0pt}}   

\bibpunct{(}{)}{;}{a}{}{,}             

\makeatletter
  \newcommandtwoopt{\citeads}[3][][]{\href{http://adsabs.harvard.edu/abs/#3}%
    {\def\hyper@linkstart##1##2{}%
     \let\hyper@linkend\@empty\citealp[#1][#2]{#3}}}
  \newcommandtwoopt{\citepads}[3][][]{\href{http://adsabs.harvard.edu/abs/#3}%
    {\def\hyper@linkstart##1##2{}%
     \let\hyper@linkend\@empty\citep[#1][#2]{#3}}}
  \newcommandtwoopt{\citetads}[3][][]{\href{http://adsabs.harvard.edu/abs/#3}%
    {\def\hyper@linkstart##1##2{}%
     \let\hyper@linkend\@empty\citet[#1][#2]{#3}}}
  \newcommandtwoopt{\citeyearads}[3][][]%
    {\href{http://adsabs.harvard.edu/abs/#3}
    {\def\hyper@linkstart##1##2{}%
     \let\hyper@linkend\@empty\citeyear[#1][#2]{#3}}}
\makeatother

\makeatletter
\renewcommand*\aa@pageof{, page \thepage{} of \pageref*{LastPage}}
\makeatother

\NewDocumentCommand{\dataset}{oov}{%
  \href{https://doi.org/#1/#3}{#2}%
}

\begin{document}

   \title{High-$z$ radio Quasars in RACS:}
   \subtitle{\texttt{I}. Selection, identification, and multi-wavelength properties}

   \author{
L. Ighina\inst{1,2}
\and
A. Caccianiga\inst{1}
\and
A. Moretti\inst{1}
\and
J. W. Broderick\inst{3}
\and
J. K. Leung\inst{4,5,6}
\and 
F. Rigamonti\inst{1,7,8}
\and 
N. Seymour\inst{9}
\and
J. Afonso\inst{10,11}
\and
T. Connor\inst{2}
\and
C. Vignali\inst{12,13}
\and
Z. Wang\inst{9}
\and
T. An\inst{14,15,16}
\and
B. Arsioli\inst{10,11}
\and
S. Bisogni\inst{17}
\and
D. Dallacasa\inst{18,12}
\and
R. Della Ceca\inst{1}
\and
Y. Liu\inst{14}
\and
A. López-Sánchez\inst{19,20}
\and
I. Matute\inst{10,11}
\and
C. Reynolds\inst{3}
\and
A. Rossi\inst{13}
\and
C. Spingola\inst{18}
\and
P. Severgnini\inst{1}
\and
F. Tavecchio\inst{1}
          }

\institute{INAF, Osservatorio Astronomico di Brera, via Brera 28, 20121, Milano, Italy\\
              \email{luca.ighina@cfa.harvard.edu}
         \and
Center for Astrophysics | Harvard \& Smithsonian, 60 Garden St., Cambridge, MA 02138, USA 
         \and
SKA Observatory, Science Operations Centre, CSIRO ARRC, 26 Dick Perry Avenue, Kensington, WA 6151, Australia 
       \and
David A. Dunlap Department of Astronomy and Astrophysics, University of Toronto, 50 St. George Street, Toronto, ON M5S 3H4, Canada
       \and
Dunlap Institute for Astronomy and Astrophysics, University of Toronto, 50 St. George Street, Toronto, ON M5S 3H4, Canada
       \and
Racah Institute of Physics, The Hebrew University of Jerusalem, Jerusalem 91904, Israel
    \and
INFN, Sezione di Milano-Bicocca, Piazza della Scienza 3, I-20126 Milano, Italy
    \and
Como Lake centre for AstroPhysics (CLAP), DiSAT, Università dell’Insubria, Via Valleggio 11, 22100 Como, Italy
        \and
International Centre for Radio Astronomy Research, Curtin University, 1 Turner Avenue, Bentley, WA, 6102, Australia
        \and
Instituto de Astrofísica e Ciências do Espaço, Universidade de Lisboa, OAL, Tapada da Ajuda, Lisboa, Portugal
        \and 
Departamento de Física, Faculdade de Ciências, Universidade de Lisboa, Lisbon, Portugal
        \and
Dipartimento di Fisica e Astronomia, Alma Mater Studiorum, Università degli Studi di Bologna, Via Gobetti 93/2, 40129 Bologna, Italy
        \and
INAF - Osservatorio di Astrofisica e Scienza dello Spazio, Via Piero Gobetti 93/3, 40129 Bologna,\, Italy
        \and
Shanghai Astronomical Observatory, Chinese Academy of Sciences (CAS), 80 Nandan Road, Shanghai 200030, China
        \and
School of Astronomy and Space Sciences, University of Chinese Academy of Sciences, No. 19A Yuquan Road, Beijing 100049, China
        \and
Key Laboratory of Radio Astronomy and Technology, CAS, A20 Datun Road, Beijing, 100101, P. R. China
        \and
INAF - Istituto di Astrofisica Spaziale e Fisica Cosmica (IASF), Via A. Corti 12, 20133, Milano, Italy
        \and
INAF - Istituto di Radioastronomia, Via Gobetti 101, I-40129 Bologna, Italy
        \and
School of Mathematical and Physical Sciences, Faculty of Science and Engineering, Macquarie University, NSW 2109, Australia        
\and
Australian Research Council Centre of Excellence for All-Sky Astrophysics in 3 Dimensions (ASTRO 3D), Australia
        \and 
INAF – Osservatorio Astronomico di Brera, Via E. Bianchi 46, I-23807 Merate, Italy
}

   \date{Received September 15, 1996; accepted March 16, 1997}

\abstract{Radio-bright, jetted quasars at $z > 5$ serve as unique laboratories for studying supermassive black hole activity in the early Universe. In this work, we present a sample of high-$z$ jetted quasars selected from the combination of the radio Rapid ASKAP Continuum Survey (RACS) with deep wide-area optical/near-infrared surveys. From this cross-match we selected 45 new high-$z$ radio quasar candidates with S$_{\rm 888MHz}>1$~mJy and mag$z$~$<21.3$ over an area of 16000~deg$^2$. Using spectroscopic observations, we confirmed the high-$z$ nature of 24 new quasars, 13 at $4.5<z<5$ and 11 at $z>5$. 
If we also consider similar, in terms of radio/optical fluxes and sky position, quasars at $z>5$ already reported in the literature, the overall $z>5$ RACS sample is composed by 33 powerful quasars, expected to be $\sim90$\% complete at mag$z$~$<21.3$ and S$_{\rm 888MHz}>1$~mJy. Having rest-frame radio luminosities in the range $\nu$L$_{\rm 1.4GHz}=10^{41.5}-10^{44.4}\ {\rm erg}\ {\rm s}^{-1}$, this sample contains the most extreme radio quasars currently known in the early Universe.
We also present all X-ray and radio data currently available for the sample, including new, dedicated {\it Chandra}, uGMRT, MeerKAT and ATCA observations for a sub-set of the sources. from the modelling of their radio emission, either with a single power law or a broken power law, we found that these systems have a wide variety of spectral shapes with most quasars (22) having a flat radio emission  (i.e., $-0.5<\alpha_{\rm r}<0.5$). At the same time, the majority of the sources with X-ray coverage present a high-energy luminosity larger than the one expected from the X-ray corona only. Both the radio and X-ray properties of the high-$z$ RACS sample suggest that many of these sources have relativistic jets oriented close to our line of sight. (i.e., blazars) and can therefore be used to perform statistical studies on the entire jetted population at high redshift.}

   \keywords{galaxies: active - galaxies: nuclei – galaxies: high-redshift - (galaxies:) quasars: general - galaxies: jets -- (galaxies:)
               }
 \titlerunning{High-$z$ radio quasars in RACS \texttt{I}}

   \maketitle
%

\section{Introduction}

The identification and the study of relativistic jets from high-$z$ active galactic nuclei (AGN) and quasars provide critical insights into the growth of the first supermassive black holes (SMBHs), their co-evolution with the host galaxies, and the properties of the intergalactic medium (IGM) in the early Universe \citep[e.g.][]{Furlanetto2006,Fabian2012,Blandford2019,Hardcastle2020,Volonteri2021,Maiolino2024}.
It has been proposed that jetted quasars could represent a viable solution to explain the extremely massive BHs ($\sim10^9$~M$_{\odot}$) already at place at $z\sim6$ (e.g. \citealt{Ghisellini2010,Connor2024}), where the presence of relativistic jets could allow for a faster BH accretion (e.g. \citealt{Jolley2008,Jolley2009}). Additionally, radio-powerful systems are normally associated with overdense environments \citep[e.g.][]{Miley2008,Gilli2019,Uchiyama2022}. Detecting such overdensities in the early Universe can provide crucial information about the early evolution of the large-scale structures that we observe in the local Universe \citep[e.g.][]{Overzier2009,Overzier2016}.
Finally, radio-bright quasars within the epoch of re-ionisation ($z\gtrsim6$) can be use as background lights to directly measure the distribution of the neutral hydrogen content along the line of sight through the absorption of the 21~cm line ($\lesssim$200~MHz in the observed frame at $z\gtrsim6$; e.g. \citealt{Carilli2002,Soltinsky2021,Niu2024}). The detection and characterisation of this feature in the radio spectrum of a high-$z$ source would have profound implications on the nature and properties of the IGM during this critical epoch \citep[e.g.][]{Ciardi2013,Soltinsky2025}.

With the release of several wide-area optical and near-infrared (NIR) photometric surveys in the last two decades, the number of high-$z$ UV/optical-bright quasars has substantially increased reaching more than $\sim$1000 objects at $z>5$ and more than $\sim300$ objects at $z>6$ (see \citealt{Fan2022} for a recent review). 
Several methods have been used in the selection of high-$z$ quasar candidates, including the Ly$\alpha$ dropout technique \citep[e.g.][]{Fan1999b,Fan200,Fan2003,Morganson2012,Venemans2013,Venemans2015,Banados2014,Banados2016,Banados2018b,Carnall2015,Matsuoka2016,Matsuoka2018a,Matsuoka2018b,Matsuoka2022,Caccianiga2019,Belladitta2020,Belladitta2023,Wolf2020,Onken2022,Yang2023}, spectral fitting of photometric measurements \citep[e.g.][]{Reed2015,Reed2017,Reed2019,Wolf2024} and machine learning approaches \citep[e.g.][]{Wenzl2021,Nanni2022,Carvajal2023,Ye2024}. Interestingly, among all these $z>5$ quasars, only $\sim$50-60 are detected in the radio band (e.g., \citealt{Banados2021,Banados2025,Caccianiga2024}), most of which were uncovered with the LOw Frequency ARray (LOFAR) Two-metre Sky Survey (LoTSS; \citealt{Shimwell2017,Shimwell2019,Shimwell2022}) by \cite{Gloudemans2021,Gloudemans2022}.\\
However, being discovered with different methods and surveys (e.g., \citealt{Li2021,Gloudemans2022,Belladitta2023}), the $z>5$ radio quasars currently known constitute a highly in-homogeneous sample. Indeed the selection completeness can be significantly different, depending on both the optical and radio data-sets/limits used. For this reason, while detailed studies on single objects were able to characterise the multi-wavelength emission of individual systems (\citealt{Spingola2020,Kappes2022,Belladitta2022,Ighina2024,Frey2024,Krezinger2024,Banados2025,Gloudemans2025}), the statistical properties of this extreme class of objects are still poorly constrained in the early Universe \citep[e.g.][]{Mao2017,Latif2024}.

In this work, we present the first part of a project aimed at studying the properties of $z>5$ radio-brightest jetted quasars from a statistical point of view. In particular, here we discuss the selection of a high-$z$ sample starting from the first data release of the Rapid ASKAP Continuum Survey at low frequency (RACS-low; \citealt{McConnell2020,Hale2021}). This survey, performed with the Australian Square Kilometre Array Pathfinder (ASKAP) telescope, covers the entire sky at Dec.~$<+41^\circ$ at 888~MHz with a resolution of $\sim$12--15\arcsec and a median RMS of $\sim$0.25~mJy~beam$^{-1}$. Moreover, as part of the RACS survey, multiple scans of the Dec.~$<+50^\circ$ sky are being performed at different frequencies (from 0.9 to 1.7~GHz; e.g., \citealt{Duchesne2023,Duchesne2024,Duchesne2025}), which provide multi-frequency and multi-epoch information for millions of radio sources. The unique large area-sensitivity combination of RACS makes it the ideal starting point for the construction of a large statistical sample of radio-bright quasars in the early Universe. Indeed, soon after the first RACS data release \citep{McConnell2020}, one of the most-distant radio quasars currently known was uncovered \citep[VIK~J2318$-$31 at $z=6.44$;][]{Ighina2021a}, showcasing the capability of this survey to explore the high-redshift Universe.
Prompted by this first detection, we combined the RACS-low survey with the Dark Energy Survey (DES; \citealt{Abbott2018,Abbott2021}) and the Panoramic Survey Telescope and Rapid Response System \citep[Pan-STARRS;][]{Chambers2016} survey in the optical/NIR band to search for new high-$z$ jetted quasars with the Ly$\alpha$ dropout technique. 
Here we present new high-$z$ quasars selected from RACS-low and we discuss their multi-wavelength properties together with the $z>5$ quasars already known from the literature with similar optical fluxes and detected in RACS-low. In particular, we provide all the radio and X-ray information currently available for these sources (from recent surveys as well as from new, dedicated observations). In future publications, this dataset will be used to statistically constrain the evolution of jetted quasars at high redshift.

\begin{figure}
   \centering
	\includegraphics[width=\hsize]{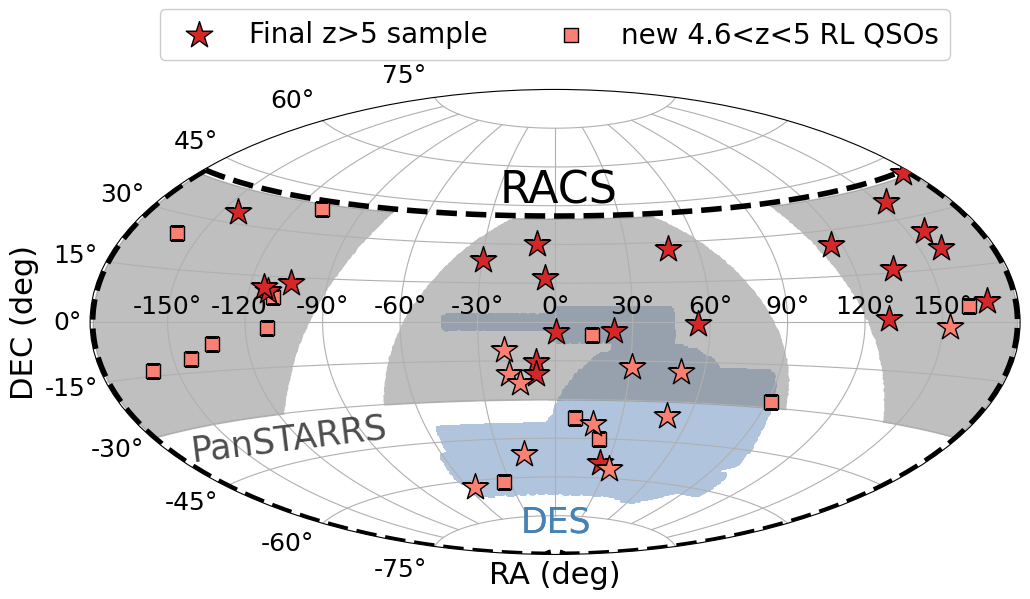}

\caption{Sky distribution of the final RACS sample together with the area covered by each optical/NIR survey considered and by the RACS-low survey (delimited by the dashed line). The overall area covers $\sim$16000~deg$^2$ at $|b|>15^{\circ}$. The stars indicate the $z>5$ sample (darker colour for the quasars already known in the literature and lighter for the new ones reported here), while the squares are the newly discovered sources at $4.5<z<5$.}
    \label{fig:sky_distr_sample}
\end{figure}

The manuscript is structured as follows: in Sec. \ref{sec:selection} we describe the optical and radio criteria used for the selection of high-$z$ quasars; in Sec. \ref{sec:identification} we present the spectroscopic observations and identification of the selected candidates; in Sec. \ref{sec:final_sample} we discuss the completeness of the final $z>5$ sample and compare it to other samples from the literature;  in Sec. \ref{sec:X-ray_properties} and \ref{sec:radio_properties} we present and discuss the X-ray and radio properties, respectively, of the high-$z$ quasars detected in RACS; finally, in Sec. \ref{sec:conclusions} we summarise our results.

Throughout the paper we assume a flat $\Lambda$CDM cosmology with $H_{0}$=70 km s$^{-1}$ Mpc$^{-1}$, $\Omega_m$=0.3 and $\Omega_{\Lambda}$=0.7. Spectral indices are given assuming S$_{\nu}\propto \nu^{-\alpha}$ and errors are reported at a 68\% confidence level, unless otherwise specified.

\section{Selection of \texorpdfstring{high-$z$}{} radio quasar candidates} 
\label{sec:selection}

In order to build a well-defined sample of $z>5$ radio quasars, we performed a systematic search of these sources in some of the deepest radio and optical/NIR wide-area surveys currently available. In particular, we considered the first data release of the RACS-low survey (low, at 888~MHz; \citealt{McConnell2020,Hale2021}) in the radio band together with the DES (2nd data release; \citealt{Abbott2021}) and the Pan-STARRS (\citealt{Chambers2016}) in the optical/NIR (both with $grizY$ filters, even though with slightly different bandpasses.
The combined footprints of these surveys are shown in Fig. \ref{fig:sky_distr_sample}.

In this section, we describe the selection of high-$z$ candidates in these surveys mainly through the use of the Ly$\alpha$ dropout technique (e.g., \citealt{Banados2016}) together with a radio association (e.g., \citealt{Caccianiga2019, Gloudemans2022}).

\begin{figure*}
   \centering
   \includegraphics[width=0.7\hsize]{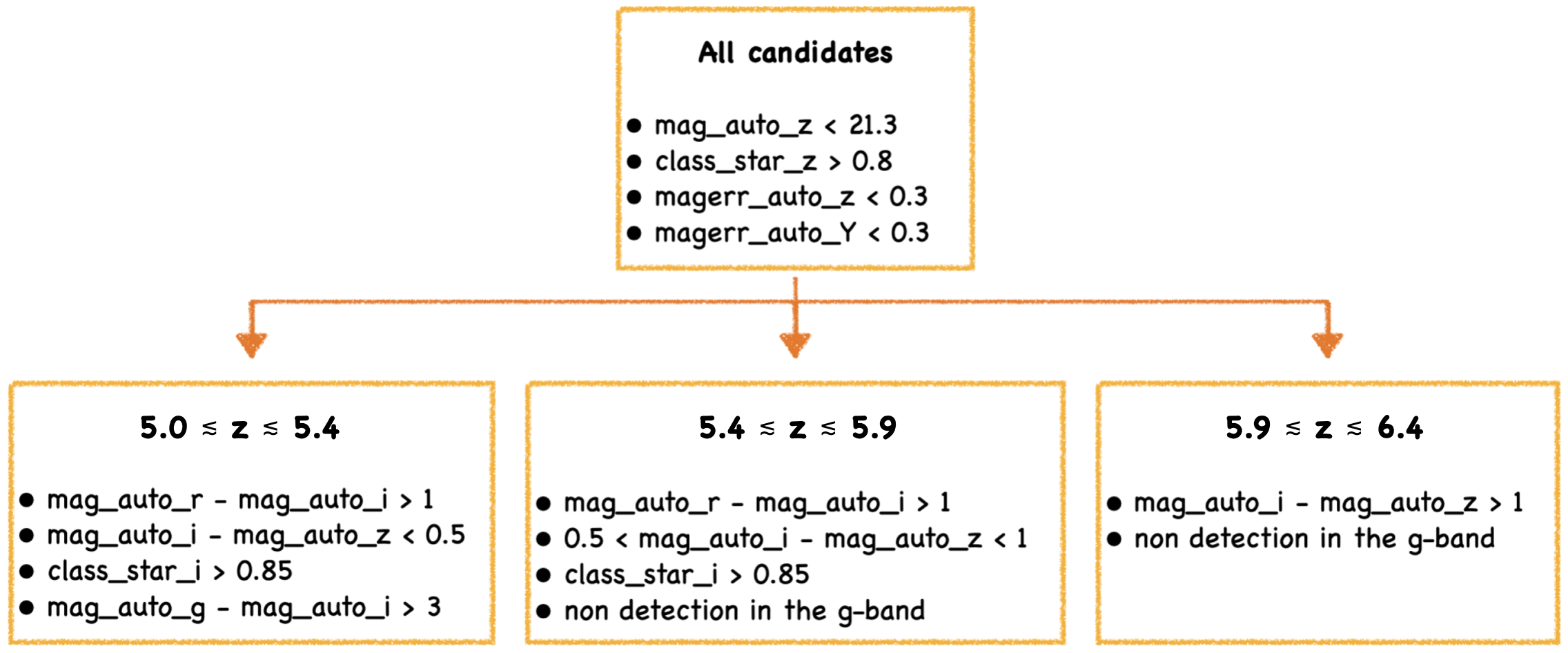}
      \caption{Criteria used for the selection of candidates from the DES survey. The criteria at the top were used for all the candidates, whereas, the criteria at the bottom were used in order to select sources at specific redshifts: (1) for $5\lesssim z \lesssim 5.4$, (2) for $5.4\lesssim z \lesssim 5.9$ and (3) for $5.9\lesssim z \lesssim 6.4$. }
         \label{fig:des_criteia}
\end{figure*}

\subsection{Radio selection in the RACS-low and EMU pilot surveys}

We started from the first data release of the RACS-low survey at 888~MHz \citep{McConnell2020,Hale2021}. 
We made use of the source lists derived from the images described in \cite{McConnell2020}\footnote{Available from the CASDA website: \url{https://data.csiro.au/domain/casdaObservation?redirected=true}.} since the final catalogue of the RACS-low survey built by \cite{Hale2021} has a much larger beam size (25\arcsec, a factor $\sim$2 larger than the un-smoothed images) and a comparable RMS (median value $\sim0.3$~mJy~beam$^{-1}$ in both cases). This larger beam size would increase the positional uncertainty, especially for faint sources. 
Given the relatively deep sensitivity of RACS, we were able to select sources with a surface brightness down to $S_\mathrm{peak}$~\textgreater~1~mJy~beam$^{-1}$, which would correspond to a S/N~$\sim4$ given the median RMS of the images. In order to reduce the error on the radio position, we only considered objects with an integrated-peak flux density ratio of $S_\mathrm{int}$/$S_\mathrm{peak}$~\textless~1.5. 
This value was chosen to select all the point-like sources in the RACS source lists for S/N as low as $\sim4$ (see, e.g., section 4.4 in \citealt{Duchesne2024}). Indeed, as shown also in Sec. \ref{sec:final_sample}, most high-$z$ quasars are expected to be point sources\footnote{While a small fraction of high-$z$ radio quasars were found to have radio jets extending up to several arcsec away from the core (e.g., \citealt{Ighina2022a}), the amount of flux in the more extended components is usually $\lesssim10$\%, well within the selection criteria.}.
Given the typical positional errors of RACS (about 2$''$ for faint sources, see section 3.4.3 in \citealt{McConnell2020}), we adopted a flat search radius of 3$''$ to match the radio positions provided in the RACS-low catalogue with the optical counterparts selected as described in the following subsection.

At the same time, we also considered the pilot of the currently on-going EMU survey. While this survey covers a smaller area ($\sim$270~deg$^2$), it has a similar central frequency (944~MHz) and angular resolution ($\sim$18$''$, considering the catalogue from the convolved images, see \citealt{Norris2021}) compared to RACS-low, with the advantage of being significantly deeper (RMS~$\sim 25-30$~\textmu Jy~beam$^{-1}$). Therefore, this catalogue is also instructive on the completeness of the radio selection from RACS. We focused again on sources with $S_\mathrm{peak}$~\textgreater~1~mJy~beam$^{-1}$, but since in this case they correspond to a S/N~$\gtrsim30$, we reduced the EMU--DES association radius to 1.5$''$.

\subsection{Optical criteria}

To select good high-$z$ quasar candidates, we considered the RACS sources that satisfy the criteria outlined above and that are detected in the deepest wide-area optical/NIR surveys currently available: DES and Pan-STARRS.

In both cases, in order to have a flux-limited sample of sources that could be spectroscopically observed in a reasonable amount of time with current ground-based optical telescopes, we restricted the search to sources with a mag$z$~\textless~21.3. {As a reference, the median depth at 5$\sigma$ of the DES and PanSTARRS catalogue considered is 23.6 and 22.1, respectively.}  We only considered regions of the sky with a Galactic latitude $|b|>15^{\circ}$ to reduce the spurious contamination from Galactic objects.
Given the wavelength coverage of the optical/NIR considered surveys, the highest redshift that could be probed is $z\sim6.8$ \citep[e.g.][]{Banados2021}, where the Ly$\alpha$ drops-out around the centre of the $Y$ filter. However, in order to have candidates with more secure measurements, we also required a detection in at least two photometric filters. This criterion restricted the search to $z\lesssim6.4$ sources. Moreover, we note that, since the wavelength coverage of the filters used for the DES and Pan-STARRS surveys are slightly different, also the criteria/threshold adopted in the selection of high-$z$ quasars from each survey is slightly different.

\subsubsection{Candidates from the DES survey}

To access the most recent data release of DES, the 2$^{\rm nd}$ \citep{Abbott2021}, we used the dedicated SQL portal\footnote{\url{https://des.ncsa.illinois.edu/desaccess/home}.}.
During the selection of new candidates, we considered three different sets of criteria based on the redshift of the potential candidates: (1) for $5\lesssim z\lesssim5.4$ sources; (2) for $5.4\lesssim z \lesssim5.9$ sources; (3) for $z\gtrsim5.9$ sources. The criteria adopted are summarised in Fig. \ref{fig:des_criteia}, where all the magnitudes are in the AB system. 

These criteria are meant to select primarily compact stellar-like objects with a dropout either in the $r$ or in $i$ the filter, hence high values of $r-i$ or $i-z$ colours. Moreover, since we expect the absorption from the IGM to be more effective at shorter wavelengths (i.e., in the $g$-band) and at higher redshift, we also applied a specific magnitude limit in the $g$-band to select only faint objects. In particular, this resulted in a colour \texttt{mag\_auto\_g}~--~\texttt{mag\_auto\_i}~$>3$, for sources selected with the criteria (1), expected to be at $5\lesssim z \lesssim 5.4$. This limit selects sources not detected in the $g$-band, or, if detected, with a $g$ magnitude much fainter with respect to the one in the $i$ filter. For candidates at $z\gtrsim5.4$, instead, we only considered candidates without a detection in the $g$ band. This condition was satisfied by selecting sources with a \texttt{mag\_auto\_g}~$>23.5$ and an error on \texttt{err\_mag\_auto\_g}~$>0.5$. Subsequently, we visually inspected the images to discard those with even a faint blue emission. On average, we were able to eliminate all the sources with an emission $\gtrsim3\sigma$. We note that the $g$-band-faintness criterion was the one that rejected most of the candidates (over 160), especially with the criteria (2), where the Ly$\alpha$ dropout is transitioning from the $r$ to the $i$ band and where stellar contamination is expected to be the predominant \citep[e.g.,][]{Yang2017}.

\begin{figure*}
   \centering
    \includegraphics[width=0.31\hsize]{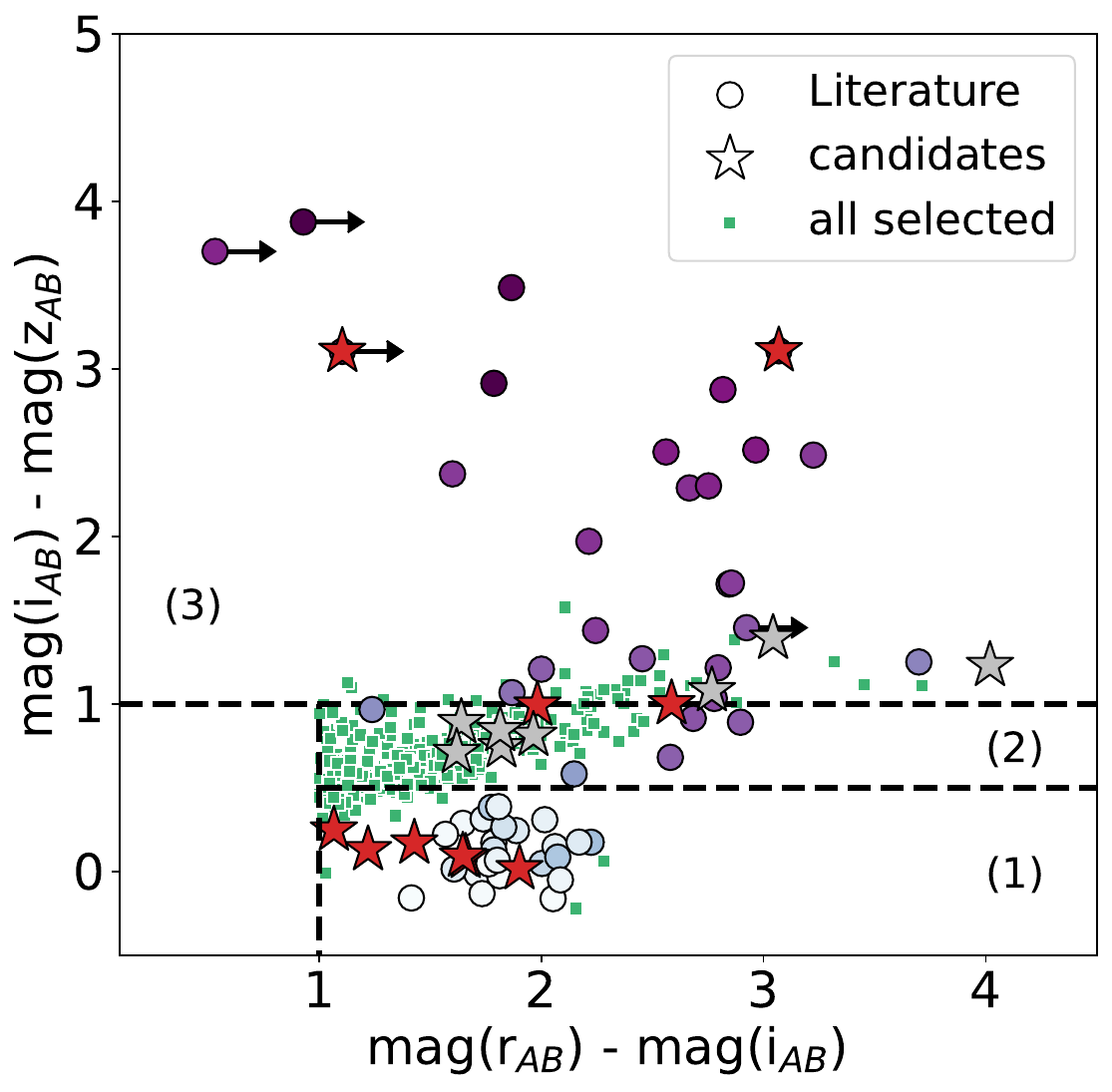}
    \includegraphics[width=0.311\hsize]{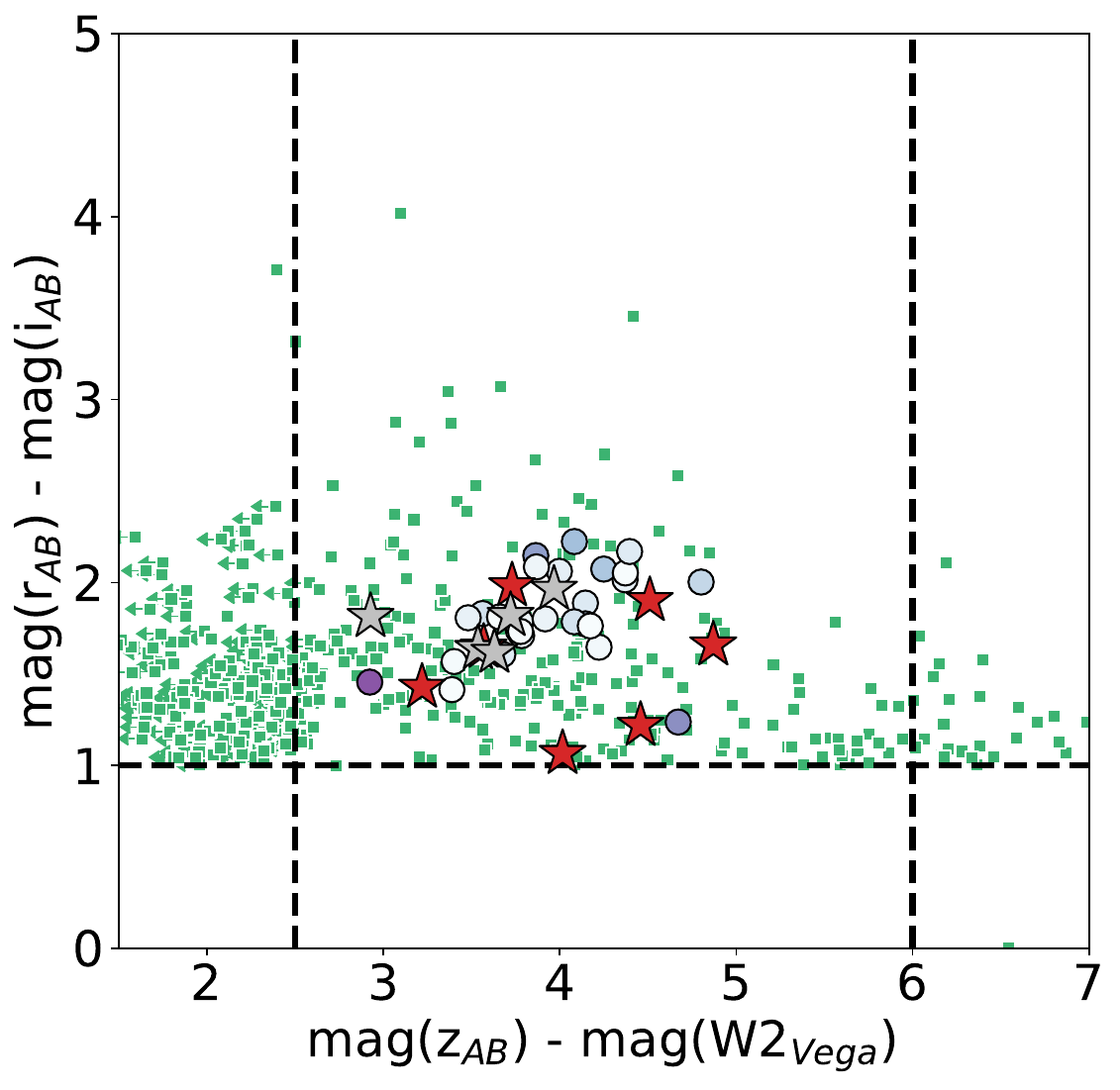}
    \includegraphics[width=0.36\hsize]{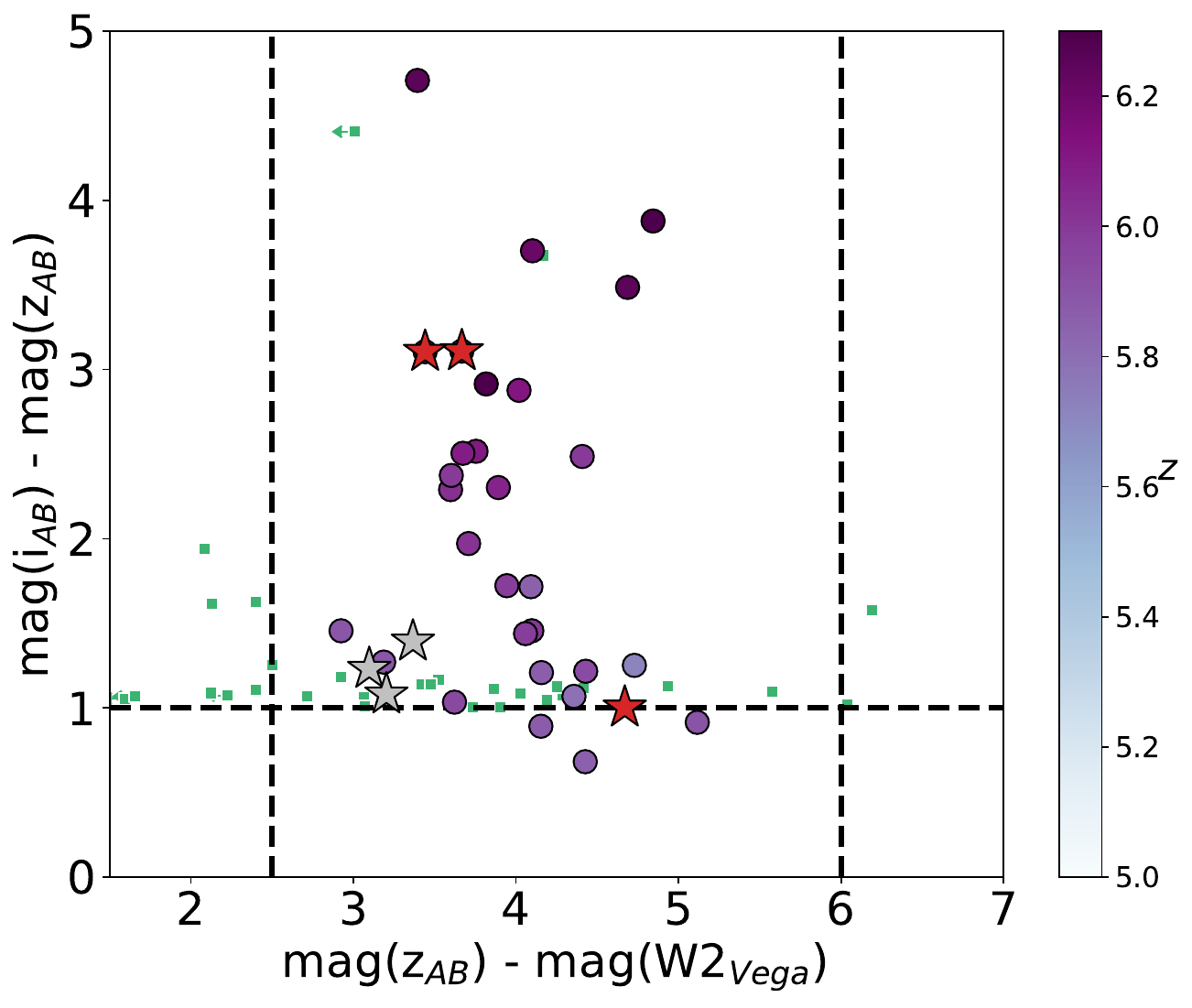}

      \caption{Colour plots used for the selection of candidates from the DES survey. In all the plots the circles represent the already known $z>5$ quasars in the DES area, colour coded based on their redshift. The dashed lines represent the threshold considered in our selection. The small green squares are all the potential candidates selected with the criteria reported in Fig. \ref{fig:des_criteia}, without the visual inspection of the $g$-band images. Stars indicate the candidates that satisfied all the selection criteria and that form the final well-defined sample: red for confirmed high-$z$ quasars and gray for not-confirmed candidates. 
      {\bf Left panel:} $i-z$ colour as a function of the $r-i$ one. The plot is divided in three regions, for which different sets of criteria were applied (see Fig. \ref{fig:des_criteia}), with the aim of selecting sources in specific redshift ranges. As clear from this plot, the region (2) is the one with the most contaminates, likely stars with a spurious radio association.
      {\bf Central panel:} $z-W2$ colour as a function of the one around the $r-i$ drop. {\bf Right panel:} $z-W2$ colour as a function of the one around the $i-z$ drop. The $W2$ magnitude is the only one in the Vega system, the rest are in the AB system. For sources not detected in the $W2$ band, we adopted an upper limit of mag$W2$=19. 
      }
    \label{fig:des_colour_plots}
\end{figure*}

After the first selection based on the DES colours, we also considered mid-IR data from catWISE \citep{Eisenhardt2020,Marocco2021}. For the cross-match, we used a radius of 2.5$''$, which corresponds to $\sim3\sigma$ the astrometric uncertainty of the faintest catWISE sources (\texttt{mag\_W2}~$\sim17$, Vega system; see fig. 13 in \citealt{Marocco2021}) and for which we only expect a spurious association chance of $\lesssim2$\%. we only kept candidates with $2.5 < \texttt{mag\_auto\_z} - \texttt{mag\_W2} < 6$ (where the {\it WISE} magnitude is in the Vega system\footnote{The conversion factors from Vega to AB magnitudes are 2.699 and 3.339 for band $W1$ and $W2$, respectively.}). This criterion is meant to exclude low-$z$ absorbed quasars and part of the elliptical galaxies at low redshift \citep[e.g.][]{Carnall2015}. After applying this filter, we were left with 254 candidates (out of the starting 373).
In Fig. \ref{fig:des_colour_plots} we show the $i-z$, $r-i$, and $z-W2$ colours for the $z>5$ quasars currently known in the DES area. At the time of writing, there are a total of 58 $z>5$ quasars with mag$z$~$<21.3$ known in the footprint of DES\footnote{ These sources were collected from different works published in the literature before the submission of this paper. A full list of objects will be presented in Caccianiga et al. in prep. Here we report a list of the works from which the majority of the quasars were discovered: \cite{Carilli2010,Banados2014,Banados2015,Banados2016,Banados2023,Wang2016,Wang2019,Yang2016,Yang2017,Yang2019,Yang2023,Gloudemans2022,Abdurrouf2022,Lai2023}. \label{refnote}}.
We note that not all of these sources have been selected from DES, but also from other surveys with different filter sets \citep[e.g.,][]{Wolf2020,Wenzl2021,Onken2022}.

\subsubsection{Candidates from the Pan-STARRS survey}

In order to access the first data release of the Pan-STARRS survey, we used the Mikulski Archive for Space Telescopes (MAST\footnote{available at: \url{https://mastweb.stsci.edu/ps1casjobs/home.aspx}}). 
In particular, we considered the `mean' values reported in the Pan-STARRS catalogue, as opposed to the `stacked' ones. This choice was made in order to have more accurate measurements of the point-like nature of the candidates (see \citealt{Chambers2016}), since, as detailed below, it is an important constraint in the selection. At the time of writing, the number of $z>5$ quasars in Pan-STARRS ($\sim$350)\footref{refnote} is significantly larger compared to DES ($\sim$60). This is due to many reasons, the main one being that this survey covers a larger fraction of the sky, including the entire northern sky, where more telescopes are available for the spectroscopic follow-up, with respect to the southern sky. 

For the Pan-STARRS selection we followed the criteria outlined in \cite{Caccianiga2019} for the selection of a complete sample of radio quasars at $z>4$. Similarly to DES, these criteria are meant to  to select point-like objects, faint in the $g$-band, with reliable measurements in the filters red-wards of the expected Ly$\alpha$ and with a drop either in the $r$ or $i$ filter. 

\begin{figure}
   \centering
   \includegraphics[width=\hsize]{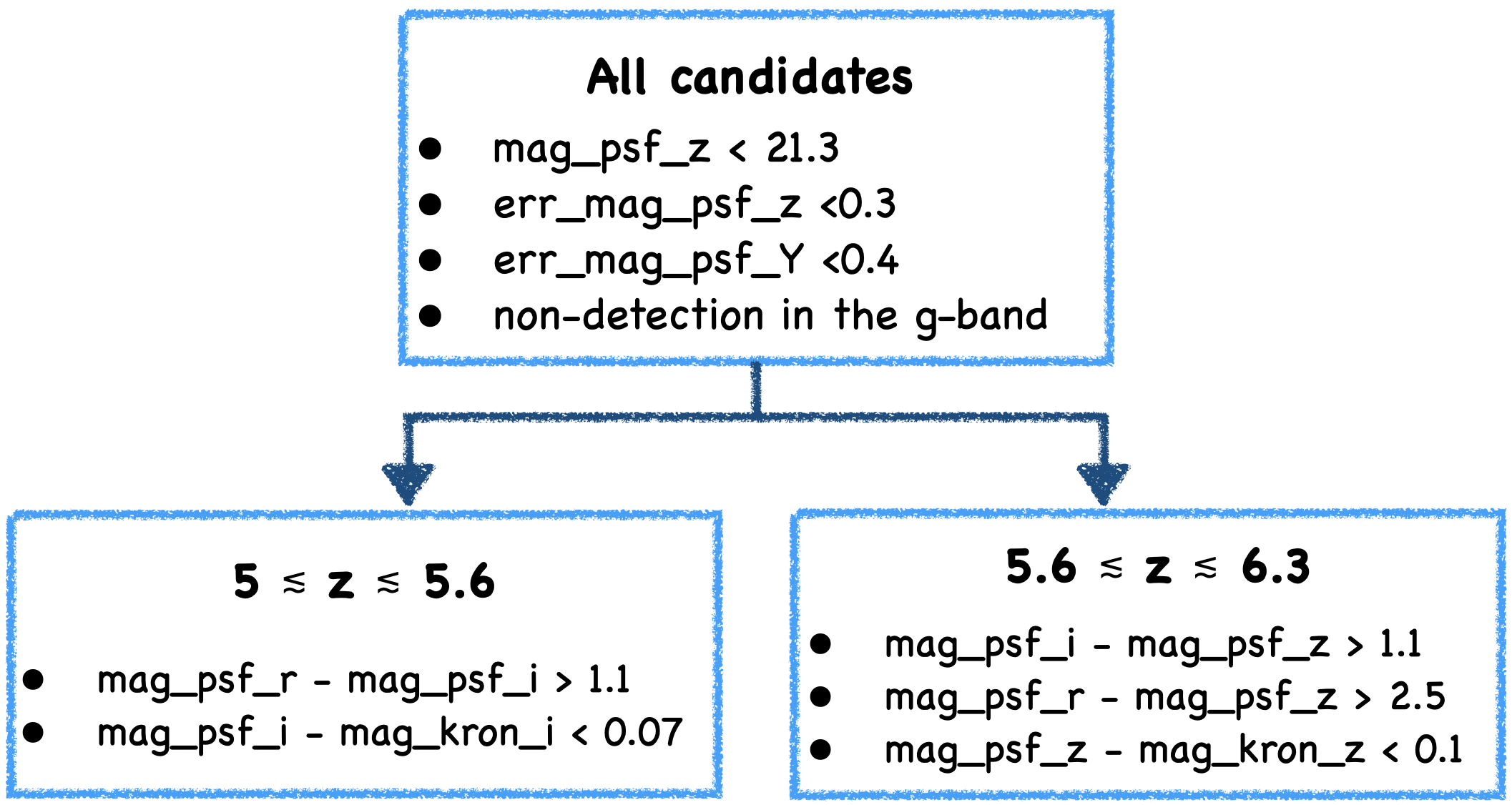}
      \caption{Criteria used for the selection of candidates from the Pan-STARRS survey. The criteria are divided in two sets, one (left) aimed at selecting $5\lesssim z\lesssim 5.6$ candidates and one (right) aimed at selecting $5.6\lesssim z\lesssim 6.3$ candidates. During the selection, we considered the `mean' magnitudes from the PanSTARRS catalogue.}
         \label{fig:pan_criteia}
\end{figure}

\begin{figure*}
   \centering
   \includegraphics[width=0.322\hsize]{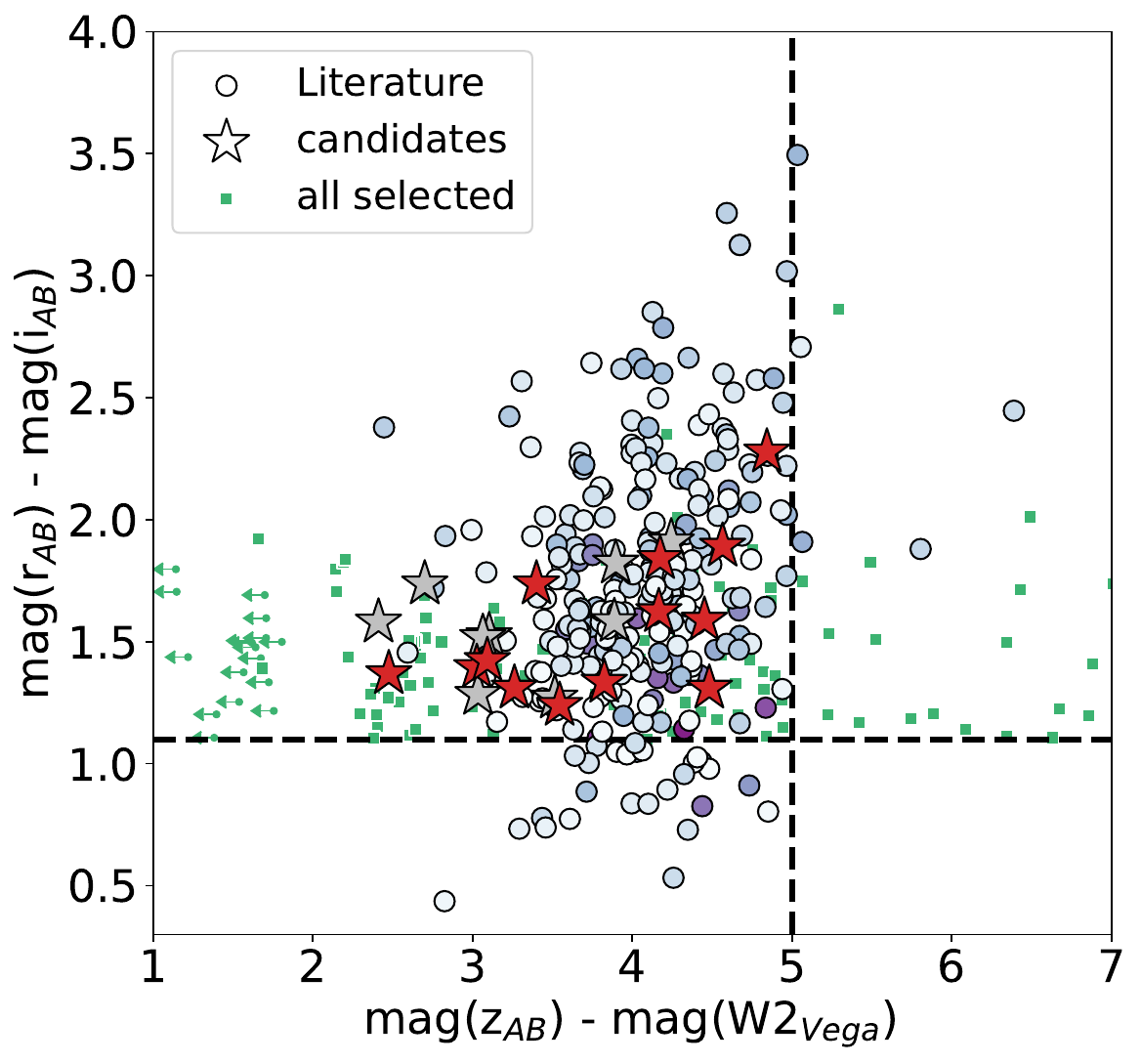}
   \includegraphics[width=0.37\hsize]{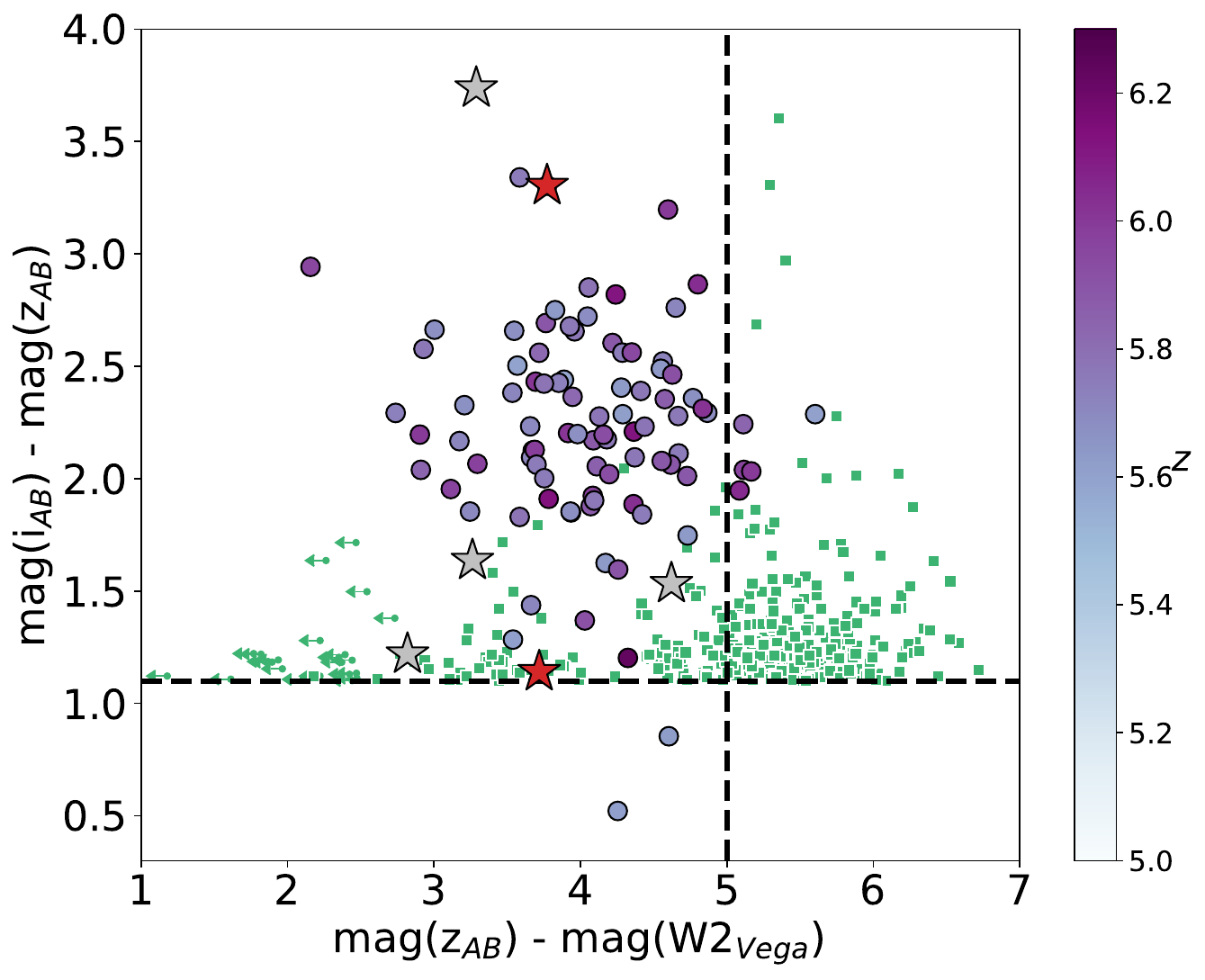}
      \caption{Criteria used for the selection of candidates from the Pan-STARRS survey. The criteria are divided in two sets, one ({\bf left panel}) aimed at selecting $5\lesssim z\lesssim 5.6$ candidates and one ({\bf right panel}) aimed at selecting $5.6\lesssim z\lesssim 6.3$ candidates. For clarity, we only show sources at $5< z<5.6$ and at $z>5.6$ in the left and right plot, respectively. For sources not detected in the $r$- or $i$-band we adopted the $5\sigma$ limits of the Pan-STARRS survey for the plot, 23.2 and 23.1, respectively. For sources not detected in the $W2$ band we adopted an upper limit of mag$W2$=19  to plot the data. The dashed lines represent the threshold considered in our selection.}
         \label{fig:pan_color_plots}
\end{figure*}

We report the criteria adopted in Fig. \ref{fig:pan_criteia}. As discussed and detailed in \cite{Caccianiga2019}, these criteria should recover about $\gtrsim90$\% of the \texttt{mag\_psf\_z}~$<21$ high-$z$ quasars (see also discussion below). Once again the criterion of a non-detection in the $g$-band was initially set to sources with a \texttt{mag\_psf\_g}~$>24$ and then, after applying the IR cut, we checked the $g$-band images of each candidate and discarded the ones where a faint emission was visible (see, e.g., Fig. \ref{fig:discard_cand}, left panel).  In this process we also checked the quality of the other images, since some fields were affected by artifacts. From the visual inspection (performed after the $z-W2$ criterion) we eliminated 27 candidates. After pre-selecting high-$z$ candidates from the Pan-STARRS catalogue with the limits reported in Fig. \ref{fig:pan_criteia}, we also considered the IR magnitudes from the catWISE survey by adopting a matching radius of $2.5''$. In this case, to maximise the completeness of the selection we used slightly different limits in the $z-W2$ colour: $z-W2<5$. Indeed, as shown in Fig. \ref{fig:pan_color_plots}, in the PanSTARRS selection there are not as many candidates with a non detection in the $W2$, with respect to the DES selection. This is likely due to the slightly different $z$-filters of the two surveys as well as the first $g$-band non detection criterion (i.e. considering a lower limit on the magnitude and its error), which  is more efficient in PanSTARRS. After applying this filter 184 candidates were left (out of the starting 439).
We show in Fig. \ref{fig:pan_color_plots}, the colour distribution of the candidates selected with the criteria just outlined. In the same plot, we also show the colours of the already known $5<z<6.3$ quasars detected in Pan-STARRS.

\subsubsection{Additional multi-wavelength information}

After selecting high-$z$ radio candidates based on the optical/IR magnitudes in the DES, PanSTARRS, and catWISE catalogues, we also used further multi-wavelength information, when available, from other public surveys to discard some contaminants that passed all the selection criteria outlined above.

In particular, we considered the `quick-look' images from the VLA Sky Survey (VLASS; \citealt{Lacy2020}). The typical RMS of these images is $\sim120$~\textmu Jy~beam$^{-1}$, which, in principle, is enough to detect at a 3$\times$RMS level even the faintest RACS-low sources we considered, assuming a typical spectral index of $\alpha_{\rm r}\sim0.7$ (S$_{\rm 3GHz} \sim 0.43$~mJy~beam$^{-1}$ for a S$_{\rm 888MHz} \sim$1~mJy~beam$^{-1}$ object). Given the better angular resolution ($\sim$4$''$) compared to RACS, this survey also has a better positional accuracy, even though it is limited to Dec.~$>-40^\circ$.
After the visual inspection of the 3~GHz VLASS images, we only kept the candidates where a $>$3$\times$RMS signal, if present, was within 1$''$ from the optical/NIR counterpart. We expect the rest to be spurious associations. In this way we discarded 113 objects.

After the selection based on colour cuts and a reliable radio association in the VLASS survey (if detected), we also considered any available NIR data from the Vista hemisphere Survey (VHS; \citealt{McMahon2021}) and the VISTA Kilo-degree Infrared Galaxy (VIKING; \citealt{Edge2013}). For sources with a NIR coverage, we compared their photometric SED to a quasar template and discarded the sources whose NIR and MIR data-points lied systematically above the emission expected from a quasar (see Fig. \ref{fig:discard_cand} for such an example). From the template comparison we eliminated 58 candidates in total, most of which are likely low-$z$ elliptical galaxies.

The total number of candidates selected with the optical/IR and radio criteria described above is 45. In particular, one source was selected from the EMU pilot survey (RACS~J2020$-$62) and another one (RACS~J2142$-$62) was selected in both RACS-low and EMU. We report the selected candidates in Tab. \ref{tab:list_confirmed} and \ref{tab:list_contaminants} together with the main optical and radio parameters considered during the selection.

Finally, we note that for one source, RACS~J0622$-$51, we obtained a more accurate radio position, with respect to the RACS-low one, based on dedicated observations with the Australia Compact Array (ATCA; see Sec. \ref{sec:radio_properties}). The new position of the radio source obtained from ATCA is not consistent with the optical counterpart (offset of 3.1\arcsec or $>10\sigma$ away; see Fig. \ref{fig:discard_cand}). For this reason, we classified this source as a non high-$z$ radio quasar.

\begin{table*}
    	\centering
 \small
 \caption{Properties of the 24 candidates confirmed to be at high redshift.}
  \addtolength{\tabcolsep}{-0.2em}
	\begin{tabular}{lcccccccccc}
	Target & R.A. & Dec. & drop & $z$ & $S_{\rm 888MHz}$ & mag-$z$ & drop & $z-W2$ & d$_{\rm ro}$ & Telescope\\
	 & (deg) & (deg) & (Survey) & & (mJy~beam$^{-1}$) & (mag AB) & & (AB -- Vega) & (arcsec) &\\
	\hline
	\hline
    RACS~J0036$-$37 & 9.035480   & $-$37.095232 & $r$-drop (D)      & 4.95$\pm$0.02 & 1.3$\pm$0.2   & 20.49$\pm$0.02 & 1.66 & 4.87 & 1.29 & AAT \\
    RACS~J0056$-$05 & 14.245799  & $-$5.140481  & $r$-drop (D)      & 4.82$\pm$0.03 & 1.5$\pm$0.2   & 21.34$\pm$0.05 & 1.22 & 4.46 & 0.67 & VLT \\
    RACS~J0110$-$39 & 17.684618  & $-$39.513237 & $r$-drop (D)      & 5.08$\pm$0.03 & 17.9$\pm$2.1  & 21.29$\pm$0.05 & 1.65 & 3.57 & 0.89 & VLT \\
    RACS~J0127$-$44 & 21.809644  & $-$44.912548 & $r$-drop (D)      & 4.96$\pm$0.02 & 17.9$\pm$2.0  & 20.67$\pm$0.03 & 1.42 & 3.22 & 1.26 & VLT \\
    RACS~J0202$-$17  & 30.618886  & $-$17.141042 & $r$-drop (P)     & 5.57$\pm$0.03 & 17.2$\pm$2.0  & 19.24$\pm$0.01 & 1.05 & 4.17 & 1.03 & AAT \\
    RACS~J0209$-$56 & 32.320529  & $-$56.447344 & $r$-drop (D)      & 5.61$\pm$0.03 & 17.8$\pm$1.8  & 20.90$\pm$0.02 & 1.98 & 3.73 & 0.39 & VLT \\
    RACS~J0320$-$35 & 50.089335  & $-$35.351148 & $i$-drop (D)      & 6.13$\pm$0.04 & 3.2$\pm$0.3   & 20.59$\pm$0.03 & 3.11 & 3.28 & 1.32 & GS \\
	RACS~J0322$-$18 & 50.560622  & $-$18.688189 & $i$-drop (D)      & 6.09$\pm$0.04 & 1.7$\pm$0.2   & 20.94$\pm$0.03 & 3.11 & 3.44 & 2.59 & GS \\
    RACS~J0606$-$27  & 91.605865  & $-$27.942459 & $r$-drop (P)     & 4.89$\pm$0.03 & 13.2$\pm$1.8  & 20.49$\pm$0.04 & 1.89 & 4.56 & 0.21 & AAT \\
    RACS~J1011$-$01  & 152.981424 & $-$1.514636  & $i$-drop (P)     & 5.56$\pm$0.03$^\dagger$ & 8.7$\pm$1.3 & 21.23$\pm$0.08 & 1.14 & 3.72 & 1.20 & GS \\
    RACS~J1042+04    & 160.642581 &   +4.210405  & $r$-drop (P)     & 4.72$\pm$0.02 & 1.5$\pm$0.3   & 21.16$\pm$0.09 & 1.74 & 3.40 & 1.21 & LBT \\
    RACS~J1321+24    & 200.260024 &   +24.896064 & $r$-drop (P)     & 4.78$\pm$0.06 & 4.6$\pm$1.0   & 21.04$\pm$0.06 & 1.31 & 3.16 & 0.53 & LBT \\
    RACS~J1322$-$13  & 200.526920 & $-$13.398532 & $r$-drop (P)     & 4.70$\pm$0.02 & 37.6$\pm$3.3  & 20.54$\pm$0.03 & 1.53 & 4.68 & 0.65 & NTT \\
    RACS~J1426$-$11  & 216.700593 & $-$11.000803 & $r$-drop (P)     & 4.73$\pm$0.05 & 11.0$\pm$0.7  & 20.91$\pm$0.05 & 1.24 & 3.55 & 0.56 & LBT \\
    RACS~J1505$-$06  & 226.286005 & $-$6.708236  & $r$-drop (P)     & 4.53$\pm$0.03 & 4.8$\pm$1.3   & 20.74$\pm$0.05 & 1.31 & 4.48 & 0.49 & LBT \\
    RACS~1634$-$01   & 248.603437 & $-$1.9576216 & $r$-drop (P)     & 4.88$\pm$0.02 & 1.8$\pm$0.3   & 20.25$\pm$0.04 & 1.37 & 2.87 & 0.40 & LBT \\
    RACS~J1638+08    & 249.729070 &   +8.324004  & $r$-drop (P)     & 4.77$\pm$0.02 & 5.2$\pm$1.5   & 20.56$\pm$0.04 & 1.62 & 4.17 & 1.26 & LBT \\
    RACS~J1645+38    & 251.363676 &   +38.131409 & $r$-drop (P)     & 4.69$\pm$0.04 & 2.0$\pm$0.2   & 21.12$\pm$0.07 & 1.39 & 3.02 & 1.21 & LBT \\
    RACS~J2020$-$62 & 305.170194 & $-$62.252554 & $i$-drop (D)      & 5.72$\pm$0.02$^\dagger$ & 1.1$\pm$0.1*  & 19.23$\pm$0.01 & 1.00 & 4.67 & 0.80 & VLT \\
    RACS~J2142$-$61 & 325.615547 & $-$61.465918 & $r$-drop (D)      & 4.74$\pm$0.03 & 4.1$\pm$0.9   & 19.85$\pm$0.01 & 1.07 & 4.02 & 0.45 & VLT \\
    RACS~J2240$-$10  & 340.007219 & $-$10.747596 & $r$-drop (P)     & 5.20$\pm$0.03 & 32.8$\pm$3.4  & 20.72$\pm$0.04 & 1.33 & 3.82 & 0.65 & LBT \\
    RACS~J2244$-$20  & 341.203286 & $-$20.130738 & $r$-drop (P)     & 5.41$\pm$0.02 & 1.5$\pm$0.2   & 20.64$\pm$0.04 & 2.28 & 4.84 & 0.42 & AAT \\
    RACS~J2252$-$51 & 343.029401 & $-$51.037899 & $r$-drop (D)      & 5.18$\pm$0.01 & 9.5$\pm$1.4   & 19.42$\pm$0.01 & 1.90 & 4.51 & 0.78 & AAT \\
    RACS~J2303$-$23 &  345.783262 & $-$23.400271 & $r$-drop (P)     & 5.03$\pm$0.03 & 10.9$\pm$1.5  & 19.51$\pm$0.02 & 1.42 & 3.34 & 0.54 & AAT \\
	\hline
	\hline
	\end{tabular}
\tablefoot{{\bf Col. (1, 2, 3):} name and coordinates of the source; {\bf Col. (4):} type of dropout and survey the source was selected from, in brackets; {\bf Col. (5):} spectroscopic redshift. The $\dagger$ symbol indicates sources whose redshift was estimated based on the fit of an emission line; {\bf Col. (6):} peak flux density at 888~MHz from the RACS source lists. The `*' indicates flux densities at 944~MHz from EMU; {\bf Col. (7):} $z$-band magnitude from DES or Pan-STARRS, depending on the survey from which the source was selected; {\bf Col. (8):} dropout $r-i$ or $i-z$ depending on the type of dropout in Col. 4; {\bf Col. (9):}  $z-W2$ colour, where $z$ is the magnitude in the $z$ filter (from DES or Pan-STARRS) in the AB system and $W2$ is in the Vega system; {\bf Col. (10):} radio-to-optical distance; {\bf Col. (11):} telescope used for the identification.}  
	\label{tab:list_confirmed}
\end{table*}

\section{Spectroscopic identification of the candidates}
\label{sec:identification} 

In order to identify the nature and the redshift of the 45 candidates selected from the cross-match of RACS-low with DES and Pan-STARRS, we performed spectroscopic observations (PI: Ighina) with different ground-based telescopes: Telescopio Nazionale Galileo (TNG; project AOT47\_3), Anglo-Australian Telescope (AAT; project O23B005), Gemini-South (GS; projects GS-2021-DD-112 and GS-2023-DD-108), Very Large Telescope (VLT; project 111.24Q5) and Large Binocular Telescope (LBT; projects 2022B\_37 and 2023B\_22).
In one case, RACS~J1322$-$13, spectroscopic observations with the ESO Faint Object Spectrograph and Camera (EFOSC2; \citealt{Buzzoni1984}) at the New Technology Telescope (NTT) were already available, but not published, (project 108.226E.002; PI Decarli). 

Almost all of the spectroscopic observations were performed using long-slit spectroscopy. The only exceptions were the observations with the AAT, for which we used the IFU KOALA instrument and, therefore, spectroscopically observed the entire region around the targets.

In order to analyse the observations obtained with different telescopes and instruments, we always followed a standard data reduction detailed in the manual or cookbook of each instrument\footnote{The links to the manuals/guides followed for the data reduction of each instrument are as follows: the Gemini--GMOS cookbook can be found at \url{https://gmos-cookbook.readthedocs.io/en/latest/};  the AAT--KOALA manual can be found at \url{https://aat.anu.edu.au/science/instruments/current/koala/manual}; the VLT--FORS2 manual can be found at \url{https://www.eso.org/sci/facilities/paranal/instruments/fors/doc}; the LBT-MODS page describing the data reduction can be found at \url{http://pandora.lambrate.inaf.it/sipgi/}.}. During the analysis, we made use of the Image Reduction and Analysis Facility (\texttt{IRAF}; \citealt{Tody1986,Tody1993}) software for the reduction of the Gemini--GMOS (with the specific \texttt{Gemini} package), TNG--DOLORES and NTT--EFOSC2 observations, the \texttt{2dfdr} software \citep{AAO2015} for AAT--KOALA spectra, the \texttt{SIPGI} \citep{Gargiulo2022} software for LBT--MODS data and the \texttt{EsoReflex} software \citep{Freudling2013} for the VLT--FORS2 observations. 
In general, for the reduction of all the observations, we performed the following steps: bias subtraction, flat fielding correction, cosmic rays removal, wavelength calibration and background subtraction. Finally, an absolute flux calibration using the observation of a standard star was applied for sources observed with the LBT and VLT telescopes. In the other cases, we normalised the extracted 1D spectra to the corresponding DES or Pan-STARRS magnitudes.

Out of the 40 candidates spectroscopically observed, 24 turned out to be high-z radio quasars (Fig. \ref{fig:pan_color_plots}). While in Fig. \ref{fig:discard_cand} we report an example of a candidate not confirmed to be at high redshift. We note that some of the new high-$z$ quasars reported in this study have also been independently identified by other projects during the course of this work: RACS~J0056$-$05 \citep{Yang2023}, RACS~J0209$-$56 \citep{Wolf2024}, RACS~J0322$-$18 \citep{Banados2023} and RACS~J2020$-$62 \citep{Wolf2024}. Moreover, the spectra of RACS~0202$-$17, RACS~J0209$-$56, RACS~J0320$-$35, RACS~J0322$-$18 and RACS~J1011$-$01 were already published in \cite{Ighina2023, Ighina2024b}.

In order to estimate the redshift of each source, we considered the composite spectra derived by \cite{Banados2016} from a sample of $z\gtrsim5.6$ quasars and performed a fit to the observed optical spectrum of the confirmed candidates. In particular, \cite{Banados2016} reports three different composite spectra (see their fig. 10), based on the intensity of the Ly$\alpha$ emission line: (i) the median spectrum of the 10\% objects with the largest rest-frame equivalent width (Ly$\alpha$+\ion{N}{V}); (ii) the median spectrum of all the $\sim120$ high-$z$ quasars they discovered; (iii) the median spectrum of the 10\% with the lowest rest-frame equivalent width (Ly$\alpha$+\ion{N}{V}).
During the fit we considered all three reference spectra and used the one with the smallest $\chi^2_{\rm red}$ to derive the best-fit redshift. The typical uncertainty of this method is about $\sim$0.03, which corresponds to the width of the  Ly$\alpha$ drop. For all the sources we obtained satisfactory fit, except in one case, RACS~J2020$-$62. The spectrum and the Ly$\alpha$ line of this quasar are strongly affected by intrinsic absorption. For this reason, we estimated the redshift from the OI1303\AA \, emission line visible at 8754~\AA, obtaining $z=5.72\pm0.02$. Moreover, in the case of RACS~J1011$-$01, we considered both the estimate derived from the template fitting as well as the \ion{C}{IV} and [\ion{C}{III}] emission lines in dedicated NIR observations (see \citealt{Ighina2024b}).

\begin{figure*}
   \centering
   \includegraphics[width=\hsize]{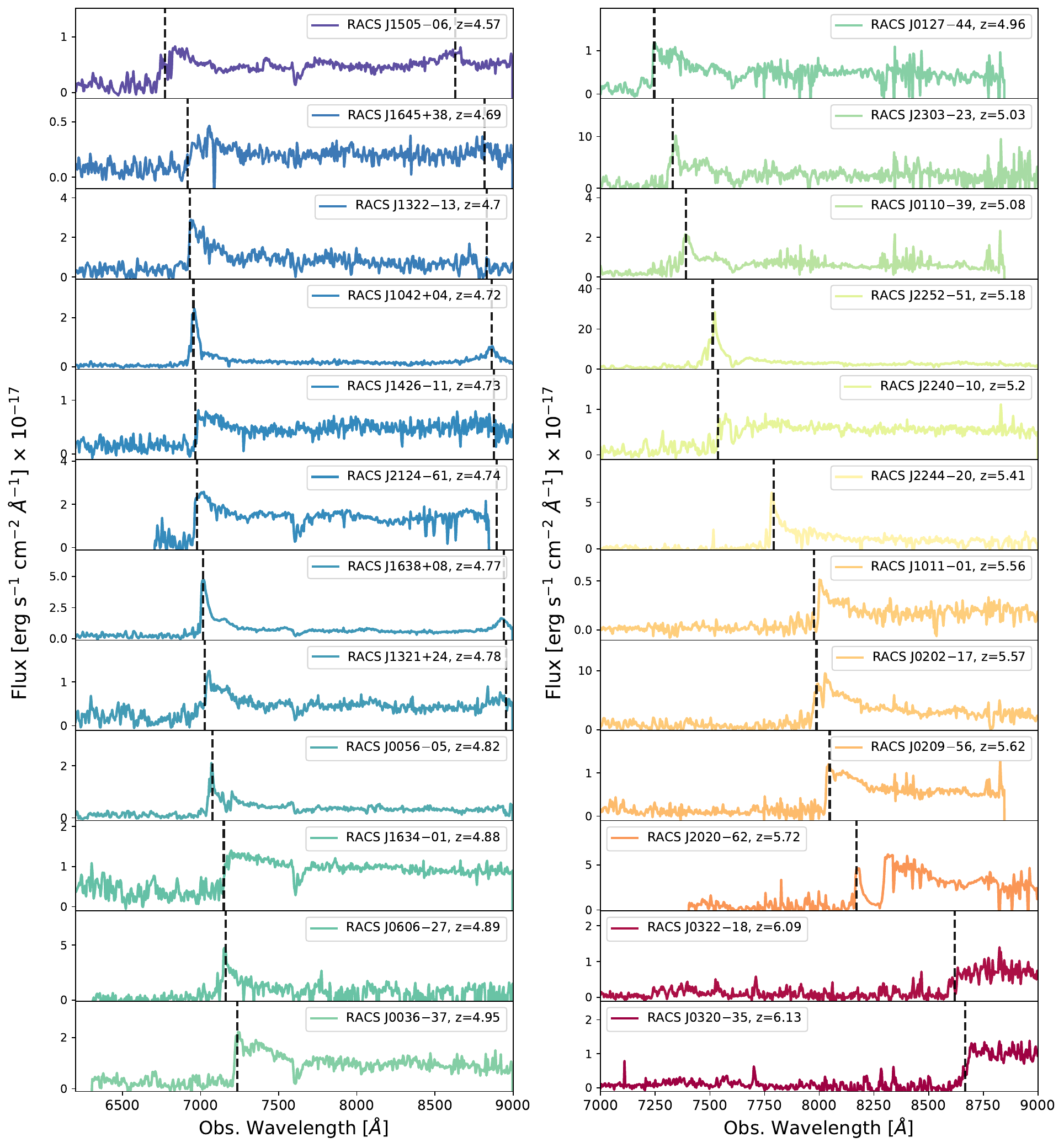}
  \caption{Optical spectra centred on the Ly$\alpha$ emission line and dropout of the newly discovered $4.53\leq z\leq6.13$ radio quasars from the RACS sample. The sources are ordered in increasing redshift and the dashed vertical line in each plot indicates the expected wavelength emission of the Ly$\alpha$1215 (and CIV1549) in the observed frame. The absorption feature observed at $\sim$7600~\AA \, in several spectra is due to the \ion{O}{II} in our atmosphere.}
    \label{fig:opt_spec_sample}
\end{figure*}

As clear from Fig. \ref{fig:opt_spec_sample}, while our selection targeted $z>5$ quasars, the majority of the newly discovered sources is at $4.5<z<5$. This is mainly due to the large variety of shapes in the rest-frame UV emission of quasars (e.g. in terms of Ly$\alpha$ strength), which results in sources at different redshift ($\Delta z\sim0.3$) having similar photometric colours and dropout. For example, RACS~J1505$-$06 at $z=4.53$ and RACS~J2240$-$10 at $z=5.20$ have a similar dropout in PanSTARRS ($r-i\sim1.3$) despite being at very different redshift. The large number of $z<5$ sources in our final sample indicates that the fixed cuts adopted in their photometric selection were conservative enough to recover the majority of the $z>5$ quasars, even the ones with small dropouts. As described below, we only focus on the $z>5$ objects when it comes to building a complete sample for future statistical works, since we expect the completeness of our selection to quickly decreases at $z\lesssim4.9-5$. Nevertheless, in this paper we also present their X-ray and radio properties since they are not discussed in other works. 

\begin{figure*}
   \centering
	\includegraphics[width=0.49\hsize]{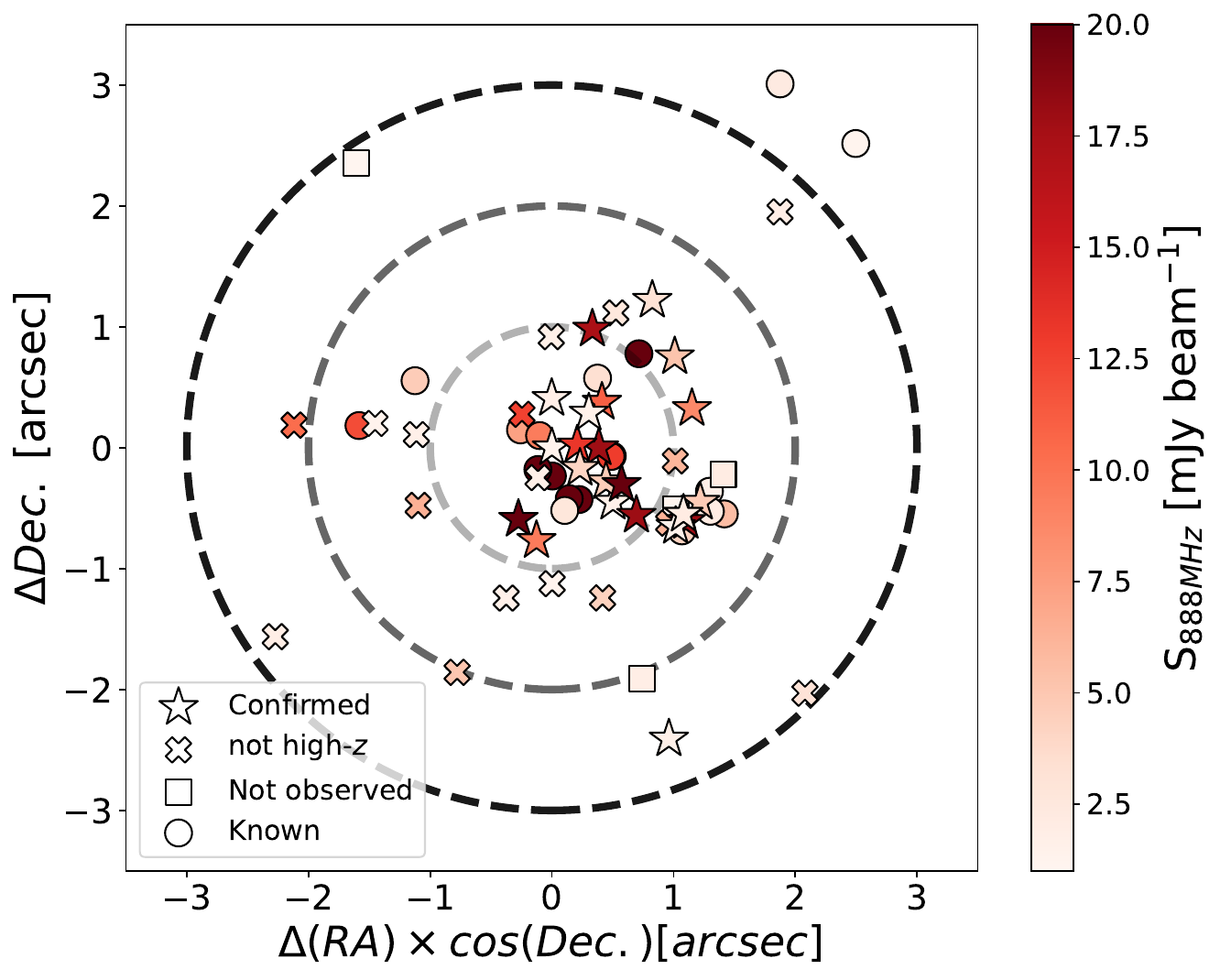}
	\includegraphics[width=0.49\hsize]{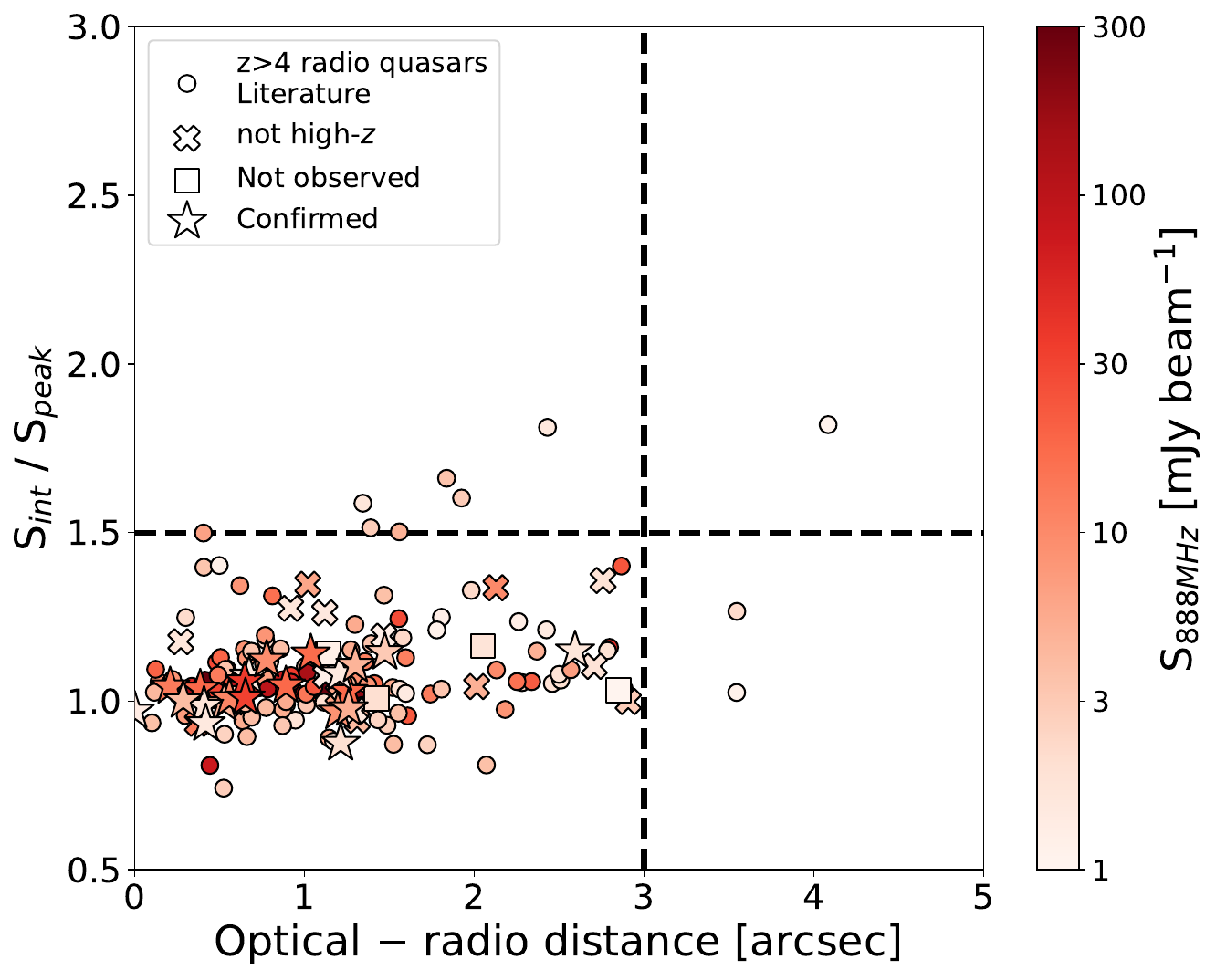}
\caption{{\bf Left panel:} Offset in declination as a function of the offset in right ascension between the optical positions from DES or Pan-STARRS and the radio positions from RACS-low for the high-$z$ candidates as well as the already known $z>5$ radio quasars with the same flux limits. The colour of the data points reflects the peak flux density of the given source in the RACS-low source lists as shown by the colour-bar. The dashed circles represent a separation of 1, 2 and 3$''$. {\bf Right panel:} Distribution of the integrated-to-peak flux density ratios from RACS-low as a function of the optical-to-RACS position for the $z>4$ radio quasars currently known in the RACS area. The colour of the markers indicates the peak flux density reported in the RACS-low source lists. The dashed black line are the threshold used during our selection.}
    \label{fig:radio_opt_distance}
\end{figure*}

Although, in principle, the radio association should have prevented stars from being selected as potential candidates (albeit some rare cases of radio emitting stars exist; e.g., \citealt{Gloudemans2023b}), many of the sources not confirmed to be at high redshift were actually stars. Indeed, at least 9 out of the 17 contaminants in our selection, turned out to be stars (mainly brown dwarfs, see, e.g., Fig. \ref{fig:radio_opt_distance}, left). The remaining sources are either low-$z$ radio galaxies, or their exact nature could not be determined, apart from excluding the high-$z$ nature due to the lack of a sharp drop in the flux. 
Therefore, in most of these cases the radio-optical association must be spurious, that is, with the stars falling within 3\arcsec of a RACS radio source by chance, although some stars have been found to be radio emitters \citep[e.g.][]{Gloudemans2023b,Driessen2024}. As expected, most of these sources are faint in the radio band (S$_{\rm 888MHz}<2$~mJy) or at Dec.~$<-40^{\circ}$ and could not be detected or observed as part of the VLASS survey. An example of spurious radio association is the object RACS~J0622$-$51, which, when observed with higher resolution radio observations, has an optical position not consistent with the associated RACS-low source (see Sec. \ref{sec:radio_properties} and appendix \ref{sec:appenA}). On the other hand, a particularly interesting example of a candidate identified as a brown dwarf with a RACS-low flux of $\sim13$~mJy~beam$^{-1}$ less than 0.4$''$ from the optical counterpart is RACS~J1521$-$21 (see spectrum in Fig. \ref{fig:discard_cand}).

If we compare the radio-to-optical distance of all the selected candidates (see Fig. \ref{fig:radio_opt_distance}, left panel), it is clear how the majority (15 out of 24) of the confirmed high-$z$ sources have a radio association within 1$''$ from their optical position. Whereas, only 3 sources out 17 that have not been confirmed to be at high redshift have a radio signal in the RACS-low survey at a distance $<1''$.

At the same time, we can also estimate the likelihood of finding a non-radio, high-$z$ quasar close, by chance, to an unrelated radio source in RACS. To this end, we considered the bright end ($M_{\rm 1450}<-27.3$; corresponding to mag$z$~$<$~21.3 at $z\sim5$) luminosity function of $z\sim$5 quasars derived by \cite{Niida2020} and estimated the expected number of bright quasars per square degree in the redshift range $z=5-6$, resulting in $\sim$0.01~deg$^{-2}$. 
Multiplying this number for the area used in our search, $\pi r^2= \pi(3'')^2 \sim 2\times 10^{-6}$~deg$^{2}$, we obtain the probability that a random position is within 3$''$ of a $z\gtrsim5$ quasar: $\sim3\times10^{-8}$. 
If we consider also the number of radio sources used for the association ($\sim1.3\times10^6$ in 16000~deg$^2$; e.g. \citealt{McConnell2020}), the probability of selecting one $z\gtrsim5$ quasar based on a spurious radio association in RACS is about 4\%. Since the majority of the confirmed quasars have an association within $1.5''$ (for which we would expect $\sim1$\% of spurious associations) and, more importantly, the association is supported by additional radio data, it is unlikely that any of the quasars discovered is not responsible for the observed radio emission.

\section{RACS sample of \texorpdfstring{high-$z$}{} radio quasars}
\label{sec:final_sample}

In this section we discuss the sample selected from the RACS radio survey. In particular, we estimated the completeness of our selection for $z>5$ sources, which is given by the combination of the optical and the radio selection processes. We also compared the distribution of the RACS sources comapred to other high-$z$ samples of radio quasars from the literature. 

\subsection{Completeness of the selection}

With the optical/NIR criteria reported above, we recovered about 96\% of the $z>5$ quasars (not necessarily radio) already known in the DES area and about 85\% of those $z>5$ in the Pan-STARRS area. The two sources not recovered in the DES selection were not selected due to their class\_star\_z parameter being $<0.8$. Whereas, in the case of Pan-STARRS quasars, the majority (51 objects, $\sim$15\%) of the quasars were not selected due to dropout (\texttt{mag\_psf\_r~--~mag\_psf\_i}~$>1.1$ or \texttt{mag\_psf\_i~--~mag\_psf\_z}~$>1.1$) and/or the point-like criteria (\texttt{mag\_psf\_i~--~mag\_kron\_i}~$<0.07$ or \texttt{mag\_psf\_z~--~mag\_kron\_z}~$<0.1$). 

Since the sources discovered in the literature cannot be considered a complete sample, as demonstrated in the previous section, the fraction of $z>5$ quasars recovered is not a good estimate to the actual completeness of our sample. In order to have a more accurate estimate of the optical completeness of our selection we considered the region of the sky given by $09 < {\rm RA} < 16$ and $0^\circ<{\rm Dec.}<60^\circ$, which is fully covered by the Sloan Digital Sky Survey (SDSS; \citealt{Abdurrouf2022}) and for which the level of completeness of $z<5.5$ bright (mag$z$~$<21$) quasar is close to 100\%. 
Indeed, there have been several studies aimed at uncovering all the $z>5$ quasars present in this region in order also to constrain the luminosity function of these objects (see, e.g., \citealt{Fan1999,Fan1999b,Fan200,Jiang2008,Wang2016,Niida2020}). Moreover, as expected, no new $z>5$ radio quasars were discovered in this same area in our selection, apart from two $z\sim4.7$ quasars with mag$z$~$>21$.
For these reasons, we considered all the bright (mag$z$~$<21$) quasars at $5<z<5.5$ available in the footprint of the SDSS between 09h$~<{\rm RA}<~$16h, a total of 82 objects, and checked their colours in the Pan-STARRS survey, since there is no overlap with the DES survey. After applying the criteria reported in Fig. \ref{fig:pan_criteia} as well as the $W2$ selection, we recover 68 sources, corresponding to the 83\% of the initial sample. We consider this value as bona-fide estimate of the optical completeness in the RACS sample, also for the selection of sources at $z>5.5$ and for sources selected in DES, for which a complete reference sample is not available yet. But we stress that, technically speaking, this is an upper limit on the optical/NIR completeness of the overall sample.

A first discussion on the completeness of the radio selection using the RACS source lists was already presented in \cite{Ighina2023} and here we report a brief summary. To assess the completeness of the RACS-low survey when it comes to point-like and relatively faint radio sources, we considered the ASKAP observations of the 23$^{\rm rd}$ GAlaxy Mass Assembly (G23; \citealt{Driver2011}) field described in \cite{Gurkan2022}. This field covers 83~deg$^2$ in the southern hemisphere (centred at RA = 23h and Dec = $-$32$^{\rm o}$) at the same frequencies of first scan of RACS-low (888~MHz) and with a similar angular resolution ($\sim$10$''$), but with a sensitivity about an order of magnitude deeper (RMS$\sim$38~\textmu Jy~beam$^{-1}$).
In order to estimate the detection completeness of the RACS-low survey, we checked how many sources with S$_\mathrm{int}$/S$_\mathrm{peak} < 1.5$ and with $1<$~S$_\mathrm{peak}<100$~mJy~beam$^{-1}$ (i.e., similar to the targets we selected) in the G23 ASKAP observation are reported in the source lists of RACS-low used in this work as a function of radio flux density. 
From this comparison, we found that the overall radio selection is about 78\% complete for S$_\mathrm{888MHz}>1$~mJy~beam$^{-1}$ increasing to 93\% for S$_\mathrm{888MHz}>2$~mJy~beam$^{-1}$.

Finally, in order to estimate the fraction of sources missed from the radio-optical association of 3$''$ and the requirement of compact radio sources (S$_\mathrm{int}$/S$_\mathrm{peak} < 1.5$), we considered all the currently known quasars at $z>4$ detected by the RACS-low survey (164 in total; to the best of our knowledge). The fraction missed due to either of these criteria is $\sim$6\% ( 10 objects)\footnote{By considering only the $z>5$ radio quasars in RACS-low (22 in total), the fraction missed due these criteria is $\sim9$\% (3 sources). Given the low statistics, we consider the estimate derived from $z>4$ quasars as more reliable.}, see Fig. \ref{fig:radio_opt_distance} for the distribution of the optical--radio distances and of the integrated/peak flux ratio. There are three sources that have been missed by both criteria and the majority of sources not included has a surface brightness $<3$~mJy~beam$^{-1}$ This suggests that the fractions of sources not selected with the different radio criteria are not independent and a larger incompleteness is expected for faint sources. For this reason, we assume that the completeness of our radio selection criteria is 75\% for $>1$~mJy~beam$^{-1}$ (90\% for $>2$~mJy~beam$^{-1}$). If the selections were completely independent to each other, these values would be only slightly smaller, by $\sim$2\% and, therefore, would not significantly change the results presented throughout this work.


\afterpage{\begin{landscape}
    
\begin{table}

 \caption{Radio quasars from the literature within the RACS+DES and RACS+Pan-STARRS area with mag$z$~\textless~21.3 and S$_{\rm 888MHz}>1$~mJy.}
 
 \addtolength{\tabcolsep}{-0.12em}
	\centering
\begin{tabular}{c c c c c c c l} 
 \hline
 Name & R.A. & Dec. & $z$ & S$_{\rm 888MHz}$  & mag$z$ & Comments & Reference\\
 & (deg) & (deg) & & mJy~beam$^{-1}$ & (mag AB) & \\
  \hline

RACS~J0001$-$03   & 0.448498  &$-$3.879857  & 5.27  & 1.2$\pm$0.4  & 20.80$\pm$0.04 & not detected in RACS & \cite{Yang2023} \\
RACS~J0131$-$03    & 22.863917  &$-$3.350053  & 5.18  & 23.2$\pm$2.5  & 18.12$\pm$0.01 &         --           & \cite{Wang2016} \\
RACS~J0141$-$54    & 25.385002  &$-$54.463861 & 5.00  & 167.4$\pm$14.4& 21.20$\pm$0.04 &          --          & \cite{Belladitta2019} \\
RACS~J0309+27      & 47.447667  & 27.299333   & 6.10  & 32.5$\pm$3.3  & 21.18$\pm$0.30 &         --           & \cite{Belladitta2020}\\
RACS~J0341$-$00    & 55.424417  &$-$0.803539  & 5.68  & 3.0$\pm$0.2   & 20.35$\pm$0.05 & not detected in RACS & \cite{Banados2015} \\ 
RACS~J0741+25      & 115.477980 & 25.34156    & 5.19  & 2.1$\pm$1.4   & 18.39$\pm$0.01 & $S_{\rm int}/S_{\rm peak}>1.5$ &  \cite{McGreer2009} \\
RACS~J0836+00	   & 129.182750 & 0.914790    & 5.82  & 2.6$\pm$1.2   & 18.82$\pm$0.02 & Opt.--Radio distance $>3''$ & \cite{Fan2001} \\
RACS~J0901+16      & 135.386042 & 16.251897   & 5.63  & 5.6$\pm$1.3   & 20.60$\pm$0.04 &         --          & \cite{Banados2015} \\
RACS~J1013+35      & 153.407793 & 35.3138347  & 5.03  & 1.2$\pm$0.2   & 20.30$\pm$0.03 & Opt.-Radio distance $>3''$ & \cite{Gloudemans2022} \\ 
RACS~J1026+25      & 156.598453 & 25.716513   & 5.25  & 290.4$\pm$24.3& 19.97$\pm$0.05 &         --          & \cite{Sbarrato2012} \\
RACS~J1034+20      & 158.577705 & 20.550059   & 5.01  & 4.3$\pm$1.0   & 19.86$\pm$0.02 &         --          &  \cite{McGreer2009} \\ 
RACS~J1111+05      & 167.798840 & 5.607160    & 5.24  & 7.10$\pm$1.3  & 20.70$\pm$0.04 & mag\_z\_psf -- mag\_z\_kron~$>0.07$ & \cite{Yang2023} \\
RACS~J1146+40      & 176.740840 & 40.619110   & 5.01  & 10.0$\pm$1.4  & 19.41$\pm$0.03 &         --          & \cite{Ghisellini2014} \\ 
RACS~J1427+33	   & 216.910792 & 33.211667   & 6.12  & 2.7$\pm$0.3   & 21.00$\pm$0.10$^*$  &   not detected in optical    & \cite{McGreer2006} \\
RACS~J1621+11      & 245.369399 & 11.557065   & 5.05  & 1.6$\pm$0.5  & 20.90$\pm$0.16 & not detected in RACS & \cite{Yang2023} \\
RACS~J1629+10	   & 247.488671 & 10.006531   & 5.00  & 67.3$\pm$6.1  & 20.47$\pm$0.05 &        --           & \cite{Caccianiga2019} \\ 
RACS~J1702+13      & 255.688789 & 13.017273   & 5.47  & 12.0$\pm$1.7  & 20.60$\pm$0.06 & mag\_i\_psf -- mag\_i\_kron~$>0.1$  & \cite{Khorunzhev2021} \\
RACS~J2201+23      & 330.281708 & 23.643875   & 5.83  & 2.1$\pm$0.8   & 20.42$\pm$0.04 & err\_mag\_Y~$>0.4$  & \cite{Gloudemans2022} \\
RACS~J2329$-$15	   & 352.403400 &$-$15.337300 & 5.84  & 23.9$\pm$2.8  & 21.08$\pm$0.01 &         --          & \cite{Banados2018b} \\
RACS~J2329+30      & 352.413750 & 30.064106   & 5.24  & 4.67$\pm$0.7  & 19.17$\pm$0.02 &         --          & \cite{Wang2016} \\
RACS~J2329$-$20    & 352.469917 &$-$20.010886 & 5.04  & 3.88$\pm$1.0  & 18.47$\pm$0.01 &         --          & \cite{Wenzl2021, Onken2022} \\
RACS~J2344+16      & 356.139583 & 16.887911   & 5.00  & 12.8$\pm$1.8  & 18.74$\pm$0.01 &         --          & \cite{Wang2016} \\

 \hline
  \end{tabular}
    \tablefoot{ {\bf Col. (1, 2, 3):} name and coordinates of the source; {\bf Col. (4):} redshift; {\bf Col. (5):} peak flux densities at 888~MHz from the RACS-low source lists; {\bf Col. (6):} $z$-band magnitude from DES or Pan-STARRS. The $^*$ indicates sources for which the magnitude was computed from other filters available in the literature; {\bf Col. (7):} reason why the specific quasar was not recovered in our selection;  {\bf Col. (8):} reference for the discovery of the source.}  
  \label{tab:list_known}
\end{table}

\end{landscape}
}

\subsection{The final RACS sample}

{As mentioned before, we only consider $z>5$ quasars as part of the final RACS sample, for which we expect the completeness of the selection to be relatively high.} By considering similar $z>5$ quasars with S$_\mathrm{888MHz}>1$~mJy~beam$^{-1}$ and mag$z$~$<21.3$, and by applying the optical and radio selection criteria listed in the previous section, we recover 12 out of the 22 radio quasars currently reported in the literature. We list in Table \ref{tab:list_known} the radio quasars already discovered in the same area and at the same optical and radio flux density limits as our selection. If an already known source is not selected by our criteria, we also indicate the reason. In order to confirm whether a known $z>5$ quasar with mag$z$~$<21.3$ was detected in RACS or not, we checked the images centred on the optical position of all such sources reported in the literature, assuring a detection of faint sources slightly above the 3$\sigma$ (i.e., not reported in the source lists). 

In particular, six high-$z$ radio quasars are not selected because of their radio association. As expected, all of them have a flux density between 1 and 3~mJy. Two of these quasars were not selected due to their large optical-to-radio offset ($\sim$3.5$''$), three because the RMS of the image was too high to select a $S_{\rm 888MHz}\sim1-3$~mJy object and the last one was not selected due to its S$_\mathrm{int}$/S$_\mathrm{peak}$, being $\sim1.6$. 

At the same time, four already known $z>5$ radio quasars were not selected from the Pan-STARRS survey due to their optical properties.
One of the sources, RACS~J1427+33 at $z=6.12$, was not selected because it is not reported in the Pan-STARRS catalogue, probably due to the presence of a contaminating nearby bright object (at $\sim3''$). Indeed, \cite{McGreer2006} discovered this quasar thanks to the FLAMINGOS Extragalactic Survey (FLAMEX; \citealt{Elston2006}), which covers a 4.1~deg$^{2}$ region in the $J$ and $K_s$ filters. For this object we report in Table \ref{tab:list_known} the magnitude in the $z$ filter computed based on the one in the filter $J$ and assuming a slope of the spectrum $\alpha_{\nu}^{\rm o} = 0.44$ \citep[e.g.][]{Vandenberk2001}. One target was not recovered since it is not detected in the $Y$-filter of Pan-STARRS (i.e., err\_mag\_Y~$>0.4$) and two high-$z$ quasars were not selected due to the point-like requirement, either in the $i$ or $z$ filter.

The overall completeness of the selection criteria adopted here, obtained from combining the radio and optical completeness derived from reference samples, is 0.83~$\times$~0.75~=0.62, which increases to 0.75\% for $S_{\rm 888MHz}>2$~mJy~beam$^{-1}$. Based on the 22 $z>5$ sources (newly discovered and from the literature) recovered from our criteria, we would expect a total of 22/0.62~$\sim35-36$ radio quasars with the same optical and radio flux limits in the area we considered. Including the eleven radio quasars already reported in the literature that were not recovered by our selection, we still expect to be missing $\sim$3-4 objects, resulting in a final completeness of the RACS $z>5$ sample of $\sim90$\%.
We also note that, while there are still 4 objects without spectroscopic identification, only a fraction is expected to be at high redshift. Indeed, by assuming the efficiency of this work in selecting $z>5$ radio quasars ($\sim$25\%) the expected number of true $z>5$ is $\sim1$. Since this corresponds to $\sim3\%$ of the final RACS sample, well within the Poisson errors, we do not include them in the following discussion.

Combining all the already known $z>5$ radio quasars from the literature (22) with the new $z>5$ ones identified in this work (11), the final sample built from the combination of RACS, DES and Pan-STARRS is composed by 33 objects that span a redshift range $5<z<6.13$. 
In Fig. \ref{fig:sky_distr_sample} we report the sky distribution of all the high-redshift sources detected in RACS sample. The $z>5$ radio quasars in the sample are distributed evenly in the Pan-STARRS (27 in 12850~deg$^2$) and DES (10 in 5000~deg$^2$) areas, with four sources covered by both, suggesting that the two selections have a similar completeness.

\subsection{Comparison with other \texorpdfstring{high-$z$}{} radio samples}

Being a radio and optically selected sample, it is valuable to compare its redshift and luminosity distribution to other samples of high-$z$ radio quasars.

\begin{figure*}
   \centering
	\includegraphics[width=0.49\hsize]{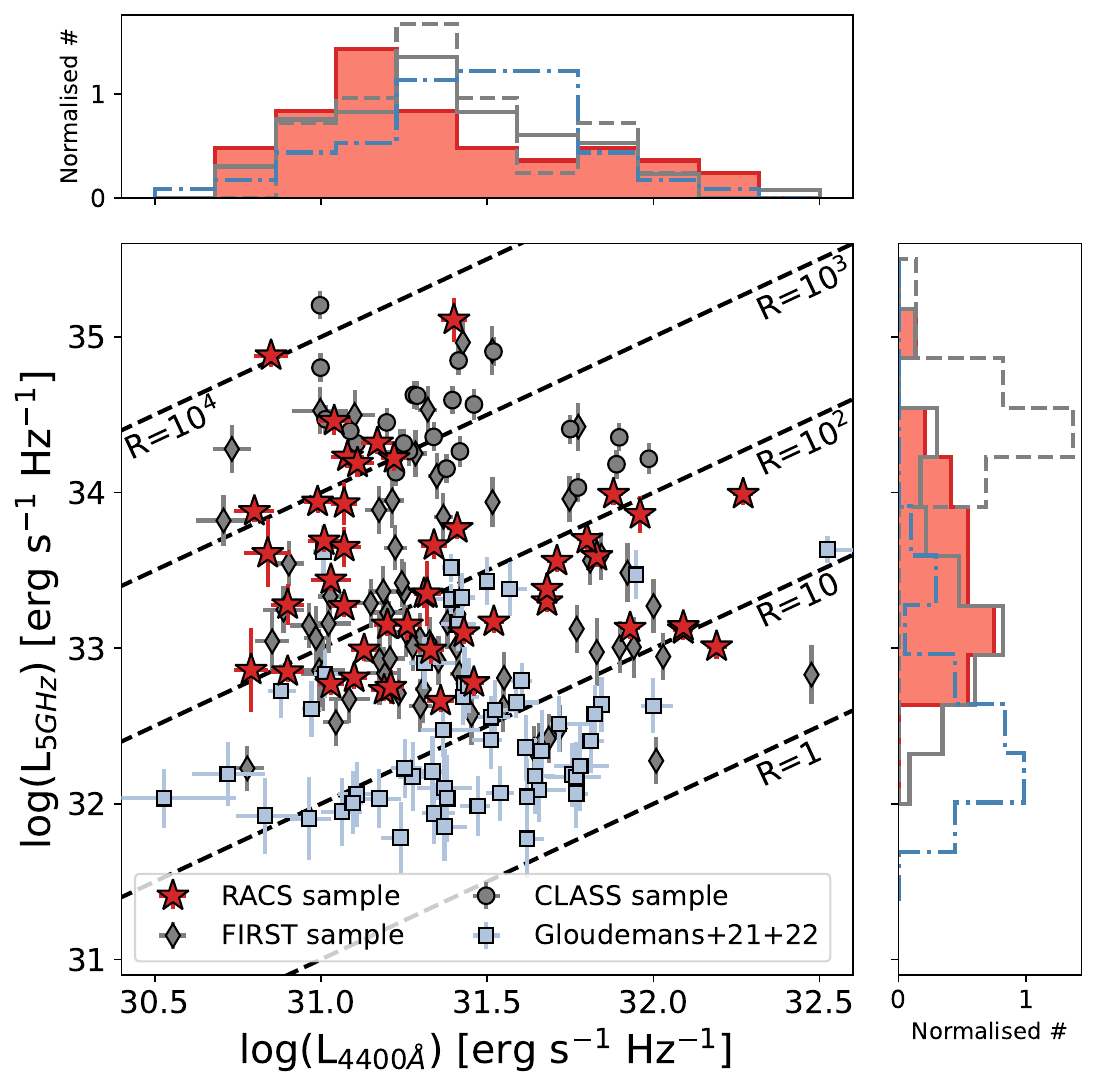}
	\includegraphics[width=0.49\hsize]{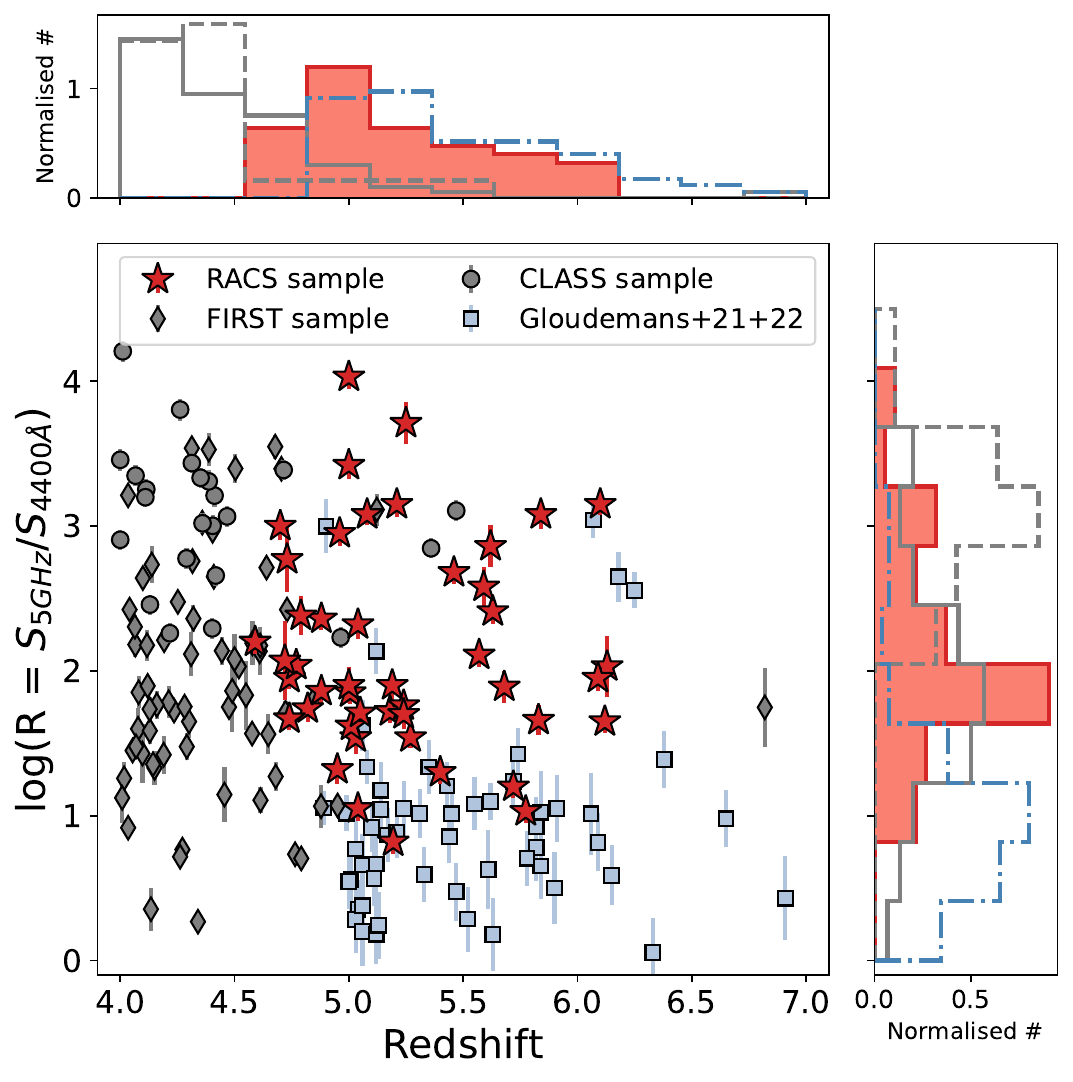}

\caption{\small {\bf Left panel:} Radio (5~GHz) and optical (4400~\AA) monochromatic luminosity distributions of the high-$z$ sample selected with RACS. The sample includes the new high-$z$ quasars discovered during this project as well as the $z>5$ radio quasar already reported in the literature. The newly discovered quasars at $z<5$ are reported in light blue, whereas the remaining $z>5$ RACS quasars are reported in dark blue. Grey diamonds and circles indicate the FIRST sample described in {\protect \cite{Caccianiga2024}} and the CLASS sample described in {\protect \cite{Caccianiga2019}}, respectively.  The light-blue squares are the high-$z$ radio quasars detected in LoTSS (from \citealt{Gloudemans2021,Gloudemans2022}). {\bf Right panel:} radio loudness, R~$=S_{\rm 5GHz}/S_{4400\AA}$, and redshift distribution of the RACS sample, together with the FIRST and CLASS ones. The symbols are the same as in the left panel. The histograms in both panels show the distribution of the given parameter for each sample (CLASS = dashed, FIRST = solid empty, LoTSS = dashed-dotted and RACS = filled histogram).} 
    \label{fig:RACS_sample_distr}
\end{figure*}

In Fig. \ref{fig:RACS_sample_distr} we report the monochromatic radio and optical luminosity distribution of the sample (left panel) as well as the redshift and radio-loudness parameter distributions (right panel). This last parameter is defined as $\rm R~=S_{\rm 5GHz}/S_{4400\AA}$ in the rest frame \citep{Kellermann89} and quantifies the strength of the radio emission compared to the optical/thermal one. In the same plots, we also show the distributions of other samples of high-$z$ radio quasars from the literature: the CLASS sample \citep{Caccianiga2019}, the FIRST sample (\citealt{Caccianiga2024}) and the radio quasars detected in LoTSS \citep{Gloudemans2021,Gloudemans2022}. The CLASS sample was built from the combination of Pan-STARRS together with the Cosmic Lens All Sky Survey \citep{myers2003}, which includes sources with $S_{\rm 5GHz}>30$~mJy and $\alpha_{1.4}^{5}<0.5$. In particular, a dedicated search for $z>4$ radio quasars led to an almost complete ($\sim95$\%) sample of 24 sources with mag~$<$21 (in the filter longwards the Ly$\alpha$ drop, which depends on the redshift) over an area of $\sim$13100~deg$^{-2}$ at $|b|>20^\circ$. Given the high radio flux limit and the criterion on the radio spectral index, the CLASS sample contains mostly blazars, as demonstrated by the X-ray analysis of the full sample \citep{Ighina2019}. 
The FIRST sample, instead, was built considering the $z>4$ quasars in the SDSS ($09 < {\rm RA} < 16$ and $0^\circ<{\rm Dec.}<60^\circ$), which, as mentioned before, should be complete for sources with mag~$<$~21, detected in the Faint Images of the Radio Sky at Twenty-Centimetres (FIRST; \citealt{Becker1995}) radio survey. The radio flux limit of this sample is 0.5~mJy~beam$^{-1}$ at 1.4~GHz (i.e., $3~\times$~RMS of the survey) and is composed by 73 radio quasars that cover an area of $\sim$5215~deg$^2$. 
The final comparison sample corresponds to all the $z>5$ quasars from the literature detected in the 2$^{\rm nd}$ data release of the LoTSS survey (38; \citealt{Gloudemans2021}) as well as 24 newly discovered $z\gtrsim4.9$ radio quasars selected from the combination of LoTSS and the DESI Legacy Imaging Surveys \citep{Dey2019}. We note that while the RACS, CLASS and FIRST sample are built with both a radio and optical flux limit, the sources detected in LoTSS do not have a defined optical threshold, even though most of them were discovered in similar optical/NIR surveys as the other samples.

To compute the optical and radio luminosities for the different samples, we proceeded as described below. For the sources detected in LoTSS, we used the measurements and the method reported in \cite{Gloudemans2021,Gloudemans2022}. In particular, we considered $\alpha_{\rm r}=0.3$ for the radio spectral index (with uncertainty $\sigma_{\alpha}=0.2$) if not reported. For the three remaining samples, we considered the optical/IR magnitudes in the $z$ and $W1$ filters in order to estimate the spectral index of the optical spectrum. Then, using each spectral index and the $W1$ magnitude, we computed the 4400~\AA \, rest-frame luminosity, while from the $z$ magnitude we computed the 2500\AA \, luminosity. We chose these two filters since all the sources have been detected in both of them, and they are the closest to the 2500 and 4400~\AA \, rest-frame wavelength, respectively. To compute the radio luminosities at 5~GHz (rest frame) of the CLASS and FIRST sources, we considered the flux density at 1.4~GHz, from either the FIRST or NVSS survey, and the spectral index computed between 1.4~GHz and 150~MHz (from either the TGSS or LoTSS survey). 
We can expect variability to potentially affect the value of the spectral index observed, especially for the most radio-luminous sources. However, \cite{Caccianiga2019} estimated the amplitude of variation for the CLASS sample  to be $\sim$14\% at 1.4~GHz in the observed frame, hence not enough to significantly change the values reported here. For sources without a 150~MHz detection (about half of the sample), we considered the median spectral index of the detected quasars, $\alpha_{0.15}^{1.4}\sim0.1$ {\bf (with uncertainty $\sigma_{\alpha}=0.2$)}. 
We note that only by assuming a significantly different value for $\alpha_{0.15}^{1.4}$ the overall distribution of the sample changes. Indeed, by assuming $\alpha_{0.15}^{1.4}\sim0.3$, the mean variation in luminosity of the sample corresponds to $\sim$5\%. In the case of the RACS sample, we used the spectral indices derived from the fit of the low-frequency radio spectrum (see Sec. \ref{sec:radio_properties}). 
The errors reported only include the uncertainty on the flux densities and do not include those associated with the spectral indices. If a source belongs to the RACS sample as well as another sample, only the marker corresponding to the RACS sample is reported in Fig. \ref{fig:RACS_sample_distr}. However, the histograms include all the sources in the given sample. In Table \ref{tab:lum_RACS_sample} we report the monochromatic luminosity and the radio-loudness parameter of the high-$z$ quasars detected in RACS.

From Fig. \ref{fig:RACS_sample_distr}, it is clear that all the samples cover a very similar range of optical luminosities. This is expected since they have been built using a similar optical limit and/or surveys. 
At the same time, different samples cover different values in radio luminosities: while the CLASS sample is mainly composed by very bright objects, the sources detected in LoTSS are mainly concentrated on the low end of the radio luminosity distribution. Again, this is also due to the radio flux limits (and frequencies) used in the selection of the different sources. Indeed, while the CLASS sample has a limit $\sim$30 times larger compared to the RACS and FIRST samples, the LoTSS survey has an RMS$\sim$90~\textmu Jy~beam$^{-1}$. By taking a deeper look, the median radio luminosity of the RACS sample (logL$_{\rm 5GHz}$~$\sim$~33.4~erg~sec$^{-1}$~Hz$^{-1}$) is slightly larger compared to that of the FIRST one (logL$_{\rm 5GHz}$~$\sim$~33.1~erg~sec$^{-1}$~Hz$^{-1}$), even though well within the dispersion of their distributions. This is likely due to the different redshift distributions of the two samples, with the RACS sample probing higher redshift and, therefore, higher luminosities for a similar radio flux limit. A similar effect can also be seen in the distribution of the optical luminosities, where the relative number of high-luminosity quasars in slightly larger in the RACS sample. Indeed, as shown in the right panel, while the majority of the FIRST sample are in the redshift range $4<z<5$, most of the RACS sources, by construction, are at $5<z<6$. At the same time, the LoTSS survey, thanks to its depth, is able to uncover many objects at $z>6$, even though many of them are relatively radio faint (L$_{\rm 5GHz}<10^{33}$~erg~sec$^{-1}$ or R~$\lesssim10$). 

Based on the plots in Fig. \ref{fig:RACS_sample_distr}, it is clear that the RACS sample contains the brightest radio quasars currently known at $z>5$. For this reason, this is the ideal starting point to identify blazars, that is, quasars with a relativistic jet oriented close to our line of sight (e.g. \citealt{Padovani1995}), and to constrain the properties and evolution of the jetted quasar population at $z>5$ \citep[see e.g.][]{Diana2022,Caccianiga2024,Banados2025}. In the following three sections we present the data/observations currently available for this sample in the X-ray and radio bands

\section{X-ray observations and detections}
\label{sec:X-ray_properties}

While the number of radio quasars has significantly increased in the last few years \citep[e.g.][]{Gloudemans2022,Belladitta2023,Banados2025}, only a small fraction of these sources have been observed at high energies, namely in the X-rays \citep[e.g.][]{Medvedev2020,Medvedev2021,Khorunzhev2021,Moretti2021,Connor2021}. In this section we report and discuss the X-ray detections of the sources within the RACS sample in the first data release of the German eROSITA All-Sky Survey (eRASS:1; \citealt{Merloni2024}. We also present the analysis of dedicated {\it Chandra}-X-ray observations on a newly discovered $z>6$ quasar (RACS~J0322$-$18).

\subsection{Information from eRASS and literature}

The eRASS:1 survey covers a total of 22 quasars from the RACS sample discussed in this work. Among them, three of the newly identify objects are detected in the soft 0.2--2.3~keV band and reported in the eRASS:1 catalogue. We present in Table \ref{tab:eRASS1_det} their main X-ray properties from catalogue. X-ray luminosities were computed assuming a photon index value of $\Gamma_{\rm X}=2.0$ (see, e.g., \citealt{Vignali2005,Nanni2017}). In particular, RACS~J2020$-$62 and RACS~J0127$-$44 (at redshift $z=5.72$ and $z=4.96$, respectively) are the most distant objects currently identified in this catalogue. The high-redshift nature of these sources, together with the detection in the eRASS:1 scan, make these objects extremely X-ray luminous.
However, the RACS~J2020$-$62 quasar is not detected in later eRASS scans and dedicated {\it Chandra} observations revealed an X-ray emission more than one order of magnitude fainter (see \citealt{Wolf2024}), suggesting that the X-ray flux reported in the \cite{Merloni2024} catalogue might have been caused by a variable or transient event.

For RACS~J0127$-$44 and RACS~J1322$-$13, the large observed X-ray luminosities indicate that the high-energy emission in these systems is dominated by relativistic effects (e.g., \citealt{Ghisellini2015c}), classifying these systems as blazars. Indeed, the corresponding $\tilde{\alpha}_{\rm ox}$ parameters, defined as $\tilde{\alpha}_{\rm ox}=0.303\times{\rm log(L_{10keV}/L_{2500{A}})}$ \citep{Ighina2019}, are consistent with the ones derived for the blazar population ($\lesssim1.36$; e.g., \citealt{Ighina2024b}). Furthermore, also the radio emission of both these systems is flat ($\alpha_{\rm r}<0.5$; see Sec. \ref{sec:radio_properties}) and strong with respect to the optical one (R$\sim$1000), hence strengthening their blazar classification \citep[e.g.][]{Caccianiga2019}.

Dedicated X-ray observations are also available for the majority of the $z>5$ sources already discussed in the literature. We report in Table \ref{tab:lum_RACS_sample} the $\tilde{\alpha}_{\rm ox}$ parameter for all the objects with X-ray information together with the corresponding references for the X-ray emission. For sources covered in eRASS:1, but not detected, we report a lower limit on their $\tilde{\alpha}_{\rm ox}$ parameter based on the $3\sigma$ upper-limit provided by the eRASS:DE consortium at any given position in the sky\footnote{\url{https://erosita.mpe.mpg.de/dr1/AllSkySurveyData_dr1/UpperLimitServer_dr1/}} and assuming again $\Gamma_{\rm X}=2.0$, as in \cite{Merloni2024}. We note that all the $\tilde{\alpha}_{\rm ox}$ values estimated from dedicated X-ray observations are consistent with the lower limits derived from the eRASS:1, when available.

\begin{figure}
    \centering
    \includegraphics[width=\linewidth]{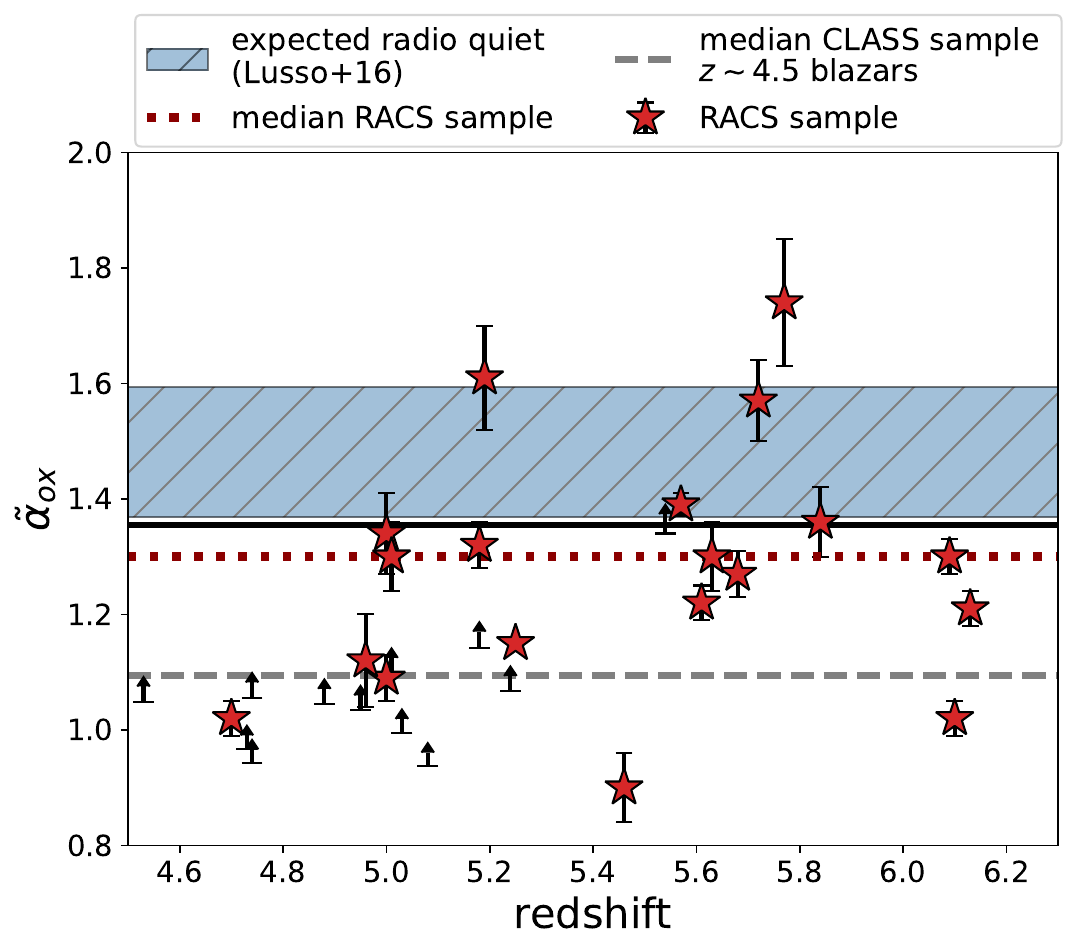}
    \caption{$\tilde{\alpha}_{\rm ox}$ of the RACS sources with X-ray coverage as a function of their redshift. The dotted red line shows the median value of the RACS sources with an X-ray detection, while the dashed gray line is the median value of the blazar sample analysed in \cite{Ighina2019}. The dashed blue area showed range of values expected from the X-ray corona (i.e., from a radio-quiet quasar) with a monochromatic UV luminosity similar to the RACS sample, logL$_{2500\AA}=30.7-32.0$~erg~sec$^{-1}$~Hz$^{-1}$, based on the relation described in \cite{Lusso2016} and assuming a photon index in the range $\Gamma=1.7-2.1$.}
    \label{fig:aox_RACS}
\end{figure}

We show in Fig. \ref{fig:aox_RACS} the $\tilde{\alpha}_{\rm ox}$ parameter as a function of redshift for the RACS sources with X-ray coverage. In the same plot, we also show the median value obtained for $z\sim4.5$ blazars from \cite{Ighina2019} ($\tilde{\alpha}_{\rm ox}=1.1$) as well as the expected values for radio-quite quasars with a similar optical luminosity, log(L$_{\rm 2500\AA}$~/~erg~sec$^{-1}$~Hz$^{-1}$)~=~30.7--32.0, based on the UV-X-ray relation derived by \cite{Lusso2016} (assuming $\Gamma_{\rm X}=1.7-2.1$). While the $\tilde{\alpha}_{\rm ox}$ values in the RACS sample are not as extreme as the radio-brighter blazars at $z=4-5.5$ from the CLASS sample, they have, on average, a stronger X-ray emission (i.e., lower $\tilde{\alpha}_{\rm ox}$ values) compared to what we would expect from thermal emission only, as in the case of radio-quiet quasars. This suggests that blazars compose a large fraction of the RACS sample, for which the high-energy emission is dominated by the relativistically boosted emission produced by the jets.

\begin{table*}
\caption{X-ray properties from the eRASS:1DE catalogue \citep{Merloni2024} of the newly discovered sources in this work.}
    \centering
	\label{tab:eRASS1_det}
    \renewcommand{\arraystretch}{1.3}
    \begin{tabular}{l c c c c c c c }
Name & $z$ & Flux[0.2-2.3keV] & Lum[2-10keV] & net counts & distance & Det. likelihood\\
    \hline
    \hline

RACS~J0127$-$44 	 & 	 4.96 	 & 	 4.2$^{+2.0}_{-1.6}$ 	 & 	 7.1$^{+3.5}_{-2.9}$ 	 & 	 6.7 	 & 	 10.2 	 & 	 8.9	\Tstrut \\
RACS~J1322$-$13 	 & 	 4.70 	 & 	 12.6$^{+3.4}_{-2.8}$ 	 & 	 18.7$^{+6.7}_{-4.7}$ 	 & 	 19.2 	 & 	 4.4 	 & 	 44.8 	 \\
RACS~J2020$-$62 	 & 	 5.72 	 & 	 6.7$^{+3.2}_{-2.4}$ 	 & 	 15.7$^{+7.1}_{-5.6}$ 	 & 	 7.4 	 & 	 17.7 	 & 	 10.0	\BBstrut\\

\hline
\hline 
    \end{tabular}
    \tablefoot{Fluxes are given in units of 10$^{-14}$ erg s$^{-1}$ cm$^{-2}$. Luminosities are given in units of 10$^{45}$ erg s$^{-1}$ and were computed assuming a photon index $\Gamma_{\rm X}=2.0$, while the uncertainties correspond to a variation of the photon index of $\sigma_{\Gamma_{\rm X}}=\pm0.3$. Net counts are in the 0.2--2.3~keV energy band. The optical-to-X-ray distance is in arcseconds. The detection likelihood is described in \cite{Merloni2024}. As a reference, the catalogue is composed mainly by sources with a detection likelihood $>5$.}  

\end{table*}

\subsection{\textit{Chandra} observations}
We obtained dedicated \textit{Chandra} observations for two of the most-distant quasars discovered in our selection, RACS~J0320$-$35 and RACS~J0322$-$18 (Proposal 24700061; PI: Ighina). 
In this section we discuss and analyse the observation of RACS~J0322$-$18, while the high-energy properties of RACS~J0320$-$35 are presented in a separate, dedicated paper (Ighina et al. in preparation).

\begin{figure}
   \centering
	\includegraphics[width=0.8\hsize]{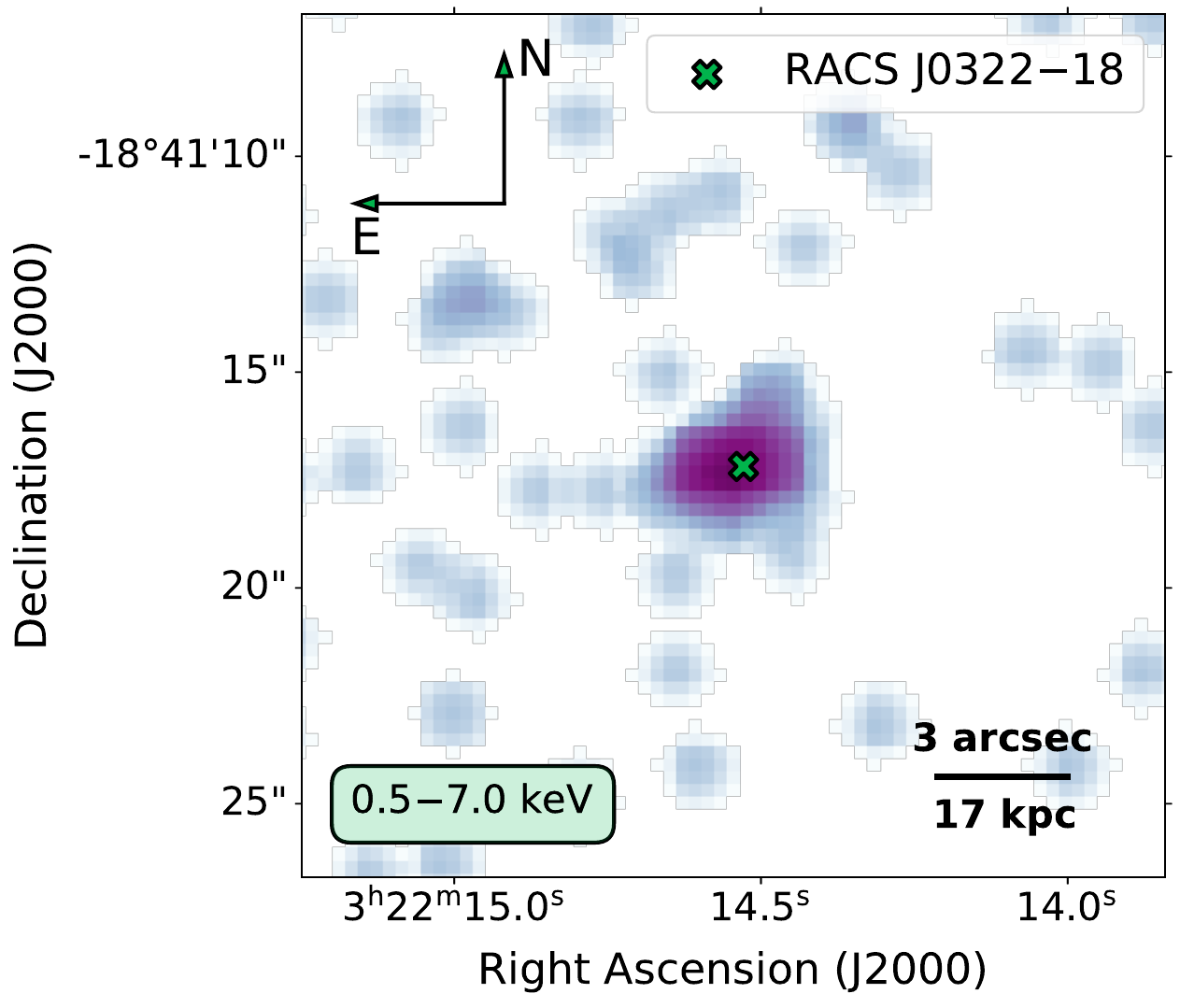}
	\includegraphics[width=0.9\hsize]{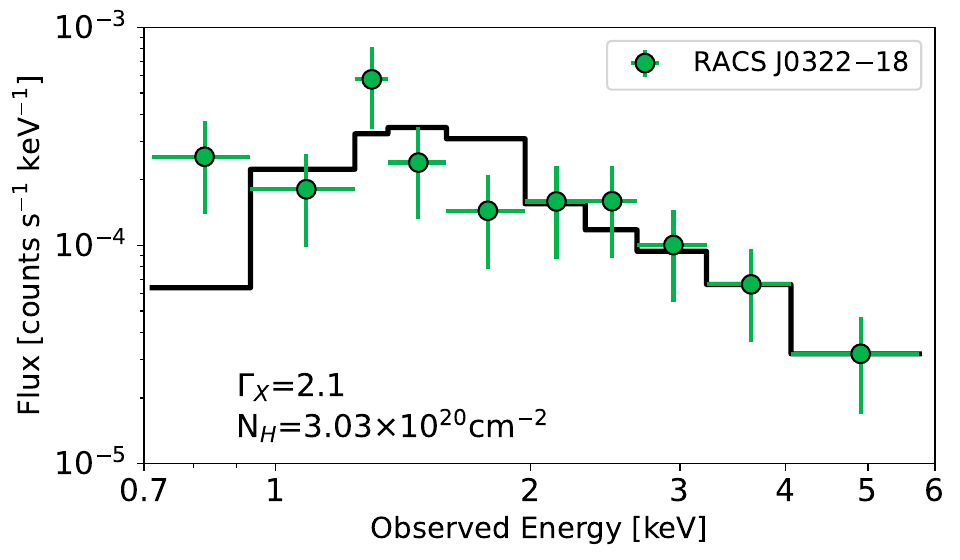}

\caption{{\bf Top panel:} 0.5--7.0~keV image of RACS~J0322$-$18. The green cross indicates the optical/NIR position of the quasar. {\bf Bottom panel:} Best-fit model, including a power-law with Galactic absorption (solid black line),  over-plotted on the X-ray {\it Chandra} spectrum of RACS~J0322$-$18.}
    \label{fig:Xray_DESJ0322-18}
\end{figure}

RACS~J0322$-$18 was observed for a total of 90~ksec,  with 30~ksec observed in 2024 February and 60~ksec observed in 2024 June, for a total of six exposure segments. Observations were performed using the Advanced CCD Imaging Spectrometer (ACIS; \citealt{Garmire2003}) in the Very Faint telemetry format and Timed Exposure mode and with the target positioned on the back-illuminated S3 chip. The observations analysed in this work are contained in the {\it Chandra} Data Collection (CDC)~\dataset[10.25574][CDC 323]{cdc.323}.
We reduced the observations using \texttt{CIAO} (v. 4.16; \citealt{Fruscinone2006}) with CALDB (v4.11.2).
We show in Fig. \ref{fig:Xray_DESJ0322-18}, top panel, the 0.5--7.0~keV image obtained from the combination of the different exposures. We used \texttt{specextract} to extract events from a 2\arcsec \, source region and from a background annulus of 10\textendash $30^{\prime\prime}$, both centred on the optical/NIR position of the quasar. An X-ray source is detected, with 52 net counts over a background of $\lesssim$1, at the optical position of the quasar.
We then analysed the extracted spectra using \texttt{XSPEC} v 12.11.1 \citep{Arnaud1996} and performed a fit minimizing the modified C-statistic \citep{Cash1979,Wachter1979}. We binned the 0.5--7.0~keV spectrum to one net count per energy bin and we modelled it with a simple power law absorbed by the Galactic column density along the line of sight (\texttt{wabs}$\times$\texttt{pow}), fixed to N$_{\rm H} =3.04\times 10^{20}$~cm$^{-2}$ \citep{HI4PI2016}. We report in Table \ref{tab:Gem_x-ray_values} the parameters derived from the best fit. In Fig. \ref{fig:Xray_DESJ0322-18}, bottom panel, we show the X-ray photon spectrum convolved with the instrumental response. We note that the discrepancy between the data and the simple power law model at energies $<1$~keV is not significant and more complex models are not statistically favoured.

Interestingly, while the best-fit photon index value has a relatively large uncertainty, $\Gamma_{\rm X}=2.1\pm0.5$, the small value of the $\tilde{\alpha}_{\rm ox}$ parameter, $\tilde{\alpha}_{\rm ox}=1.30\pm0.03$, is consistent with what is normally found in blazars (e.g., \citealt{Ighina2019,Sbarrato2022}). This would make RACS~J0322$-$18 one of the highest-redshift blazars currently known, with only other two identified at $z>6$ (see \citealt{Belladitta2020,Banados2025}). Very long baseline interferometric (VLBI) observations are needed to strengthen this classification (e.g., \citealt{Coppejans2016}).

\begin{table}
	\caption{Results from the analysis of the X-ray spectrum of RACS~J0322$-$18.}
	\label{tab:Gem_x-ray_values}
    
\centering
\begin{tabular}{ccccccc}
\\
$\Gamma_{\rm X}$    &   Flux[0.5-7keV] &  
Lum[2-10keV]  & cstat/d.o.f.\\
\hline
\hline
2.1$\pm0.5$  & 	1.07$_{-0.04}^{+0.13}$  &   3.1$_{-1.3}^{+2.5}$ & 35 / 51 \BBstrut \Tstrut\\
\hline
\hline
\end{tabular}

\tablefoot{Col. (1) best-fit photon index; col. (2) un-absorbed flux in the energy band 0.5--7~keV in units of 10$^{-14}$ erg s$^{-1}$ cm$^{-2}$; col. (3) rest-frame luminosity in the energy range 2--10~keV in units of 10$^{45}$ erg s$^{-1}$; col. (4) c-statistic and degrees of freedom of the fit.
Errors are reported at 90\% level of confidence.
}
\end{table}

\section{Radio properties}
\label{sec:radio_properties}

In this section we describe the radio properties of the high-$z$ quasars discussed in this work.
The analysis is based on measurements available from public surveys, as well as dedicated radio observations performed on a sub-set of the sample. We report the data analysis of dedicated radio observations in appendix \ref{sec:radio_observations}.

\subsection{Surveys and dedicated observations.}
Thanks to the large variety of radio surveys performed in the last few years, many data-points covering frequencies in the observed range 70~MHz--3GHz are available for a given high-$z$ radio-bright quasar (depending on its sky position and intensity). 
In particular, we considered the measurements available from the following surveys: the Galactic and Extra-Galactic All-Sky MWA survey (GLEAM); \citealt{Wayth2015,Wayth2018,Hurley-Walker2017}, the specific survey covering the south Galactic pol (GLEAM-SGP; \citealt{Franzen2021}) and the extended survey (GLEAM-X; \citealt{Hurley-Walker2022,Ross2024}), at frequencies between 70 and 230~MHz; the LoTSS (\citealt{Shimwell2017,Shimwell2019,Shimwell2022}), including the deep fields \citep{Tasse2021,Sabater2021}, at 144~MHz; the TIFR GMRT Sky Survey (TGSS; \citealt{Intema2017}) at 150~MHz; the Westerbork In the Southern Hemisphere (WISH; \citealt{DeBreuck2002}) survey at 352~MHz; the Sydney University Molonglo Sky Survey (SUMSS; \citealt{Mauch2003}) at 843~MHz; the RACS-low survey \citep{McConnell2020,Hale2021}; the continuum measurements from the MeerKAT Absorption Line Survey (MALS; \citealt{Deka2024}) at 1~GHz; the first scan of the RACS-mid survey \citep{Duchesne2023,Duchesne2024} at 1.37~GHz; the FIRST (\citealt{Becker1995}) survey at 1.4~GHz, the NRAO VLA Sky Survey (NVSS; \citealt{Condon98}) at 1.4~GHz; the quick-look images from the VLA Sky Survey (VLASS; \citealt{Lacy2020,Gordon2020}) at 3GHz. 

For all the targets, we first considered the radio detections as reported in each official catalogue. If no detections were reported, we checked the images and considered as detections all those sources with a signal $>3\times$~the off-source RMS otherwise we considered this value as upper limit. 
For sources detected in VLASS, we applied a 15\% (to the 1.1 epoch) or 8\% correction (to the other epochs) to increase the peak flux densities derived from the images and considered an additional 10\% of the flux density when computing the corresponding uncertainty, added in quadrature\footnote{As suggested in the user guide: \url{https://science.nrao.edu/science/surveys/vlass/vlass-epoch-1-quick-look-users-guide}.}

Moreover, we included the flux density measurements as reported in the source lists of the ASKAP telescope observations taken as part of the following projects\footnote{Available from the CASDA data archive under the codes AS110 for RACS and AS201 for EMU.}: the 3$^{\rm rd}$ data release of RACS-low at 944~MHz, the 1$^{\rm st}$ data release of RACS-high at 1.67~GHz \citep{Duchesne2025} and the Evolutionary Map of the Universe (EMU; \citealt{Norris2021}) at 944~MHz. Since these estimates are not part of a final catalogue, we added in quadrature a further 10\% to their uncertainties. All of the radio quasars discussed here are compact on $\sim10$\arcsec scales, based on their the VLASS radio images, which is expected for these high-$z$ sources (e.g. \citealt{Capetti2024}). The only exception is RACS~J0309+27, discussed in \cite{Ighina2022a} which shows an extended component consisting of only $\sim$4\% of the core flux. For these reasons, we do not expect the different angular resolution to play a significant role in the radio flux density recovered in each survey.

Finally, we also considered dedicated radio observations discussed in the literature as well as data newly obtained as part of this project (with the upgraded Giant Metrewave Radio Telescope, uGMRT, MeerKAT and ATCA). We provide the details of these observations in appendix \ref{sec:radio_observations}.

\subsection{Radio spectra}

We present in Fig. \ref{fig:radio_spectra_pt1}, \ref{fig:radio_spectra_pt2} the radio spectra of the RACS--selected $z>5$ quasars based on the data-sets described in the previous sub-section. Moreover, we show the radio spectra of the newly discovered $4.5<z<5$ radio quasars in appendix \ref{sec:radio_zless5}.
Thanks to the low-frequency coverage of the GLEAM-X, TGSS and LoTSS surveys as well as the high-frequency coverage of the VLASS survey and the dedicated ATCA observations, we are able to constrain the radio emission of most systems over a very large range of frequencies. 
Although some radio sources may present a very complex spectral shape with multiple components (see, e.g., \citealt{Shao2022,Ighina2022b}), in this work we limit the analysis of the radio spectra to simple power laws. In particular, for each object we performed a fit to all the data-points available with a simple power law:

\begin{equation}
    S_\nu = N \times \nu^{-\alpha} 
      \label{eq:single_pl}
\end{equation}

and a broken power law:
\begin{equation}
    S_\nu =
  \begin{cases}
     N \times (\nu/\nu_{\rm break})^{-\alpha_{\rm low}}  & \text{if $\nu \leq \nu_{\rm break}$} \\
     N \times (\nu/\nu_{\rm break})^{-\alpha_{\rm high}}  & \text{if $\nu > \nu_{\rm break}$}  \\
  \end{cases}
  \label{eq:broken_pl}
\end{equation}

To perform the fit we used the \texttt{MrMOOSE} code \citep{Drouart2018a,Drouart2018b}, which also takes the upper-limits into account. This is especially important for sources with a non-detection at low frequencies, for which the limited frequency coverage of the detections would result in a highly uncertain estimate. During the fit, we added a further 10\% to the flux density measurements in order to account for the typical radio variability of quasars (e.g. \citealt{Barvainis2005,Tinti2005,Liu2009}). We do note that strong variability ($>$10\%; as expected in some blazars; \citealt{Sotnikova2024}) can significantly affect the observed shape of the emission, for example introducing a peak in the spectrum \citep[][]{Mingaliev2012}.

In order to choose the model that better reproduces the observed data-points, we compared their relative likelihood with the Akaike Information Criteria (AICc; \citealt{Akaike1974,Burnham2002}) defined as:

\begin{equation}
    {\rm AICc} = 2k  - 2{\rm ln}(L) - \frac{2k(k+1)}{n- k -1}
    \label{eq:AICc}
\end{equation}

where $n$ the number of data points, $k$ the number of free parameters (two in eq. \ref{eq:single_pl} and four in eq. \ref{eq:broken_pl}), and ln($L$) the maximum of the likelihood function (calculated with \texttt{MrMoose}). The lower the value of the AICc parameter is, the better the model describes the observations. We note that the $2k$ term in eq. \ref{eq:AICc} penalises the addition of free parameters and the last term is a further correction in case of a limited number of data points.
We report the best-fit spectral index and break frequency values obtained from the fit in Tab. \ref{tab:lum_RACS_sample} and we show the best-fit models in Fig. \ref{fig:radio_spectra_pt1} and \ref{fig:radio_spectra_pt2} 
(Tab. \ref{tab:lum_RACS_zless5} and Fig. \ref{fig:radio_spectra_zless5} for sources at $z<5$).

\begin{figure*}
   \centering
	\includegraphics[width=0.97\hsize]{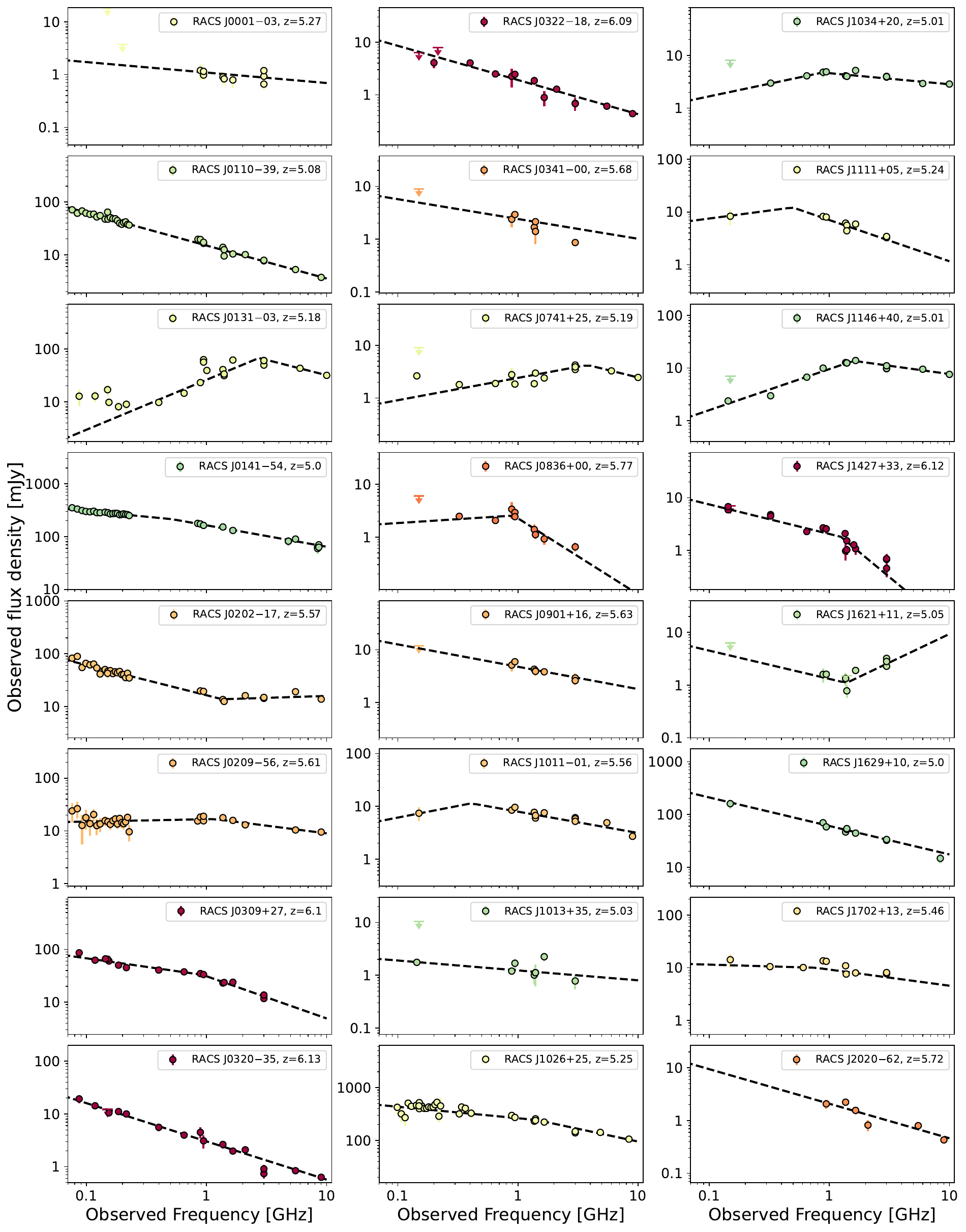}
\caption{Radio spectra of the newly discovered sources in the RACS sample in order of increasing redshift (colour-coded with the same scale of Fig. \ref{fig:opt_spec_sample}). Radio data-points are mainly from publicly available surveys, as described in the text, as well as from dedicated ATCA, uGMRT and MeerKAT observations on a subset of objects. The dashed black line show the best-fit power law (or broken power law) function.}
    \label{fig:radio_spectra_pt1}
\end{figure*}

\begin{figure*}
   \centering
	\includegraphics[width=0.97\hsize]{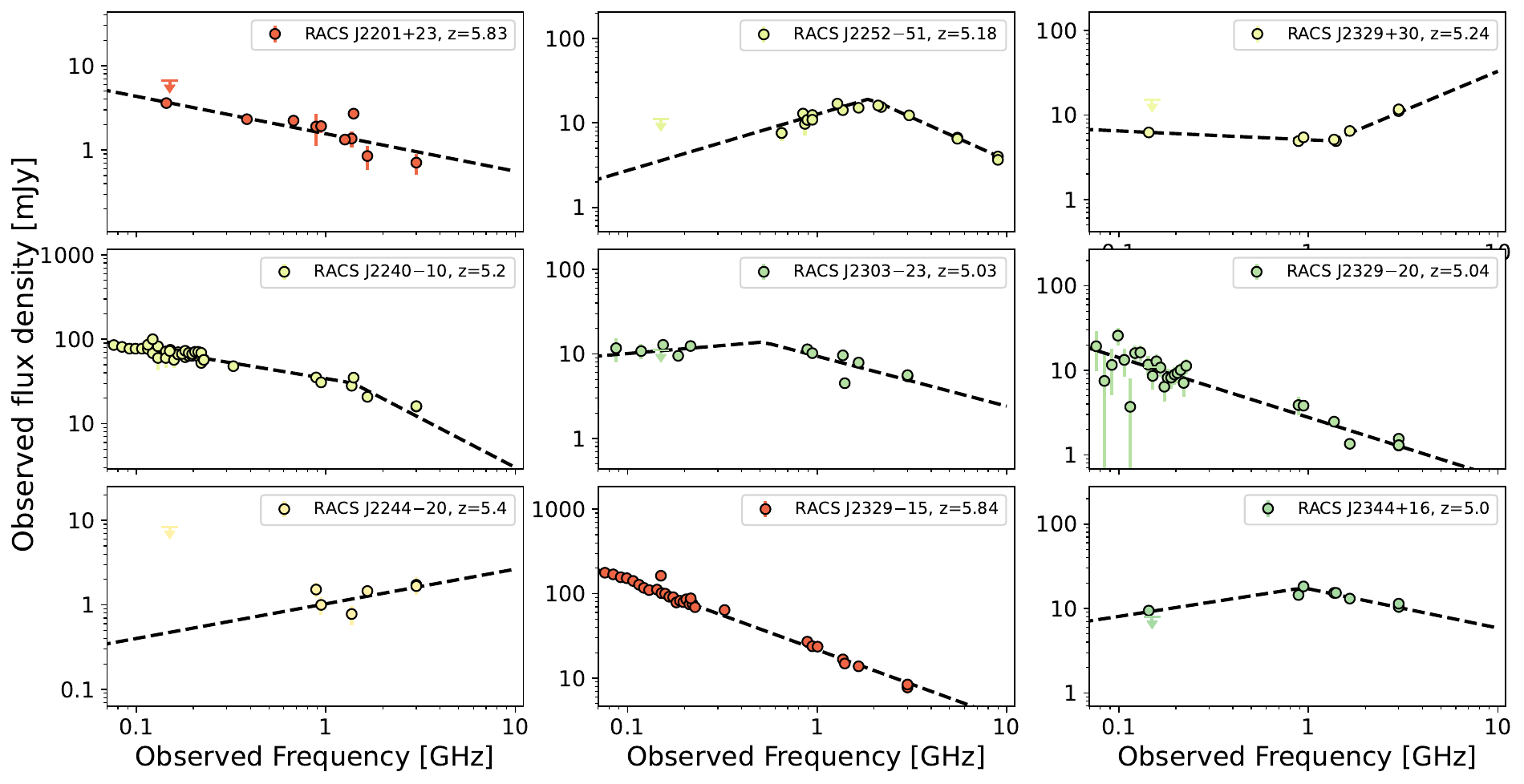}
\caption{Continuation of Fig. \ref{fig:radio_spectra_pt1}}
    \label{fig:radio_spectra_pt2}
\end{figure*}

Despite the simple models assumed, a power law effectively describes the radio emission in a given frequency range for the majority of the sources. In total, a simple power law model was favoured for half of the sample (23 out of 46 objects; 15 out of 33 at $z>5$). As expected, most of these objects have a limited frequency coverage (namely lack of detections at $<800$~MHz or $>2$~GHz).

\begin{figure}
   \centering
	\includegraphics[width=\hsize]{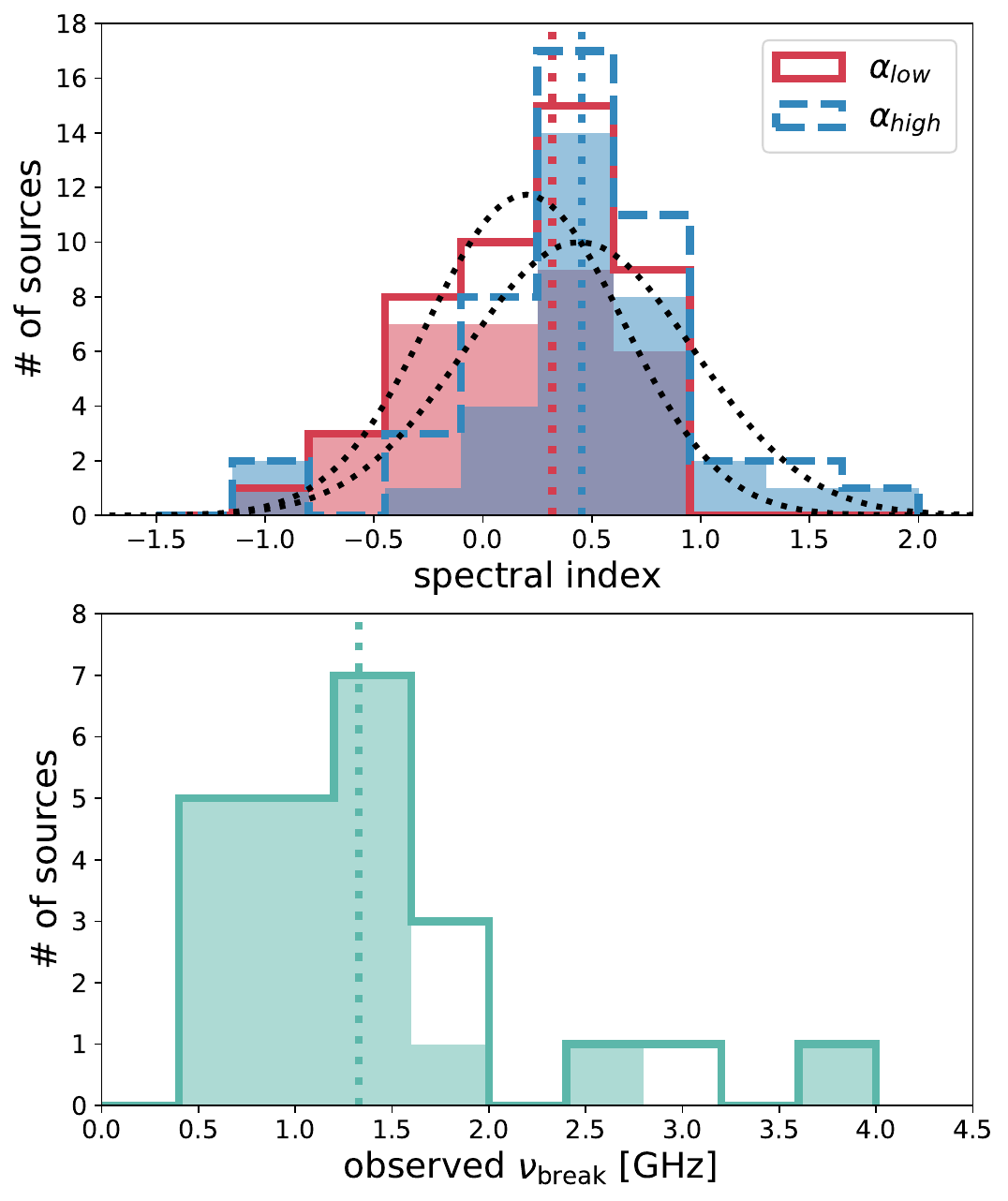}
\caption{{\bf Top panel:} Distribution of the $\alpha_{\rm low}$ and $\alpha_{\rm high}$ parameters derived for all the radio quasars discussed in this work. In case of a single power law fit, the spectral index is considered in both distributions. Dotted vertical lines indicate the median values of each spectral index distribution (0.32 and 0.45 for $\alpha_{\rm low}$ and $\alpha_{\rm high}$, respectively). The best-fit Gaussian distributions are shown with dotted black lines. {\bf Bottom panel:} Distribution of the break frequency in the observed frame for all the sources were a double power law was favoured. Dotted vertical line is the median value (1.3~GHz). In both panels, the distribution of $z>5$ sources is highlighted with filled histograms. }
    \label{fig:radio_indices_hist}
\end{figure}

In Fig. \ref{fig:radio_indices_hist} we show the best-fit distribution of the radio spectral indices and break frequencies derived with \texttt{MrMOOSE}. The majority of the RACS sample presents a spectral index, both at low and high frequencies, in the range $\sim0.0-0.8$. On average, the high-frequency emission is steeper (median value $=$0.45) than the one at low frequency (median value $=$0.32). 
We note that, by considering the $z>5$ quasars only, the distributions and the median values change only slightly (0.48 and 0.22, for ${\alpha}_{\rm high}$ and ${\alpha}_{\rm low}$, respectively) . These values are on the lower end of the distribution of radio spectral indices found at lower redshift. For example, \cite{CalistroRivera2017}, considering $z<2.5$ AGN selected from the SDSS mean value $\bar{\alpha}=0.7$ with a dispersion of $\sigma=0.3$ in the 0.1--1.4~GHz observed range.
This is likely due to a selection effect; with the RACS-low survey we are only sensitive to relatively luminous high-$z$ quasars (L$_{\rm 5GHz}\gtrsim10^{33}$~erg~s$^{-1}$~Hz$^{-1}$) at a rest-frame frequency of $\sim$5~GHz and, therefore, we expect a significant fraction of the radio emission to be relativistically boosted. Similar values were found when considering the $z>4$ quasars detected in FIRST (0.1 and 0.4 at low and high frequencies, respectively;  \citealt{Caccianiga2024}) as well as very radio-luminous (L$_{\rm 5GHz}\gtrsim10^{34}$~erg~s$^{-1}$~Hz$^{-1}$) $3<z<4$ quasars (0.0 and 0.4 at low and high frequencies, respectively; \citealt{Sotnikova2021}).

Based on a Gaussian fit of the spectral index distributions we find that low-frequency spectral indices have a mean value $\bar{\alpha}_{\rm low}=0.20$ and a standard deviation $\sigma_{\rm low}=0.45$, while high-frequency spectral indices have $\bar{\alpha}_{\rm high}=0.44$ and $\sigma_{\rm high}=0.52$.

\begin{figure}
   \centering
	\includegraphics[width=\hsize]{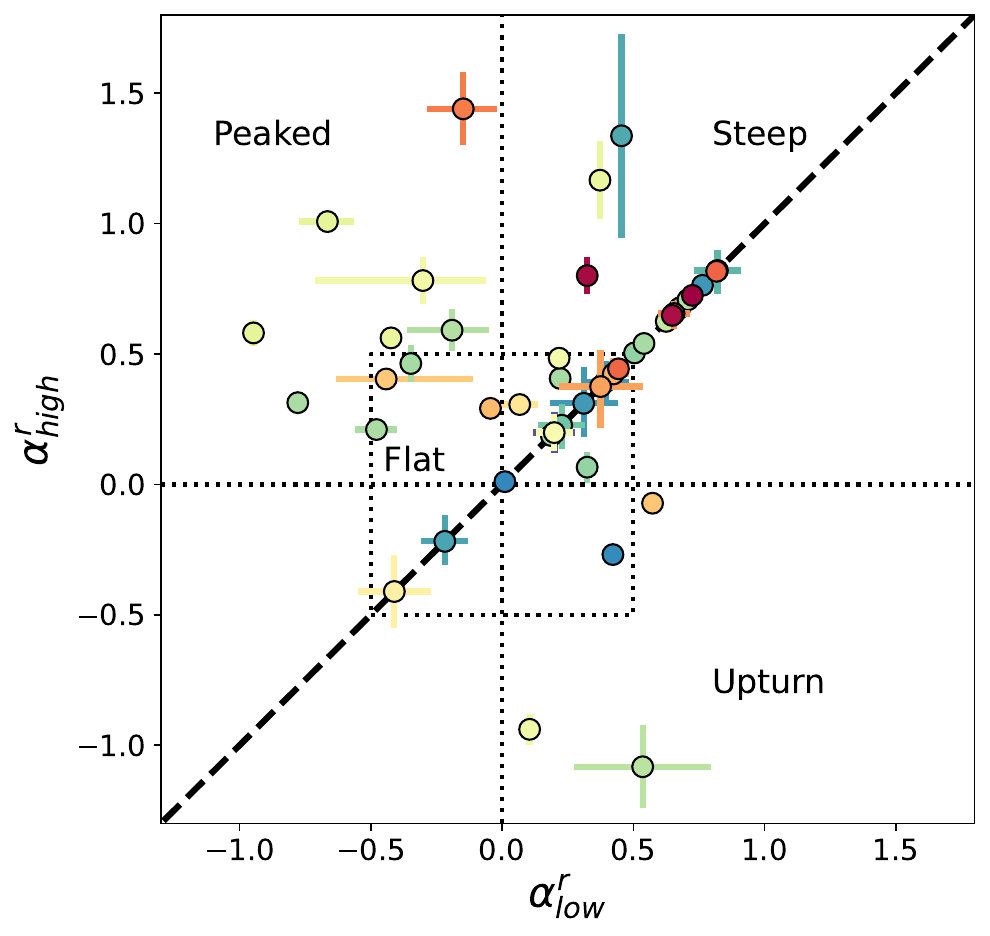}
\caption{Low-frequency ($\alpha_{\rm low}$) and high-frequency ($\alpha_{\rm high}$) radio spectral indices of the radio quasars discussed in this work. For clarity reasons, we only show values with an uncertainty on both spectral indices $\sigma<0.7$. Sources with a single power law fit are shown on the 1:1 relation (dashed black line). Objects with a single power law fit are reported in this line. The dotted lines at $\alpha=0,\pm0.5$ divide the space in different regions, as discussed in the text.}
    \label{fig:radio_indices}
\end{figure}

If we compare the values derived for the spectral indices at low and high frequencies for each source (Fig. \ref{fig:radio_indices}) the quasars in the RACS sample occupy different regions of the plane, which can be interpreted as different physical properties.
A large fraction of the sample lies within the region delimited by a spectral index $=\pm0.5$ (23 in total, 15 of which at $z>5$). This is the threshold commonly used in the literature to identify flat-spectrum radio quasars (i.e. blazars; e.g. \citealt{Sotnikova2021}), for which we expect the radiation to be dominated by strong Doppler boosting due to jet being oriented close to our line of sight. Indeed, two of the newly discovered quasars detected in the eRASS:1 (see Sec. \ref{sec:X-ray_properties}) are in this region.
Identifying blazars in a well-defined sample, as the one discussed here, is extremely important to constrain the evolution of the entire jetted population, especially at high redshift (see, e.g., \citealt{Diana2022,Sbarrato2022}).
However, further observations (especially X-rays and VLBI) are needed to accurately constrain the nature of these systems, since variability might affect the observed shape of their radio spectra.

A small fraction of the RACS quasars shows typical $\alpha_{\rm low}\sim0.5$ values with a significant steepening at higher frequencies, $\Delta\alpha\gtrsim0.4$. This type of steeping at high frequencies has already been observed in other high-$z$ radio sources (see, e.g., \citealt{Drouart2020,Spingola2020}) and is typically associated with the radiative cooling of the most energetic electrons in the extended regions of the jets \citep[e.g.][]{Harwood2013,Morabito2016}.

There are also significant outliers from the one-to-one relation, having opposite spectral indices at low and high frequencies (top left and bottom right regions in Fig. \ref{fig:radio_indices}). In particular, sources with $\alpha_{\rm low}<0$ and $\alpha_{\rm high}>0$ present a peak in their radio spectrum, typically around $\sim1$~GHz in the observed frame (see HZQ~J1146+40 and RACS~J2252$-$51), which corresponds to frequencies $>6$~GHz in the rest frame.
These `peaked spectrum' radio sources are normally associated with young and compact relativistic jets that are still expanding in the surrounding medium \citep[e.g.,][]{Dallacasa2000,Orienti2012,Orienti2014,Odea2021}.
In particular, we note that in the case of RACS~J2252$-$51 the presence of a peak at $\sim2$~GHz in the observed frame is confirmed by simultaneous MeerKAT observations covering the 1--3~GHz frequency range. Dedicated low-frequency and VLBI observations on this high-$z$ system are crucial to constrain the origin of this turnover and therefore to constrain intrinsic properties of the jet (e.g., magnetic field) or the external environment (e.g., electron density; \citealt{Gloudemans2023}) via spectral modelling and morphology studies (e.g.; \citealt{Keim2019,Shao2022}).

On the other side of the plot, objects with $\alpha_{\rm low}>0$ and $\alpha_{\rm high}<0$ present a so-called `up-turn'. The radio spectrum of these objects is probably composed of the radiation produced by different jet components. For example, it can be caused by a steep emission of an extended, not resolved component and a flat/inverted emission of the core (see e.g. \citealt{Behiri2025})or also by the presence of a double peak in the radio spectrum (see, e.g., fig. 2 in \citealt{Shao2022}).

\section{Summary and conclusions}
\label{sec:conclusions}

We summarise here the main results reported in this work.

\begin{itemize}
    \item Sample selection: from the combination of the RACS-low, DES, and Pan-STARRS surveys, we selected a sample of 45 high-$z$ radio quasar candidates, well-defined in radio, optical, and NIR flux limits and colours, with S$_{\rm 888MHz}>1$~mJy and mag$z$~$<21.3$ over an area of 16000~deg$^2$ (see Sec. \ref{sec:selection}). Based on the number of $5<z<5.5$ quasars selected in the deeply explored SDSS area, we estimate our optical/NIR completeness to be $\sim$83\%. At the same time, based on the comparison of the RACS-low survey with similar, deeper ASKAP observations and based on the selection of already known $z>4$ radio quasars, we estimate the completeness of our radio selection to be $\sim75\%$ for surface brightness S$_{\rm 888MHz}>1$~mJy~beam$^{-1}$ and $\sim90\%$ for S$_{\rm 888MHz}>2$~mJy~beam$^{-1}$. The overall completeness of our selection is therefore $\sim62\%$ at S$_{\rm 888MHz}>1$~mJy~beam$^{-1}$ and $\sim75\%$ at S$_{\rm 888MHz}>2$~mJy~beam$^{-1}$ (see Sec. \ref{sec:final_sample}).\\
    
    \item Identification and final sample: we obtained spectroscopic observations for the majority of the selected candidates (40 out of 44) with the AAT, Gemini-South, LBT, NTT, TNG and VLT. We confirmed the high-$z$ quasar nature of 24 of the candidates, including 11 at $z>5$ (see Fig. \ref{fig:opt_spec_sample}). Considering the quasars already known from the literature that are also recovered from our selection criteria, we built a sample of 22 radio quasars at $z>5$ with S$_{\rm 888MHz}>1$~mJy and mag$z$~$<21.3$.
    Given the estimated level of completeness (62\% above 1~mJy~beam$^{-1}$), we expect that in the surveyed sky area and at the radio/optical limits of our sample, there are $\sim$35 $z>5$ quasars, $\sim$13 of which were missed by our selection. However, we recovered many (11) of these missing sources from the literature, thus obtaining a final sample of 33 $z>5$ radio quasars which we expect to be $\sim$90\% complete (see Table \ref{tab:lum_RACS_sample}).\\

    \item X-ray band: detections or strong upper-limits\textemdash are available for $\sim$55\% of the $z>5$ quasars (18 out of 33, or 20 out of 46 including sources at $z<5$). This information mainly comes from dedicated X-ray observations available from the literature \citep[e.g.][]{Moretti2021,Connor2021,Zuo2024} or obtained as part of this project (e.g., \citealt{Ighina2024b}). In particular, from dedicated \textit{Chandra} observations of RACS~J0322$-$18 at $z=6.09$ (see Fig. \ref{fig:Xray_DESJ0322-18}) revealed a relatively bright X-ray emission, compared to the optical/UV one ($\tilde{\alpha}_{\rm ox}=1.30\pm0.03$), making this one of the highest redshift blazars currently known. At the same time, four sources in the RACS sample have also been detected as part of the eRASS:1 survey \citep[e.g.][]{Khorunzhev2021,Wolf2024}, three of which were discovered as part of this work, including one at $z>5$ (RACS~J2020$-$62; independently discovered by \citealt{Wolf2024}). These sources represent the highest-redshift quasars currently detected in this first scan. In general, the X-ray emission of the selected sources is systematically larger compared to that expected from radio-quiet quasars, having an $\tilde{\alpha}_{\rm ox}$ parameter larger than the one expected for radio-quite quasars (based on the \citealt{Lusso2016} relation). The strong X-ray emission observed in our sources is likely associated with the radiation produced by the relativistic jets, indicating that many of the radio-bright quasars we selected are blazars.\\

    \item Radio band: combining measurements and upper-limits from wide-area radio surveys (e.g., GLEAM-X, LoTSS, TGSS, NVSS, RACS, VLASS) together with newly obtained dedicated observations (with uGMRT, MeerKAT and ATCA), we were able to constrain the radio shape of the high-$z$ quasars detected in RACS-low over a very wide range of frequencies, $\sim0.1-10$~GHz (observed frame; see Fig. \ref{fig:radio_spectra_pt1}, \ref{fig:radio_spectra_pt2} and \ref{fig:radio_spectra_zless5}). We do note that simultaneous radio observations are available for only a small fraction of the entire sample and variability can still play a significant role in these measurements.  From the fit of the radio data-points below and above 1.4~GHz in the observed frame, we found that most of the sources have a radio slope in the range $0.0-0.7$, with a median spectral index of 0.31 and 0.46 at low and high frequencies, respectively (see Fig. \ref{fig:radio_indices_hist}). However, some notable outliers are also present, showing clear signs of a peak or of an upturn in the radio spectrum. At the same time, a third of the $z>5$ radio quasars (12 out of 33) presents a flat radio spectrum (i.e. $-0.5<\alpha_{\rm r}<0.5$ at both low and high frequencies; 22 including $z<5$ sources). Once again, the large number of quasars with flat radio spectrum, compared to the population of quasars at lower redshifts (e.g., \citealt{CalistroRivera2017}), suggests that many of the sources in the RACS sample hinting at the blazar nature of these systems. \\

    \item Future prospects: in this first paper on the high-$z$ RACS quasars we presented the largest flux-limited and well-defined (i.e., with a completeness estimate) of radio quasars at $z>5$ currently available. Moreover, this work also paves the way for the search of even more high-$z$ quasars with the next generation of radio and optical/NIR wide-area survey expected to be completed in the next few years: the Evolutionary Map of the Universe (EMU; \citealt{Norris2021}) in the radio band, the Legacy Survey of Space and Time (LSST; \citealt{Ivezic2019}) performed with the Vera C. Rubin Observatory in the optical band and the EUCLID-Wide survey \citep{Euclid2022} in the NIR band. Indeed, thanks to the large area covered by RACS, we were able to significantly increase the number of radio-bright quasars at high redshift, even though some regions of the sky still remain unexplored due to the lack of deep optical/NIR coverage (see Fig. \ref{fig:sky_distr_sample}). The existence of this population at $z\sim5-6$ (see also \citealt{Drouart2020}) suggests the presence of similarly radio bright systems even at higher redshifts (see, e.g., \citealt{Banados2025}).
    Next-generation surveys will be crucial in the exploration of the jetted quasar $z>7$ (see, e.g., fig. 5 in \citealt{Ighina2023}).

    The $z>5$ sources discussed in this work represent the largest well-defined sample of radio-powerful quasars. For this reason, in a future work, we will use the multi-wavelength information available to identify all the blazars in the sample and constrain the radio luminosity function of this class of objects at $z>5$ for the first time (e.g. \citealt{Mao2017}). 
    
\end{itemize}

\begin{acknowledgements}
L.I. would like to thank all the staff working at the ATCA/Paul Wild and AAT/Siding Spring observatories for the great support and the amazing experience provided during the observations. Moreover, we would also like to thank the staff at the LBT, Gemini-South and ESO-VLT telescopes for their help and support in the preparation, execution and reduction of the observations.\\
We would like to thank the referee for their input and suggestions, which resulted in a clearer manuscript. We also want to thank S. Belladitta for her help in the preparation of some of the observing proposals used in this work.\\
We acknowledge financial support from INAF under the projects ``Quasar jets in the early Universe'' (Ricerca Fondamentale 2022), ``Collaborative research on VLBI as an ultimate test to LambdaCDM model'' (Ricerca fondamentale 2022) and ``Testing the obscuration in the early Universe'' (Ricerca Fondamentale 2023).
A.R. acknowledges the INAF project Supporto Arizona \& Italia.\\
This scientific work uses data obtained from Inyarrimanha Ilgari Bundara / the Murchison Radio-astronomy Observatory. We acknowledge the Wajarri Yamaji People as the Traditional Owners and native title holders of the Observatory site. CSIRO’s ASKAP radio telescope is part of the Australia Telescope National Facility (\url{https://ror.org/05qajvd42}). Operation of ASKAP is funded by the Australian Government with support from the National Collaborative Research Infrastructure Strategy.\\
The Australia Telescope Compact Array is part of the Australia Telescope National Facility (\url{https://ror.org/05qajvd42}) which is funded by the Australian Government for operation as a National Facility managed by CSIRO. We acknowledge the Gomeroi people as the Traditional Owners of the Observatory site.\\
This project has received funding from the European Union's Horizon 2020 research and innovation programme under grant agreement No 101004719. This material reflects only the authors views and the Commission is not liable for any use that may be made of the information contained therein.\\
This work was supported by resources provided by the Pawsey Supercomputing Research Centre with funding from the Australian Government and the Government of Western Australia.\\
The MeerKAT telescope is operated by the South African Radio Astronomy Observatory, which is a facility of the National Research Foundation, an agency of the Department of Science and Innovation.\\
This paper employs a list of Chandra datasets, obtained by the Chandra X-ray Observatory, contained in the Chandra Data Collection (CDC) \dataset[10.25574][CDC 323]{cdc.323}.\\
We acknowledge the support from the LBT-Italian Coordination Facility for the execution of the observations. The LBT is an international collaboration among institutions in the United States, Italy and Germany. LBT Corporation partners are: The University of Arizona on behalf of the Arizona Board of Regents; Istituto Nazionale di Astrofisica, Italy; LBT Beteiligungsgesellschaft, Germany, representing the Max-Planck Society, The Leibniz Institute for Astrophysics Potsdam, and Heidelberg University; The Ohio State University, representing OSU, University of Notre Dame, University of Minnesota and University of Virginia. This research used the facilities of the Italian Center for Astronomical Archive (IA2) operated by INAF at the Astronomical Observatory of Trieste.\\
This publication makes use of data products from the Wide-field Infrared Survey Explorer, which is a joint project of the University of California, Los Angeles, and the Jet Propulsion Laboratory/California Institute of Technology, funded by the National Aeronautics and Space Administration.
This research has made use of the \texttt{CIRADA} cutout service at \url{cutouts.cirada.ca}, operated by the Canadian Initiative for Radio Astronomy Data Analysis (\texttt{CIRADA}). \texttt{CIRADA} is funded by a grant from the Canada Foundation for Innovation 2017 Innovation Fund (Project 35999), as well as by the Provinces of Ontario, British Columbia, Alberta, Manitoba and Quebec, in collaboration with the National Research Council of Canada, the US National Radio Astronomy Observatory and Australia’s Commonwealth Scientific and Industrial Research Organisation.\\
We acknowledge the use of the ilifu cloud computing facility - \url{www.ilifu.ac.za}, a partnership between the University of Cape Town, the University of the Western Cape, Stellenbosch University, Sol Plaatje University, the Cape Peninsula University of Technology and the South African Radio Astronomy Observatory. The ilifu facility is supported by contributions from the Inter-University Institute for Data Intensive Astronomy (IDIA - a partnership between the University of Cape Town, the University of Pretoria and the University of the Western Cape), the Computational Biology division at UCT and the Data Intensive Research Initiative of South Africa (DIRISA).
This work made use of the CARTA (Cube Analysis and Rendering Tool for Astronomy) software (DOI 10.5281/zenodo.3377984 – \url{https://cartavis.github.io}). \\

\end{acknowledgements}

\afterpage{\begin{landscape}
   
\begin{table}
\centering
    \caption{Multi-wavelength properties of the $z>5$ RACS quasars.}
    \renewcommand{\arraystretch}{1.15}
     \addtolength{\tabcolsep}{0.2em}

    \begin{tabular}{l c c c c c c c c c c }
Name & $z$ & log(L$_{2500\text{\normalfont\AA}}$) & log(L$_{4400\text{\normalfont\AA}}$) & log(L$_{\rm 5GHz}$) & log($\nu$L$_{\rm 1.4GHz}$) & log(R) & ${\alpha}_{\rm low}^{\rm r}$ & ${\alpha}_{\rm high}^{\rm r}$ & ${\nu}_{\rm break}$ & $\tilde{\alpha}_{\rm ox}$\Bstrut\\
 & & erg~s$^{-1}$~Hz$^{-1}$ & erg~s$^{-1}$~Hz$^{-1}$ & erg~s$^{-1}$~Hz$^{-1}$ & erg~s$^{-1}$ & & & & GHz & \\
    \hline
    \hline
        {RACS~J0001-03}  	   & 5.27 & 31.04$\pm$0.03 & 31.19$\pm$0.04 & 32.73$\pm$0.09 & 41.99$\pm$0.09 & 1.54$\pm$0.10 & 0.20$_{- 0.07 }^{+ 0.07 }$ 	 & 	 -- 	 & 	 -- &  -- \Tstrut \\
        {\bf RACS~J0110$-$39}  & 5.08 & 30.74$\pm$0.04 & 30.80$\pm$0.06 & 33.88$\pm$0.07 & 43.37$\pm$0.07 & 3.08$\pm$0.09 	&	 0.63 $_{- 0.01 }^{+ 0.01 }$ 	 & 	 -- 	 & 	 --  & $>0.94^1$ \\ 
        {\bf RACS~J0131$-$03}  & 5.18 & 32.11$\pm$0.03 & 32.27$\pm$0.03 & 33.99$\pm$0.07 & 42.61$\pm$0.07 & 1.72$\pm$0.08 	&	 -0.95 $_{- 0.02 }^{+ 0.02 }$ 	 & 	 0.58 $_{- 0.05 }^{+ 0.05 }$ 	 & 	 2.73 $_{- 0.10 }^{+ 0.10 }$    & 1.32$\pm$0.04$^2$\\ 
        {\bf RACS~J0141$-$54}  & 5.00 & 30.77$\pm$0.03 & 30.85$\pm$0.05 & 34.88$\pm$0.07 & 44.19$\pm$0.07 & 4.03$\pm$0.0 &	 0.22 $_{- 0.01 }^{+ 0.01 }$ 	 & 	 0.41 $_{- 0.01 }^{+ 0.01 }$ 	 & 	 0.52 $_{- 0.06 }^{+ 0.06 }$    & 1.34$\pm$0.07$^3$ \\ 
        {\bf RACS~J0202$-$17}  & 5.57 & 31.82$\pm$0.03 & 31.94$\pm$0.03 & 33.99$\pm$0.07 & 43.45$\pm$0.07 & 2.11$\pm$0.08 & 0.57 $_{- 0.01 }^{+ 0.02 }$ 	 & 	 $-$0.07 $_{- 0.02 }^{+ 0.02 }$ 	 & 	 1.35 $_{- 0.07 }^{+ 0.06 }$    & 1.39$\pm$0.02$^4$ \\ 
        {\bf RACS~J0209$-$56}  & 5.61 & 31.04$\pm$0.03 & 30.93$\pm$0.04 & 33.93$\pm$0.07 & 43.05$\pm$0.07 & 2.86$\pm$0.08 &-0.04 $_{- 0.03 }^{+ 0.03 }$ 	 & 	 0.29 $_{- 0.03 }^{+ 0.03 }$ 	 & 	 1.18 $_{- 0.16 }^{+ 0.16 }$ & 1.22$\pm$0.03$^4$\\ 
        {\bf RACS~J0309+27}    & 6.10 & 31.02$\pm$0.13 & 31.17$\pm$0.05 & 34.32$\pm$0.07 & 43.65$\pm$0.07 & 3.15$\pm$0.09 & 0.32 $_{- 0.03 }^{+ 0.02 }$ 	 & 	 0.8 $_{- 0.07 }^{+ 0.07 }$ 	 & 	 0.93 $_{- 0.08 }^{+ 0.07 }$  & 1.02$\pm$0.03$^3$   \\ 
        {\bf RACS~J0320$-$35}  & 6.13 & 31.15$\pm$0.03 & 31.16$\pm$0.04 & 33.35$\pm$0.07 & 42.91$\pm$0.07 & 2.03$\pm$0.08 & 0.72 $_{- 0.02 }^{+ 0.02 }$ 	 & 	 -- 	 & 	 --    & 1.21$\pm$0.03$^6$  \\ 
        {\bf RACS~J0322$-$18}  & 6.09 & 31.13$\pm$0.03 & 31.14$\pm$0.04 & 33.15$\pm$0.07 & 42.65$\pm$0.07 & 1.95$\pm$0.08 & 0.65 $_{- 0.02 }^{+ 0.02 }$ 	 & 	 -- 	 & 	 --    & 1.30$\pm$0.03$^7$ \\ 
        RACS~J0341$-$00        & 5.68 & 31.20$\pm$0.04 & 31.26$\pm$0.04 & 33.15$\pm$0.14 & 42.50$\pm$0.14 & 1.89$\pm$0.15 & 0.38 $_{- 0.16 }^{+ 0.14 }$ 	 & 	 -- 	 & 	 --    & 1.27$\pm$0.04$^8$ \\ 
        RACS~J0741+25          & 5.19 & 32.01$\pm$0.03 & 32.20$\pm$0.03 & 33.01$\pm$0.07 & 41.92$\pm$0.07 & 0.82$\pm$0.08 &	 -0.42 $_{- 0.02 }^{+ 0.02 }$ 	 & 	 0.56 $_{- 0.02 }^{+ 0.02 }$ 	 & 	 3.8 $_{- 0.08 }^{+ 0.08 }$    & 1.61$\pm$0.09$^9$ \\
        RACS~J0836+00          & 5.77 & 31.92$\pm$0.03 & 32.09$\pm$0.03 & 33.12$\pm$0.13 & 42.19$\pm$0.13 & 1.03$\pm$0.13 &	 -0.15 $_{- 0.14 }^{+ 0.13 }$ 	 & 	 1.44 $_{- 0.14 }^{+ 0.14 }$ 	 & 	 0.92 $_{- 0.04 }^{+ 0.04 }$   & 1.74$\pm$0.11$^{10}$ \\  
        {\bf RACS~J0901+16}    & 5.63 & 31.03$\pm$0.03 & 31.03$\pm$0.06 & 33.44$\pm$0.08 & 42.82$\pm$0.08 & 2.41$\pm$0.10 &	 0.42 $_{- 0.06 }^{+ 0.06 }$ 	 & 	 -- 	 & 	 --   & 1.30$\pm$0.06$^{11}$   \\  
        {\bf RACS~J1011$-$01}  & 5.56 & 30.81$\pm$0.04 & 30.92$\pm$0.05 & 33.65$\pm$0.22 & 42.78$\pm$0.22 & 2.58$\pm$0.23 &	 -0.44 $_{- 0.19 }^{+ 0.33 }$ 	 & 	 0.4 $_{- 0.03 }^{+ 0.03 }$ 	 & 	 0.41 $_{- 0.26 }^{+ 0.28 }$   & $>1.32^4$ \\ 
        RACS~J1013+35          & 5.03 & 31.14$\pm$0.03 & 31.21$\pm$0.04 & 32.75$\pm$0.07 & 42.01$\pm$0.07 & 1.54$\pm$0.08 &	 0.19 $_{- 0.04 }^{+ 0.04 }$ 	 & 	 -- 	 & 	 --   &  $>0.99^1$ \\ 
        {\bf RACS~J1026+25}    & 5.25 & 31.31$\pm$0.04 & 31.40$\pm$0.04 & 35.11$\pm$0.07 & 44.38$\pm$0.07 & 3.71$\pm$0.08 &	 0.22 $_{- 0.02 }^{+ 0.02 }$ 	 & 	 0.48 $_{- 0.03 }^{+ 0.03 }$ 	 & 	 1.43 $_{- 0.11 }^{+ 0.22 }$    & 1.15$\pm$0.01$^{12}$ \\
        {\bf RACS~J1034+20}    & 5.01 & 31.44$\pm$0.03 & 31.68$\pm$0.03 & 33.30$\pm$0.09 & 42.18$\pm$0.09 & 1.62$\pm$0.09 &	 -0.48 $_{- 0.08 }^{+ 0.08 }$ 	 & 	 0.21 $_{- 0.02 }^{+ 0.02 }$ 	 & 	 0.88 $_{- 0.07 }^{+ 0.08 }$   & $>1.10^1$ \\ 
        RACS~J1111+05          & 5.24 & 31.35$\pm$0.03 & 31.83$\pm$0.03 & 33.59$\pm$0.27 & 42.80$\pm$0.27 & 1.76$\pm$0.27 &	 -0.3 $_{- 0.41 }^{+ 0.24 }$ 	 & 	 0.78 $_{- 0.09 }^{+ 0.09 }$ 	 & 	 0.50 $_{- 0.12 }^{+ 0.33 }$ & $>1.07^1$ \\ 
        {\bf RACS~J1146+40}    & 5.01 & 31.55$\pm$0.03 & 31.71$\pm$0.03 & 33.56$\pm$0.07 & 42.27$\pm$0.07 & 1.85$\pm$0.08 &	 -0.78 $_{- 0.02 }^{+ 0.02 }$ 	 & 	 0.31 $_{- 0.03 }^{+ 0.03 }$ 	 & 	 1.59 $_{- 0.06 }^{+ 0.07 }$   & 1.30$\pm$0.06$^{13}$ \\ 
        RACS~J1427+33          & 6.12 & 31.23$\pm$0.05 & 31.52$\pm$0.03 & 33.17$\pm$0.07 & 42.62$\pm$0.07 & 1.65$\pm$0.08 &	 0.56 $_{- 0.03 }^{+ 0.03 }$ 	 & 	 1.89 $_{- 0.72 }^{+ 0.72 }$ 	 & 	 1.25 $_{- 0.14 }^{+ 0.09 }$    & -- \\ 
        {RACS~J1621+11}  	   & 5.05 & 30.96$\pm$0.07 & 31.10$\pm$0.04 & 32.81$\pm$0.21 & 42.25$\pm$0.21 & 1.71$\pm$0.21 &	 0.54 $_{- 0.26 }^{+ 0.26 }$ 	 & 	 -1.08 $_{- 0.16 }^{+ 0.16 }$ 	 & 	 1.39 $_{- 0.14 }^{+ 0.1 }$ & --\\
        {\bf RACS~J1629+10}    & 5.00 & 31.02$\pm$0.04 & 31.04$\pm$0.05 & 34.46$\pm$0.07 & 43.91$\pm$0.07 & 3.42$\pm$0.09 &	 0.54 $_{- 0.03 }^{+ 0.03 }$ 	 & 	 -- 	 & 	 --   & 1.09$\pm$0.04$^{12}$ \\
        RACS~J1702+13          & 5.46 & 31.01$\pm$0.04 & 31.01$\pm$0.05 & 33.69$\pm$0.10 & 42.88$\pm$0.10 & 2.68$\pm$0.11 &	 0.07 $_{- 0.13 }^{+ 0.07 }$ 	 & 	 0.31 $_{- 0.03 }^{+ 0.03 }$ 	 & 	 0.78 $_{- 0.21 }^{+ 0.11 }$     & 0.90$\pm$0.06$^{14}$  \\
        RACS~J2020$-$62        & 5.72 & 31.76$\pm$0.03 & 31.93$\pm$0.03 & 33.13$\pm$0.08 & 42.64$\pm$0.08 & 1.20$\pm$0.09 &	 0.66 $_{- 0.06 }^{+ 0.05 }$ 	 & 	 -- 	 & 	 --   & 1.57$\pm$0.07$^{15}$ \\ 
        RACS~J2201+23          & 5.83 & 31.23$\pm$0.03 & 31.33$\pm$0.04 & 32.99$\pm$0.07 & 42.38$\pm$0.07 & 1.66$\pm$0.08 &	 0.44 $_{- 0.04 }^{+ 0.04 }$ 	 & 	 -- 	 & 	 --   & --\\ 
        {\bf RACS~J2240$-$10}  & 5.20 & 31.00$\pm$0.03 & 31.08$\pm$0.05 & 34.23$\pm$0.07 & 43.58$\pm$0.07 & 3.15$\pm$0.09 &    0.37 $_{- 0.01 }^{+ 0.01 }$ 	 & 	 1.17 $_{- 0.15 }^{+ 0.15 }$ 	 & 	 1.39 $_{- 0.04 }^{+ 0.04 }$   & -- \\ 
        {\bf RACS~J2244$-$20}  & 5.41 & 31.16$\pm$0.03 & 31.36$\pm$0.04 & 32.66$\pm$0.13 & 41.58$\pm$0.13 & 1.30$\pm$0.14 &	 -0.41 $_{- 0.14 }^{+ 0.14 }$ 	 & 	 -- 	 & 	 -- & -- \\ 
        {\bf RACS~J2252$-$51}  & 5.18 & 31.61$\pm$0.03 & 31.80$\pm$0.03 & 33.70$\pm$0.11 & 42.48$\pm$0.11 & 1.90$\pm$0.11 &	 -0.67 $_{- 0.11 }^{+ 0.10 }$ 	 & 	 1.01 $_{- 0.03 }^{+ 0.03 }$ 	 & 	 1.9 $_{- 0.08 }^{+ 0.09 }$ & $>1.14^1$ \\ 
        {\bf RACS~J2303$-$23}  & 5.03 & 31.37$\pm$0.03 & 31.34$\pm$0.04 & 33.66$\pm$0.14 & 42.86$\pm$0.14 & 2.32$\pm$0.15 &-0.19 $_{- 0.17 }^{+ 0.14 }$ 	 & 	 0.59 $_{- 0.08 }^{+ 0.08 }$ 	 & 	 0.52 $_{- 0.12 }^{+ 0.17 }$    & -- \\ 
        {\bf RACS~J2329$-$15}  & 5.84 & 30.99$\pm$0.05 & 31.11$\pm$0.05 & 34.19$\pm$0.07 & 43.78$\pm$0.07 & 3.08$\pm$0.09 &	 0.82 $_{- 0.01 }^{+ 0.01 }$ 	 & 	 -- 	 & 	 --   & 1.36$\pm$0.06$^{16}$ \\ 
        {\bf RACS~J2329+30 }   & 5.24 & 31.61$\pm$0.03 & 31.68$\pm$0.03 & 33.38$\pm$0.07 & 42.59$\pm$0.07 & 1.70$\pm$0.08 &	 0.10 $_{- 0.02 }^{+ 0.03 }$ 	 & 	 $-$0.94 $_{- 0.06 }^{+ 0.06 }$ 	 & 	 1.33 $_{- 0.10 }^{+ 0.09 }$ & -- \\ 
        {\bf RACS~J2329$-$20 } & 5.04 & 31.93$\pm$0.03 & 32.09$\pm$0.03 & 33.14$\pm$0.07 & 42.68$\pm$0.07 & 1.05$\pm$0.08 &	 0.71 $_{- 0.04 }^{+ 0.04 }$ 	 & 	 -- 	 & 	 --   & -- \\ 
        {\bf RACS~J2344+16 }   & 5.00 & 31.81$\pm$0.03 & 31.96$\pm$0.03 & 33.86$\pm$0.07 & 42.82$\pm$0.07 & 1.90$\pm$0.08 &	 -0.35 $_{- 0.04 }^{+ 0.03 }$ 	 & 	 0.46 $_{- 0.07 }^{+ 0.07 }$ 	 & 	 0.94 $_{- 0.10 }^{+ 0.07 }$    & -- \Bstrut\\ 

\hline
\hline 

    \end{tabular}
    \tablefoot{{\bf (Col. 1)} Name of the source; {\bf (Col. 2)} spectroscopic redshift; {\bf (Col. 3 and 4)} UV (25500\AA) and optical (4400\AA) rest-frame luminosity computed from the observed magnitudes in the $z$ and $W1$ filters;  {\bf (Col. 5 and 6)} Rest-frame radio luminosity at 5~GHz and integrated radio luminosity at 1.4~GHz computed from the best-fit radio spectrum of each source; {\bf (Col. 7)} radio-loudness parameter, defined as R~$=S_{\rm 5GHz}/S_{4400\AA}$; {\bf (Col. 8 and 9)} Best-fit spectral indices before and after the break frequency. Only value is shown for objects where a single power law was preferred; {\bf (Col. 10)} best-fit break frequency, in GHz and in the observed frame, for systems where a broken power law was preferred; {\bf (Col. 11)}  $\tilde{\alpha}_{\rm ox}$ parameter, defined as $\tilde{\alpha}_{\rm ox}=0.303\times{\rm log(L_{\rm 10keV}/L_{2500\AA})}$. Names marked in boldface indicate sources selected using the criteria described in the text. The references for the X-ray information are as follows: (1) \cite{Merloni2024}; (2) \cite{An2020b}; (3) \cite{Belladitta2019}; (4) \cite{Ighina2024b}; (5) \cite{Moretti2021}; (6) Ighina et al. in preparation; (7) This work; (8) \cite{Zuo2024}; (9) \cite{Wu2013}; (10) \cite{Wolf2021}; (11) \cite{Caccianiga2024}; (12) \cite{Ighina2019}; (13) \cite{Ghisellini2013}; (14) \cite{Khorunzhev2021}; (15) \cite{Wolf2024}; (16) \cite{Connor2021}.}  

\label{tab:lum_RACS_sample}
\end{table}
\end{landscape}
}

{\small
%
\bibliographystyle{aa} 
\bibliography{referenze} 
%
}

\begin{appendix}
\onecolumn

\section{Candidates not confirmed}
\label{sec:appenA}
\begin{figure}[th]
\centering
    \includegraphics[width=0.365\hsize]{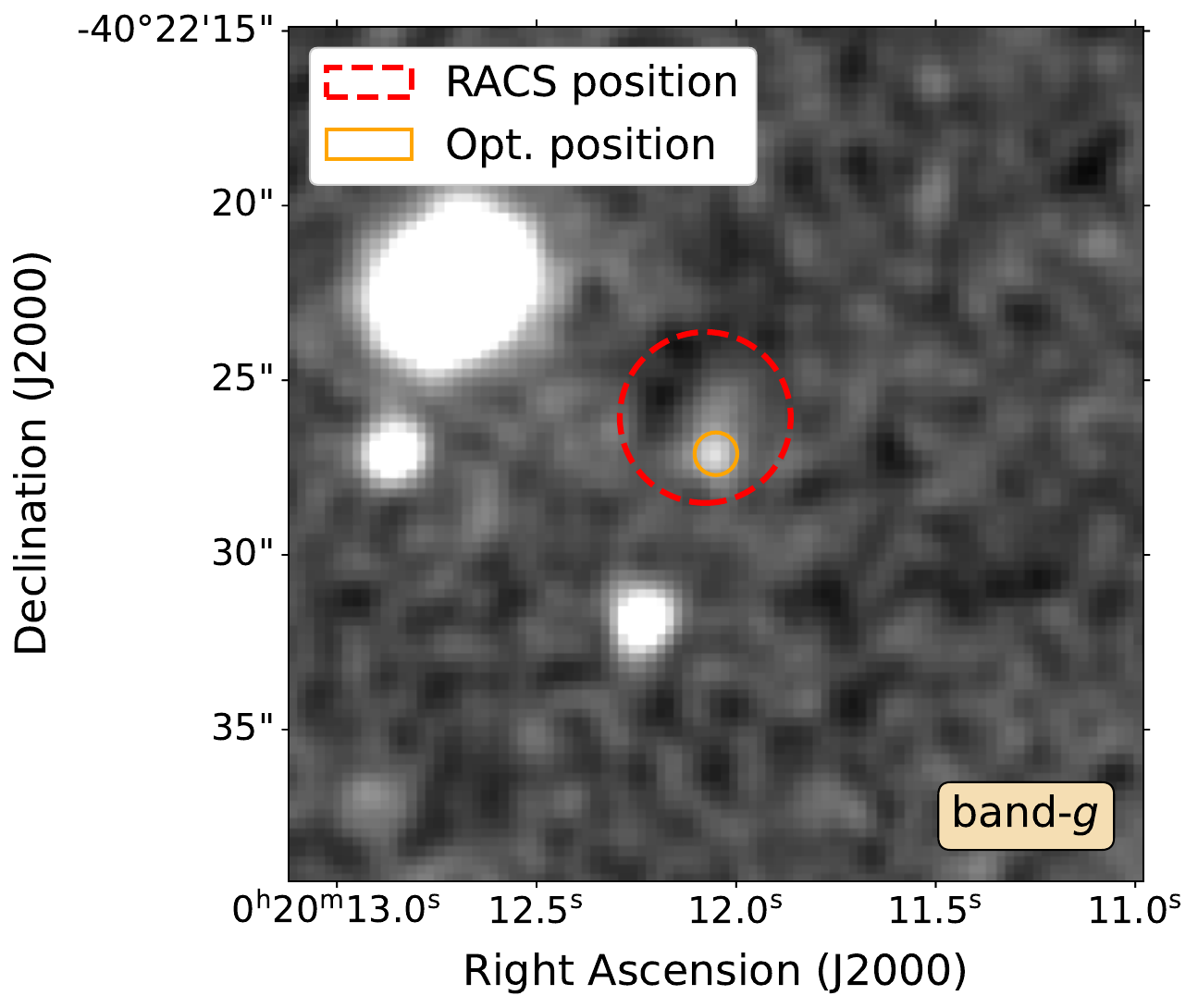}
    \includegraphics[width=0.625\hsize]{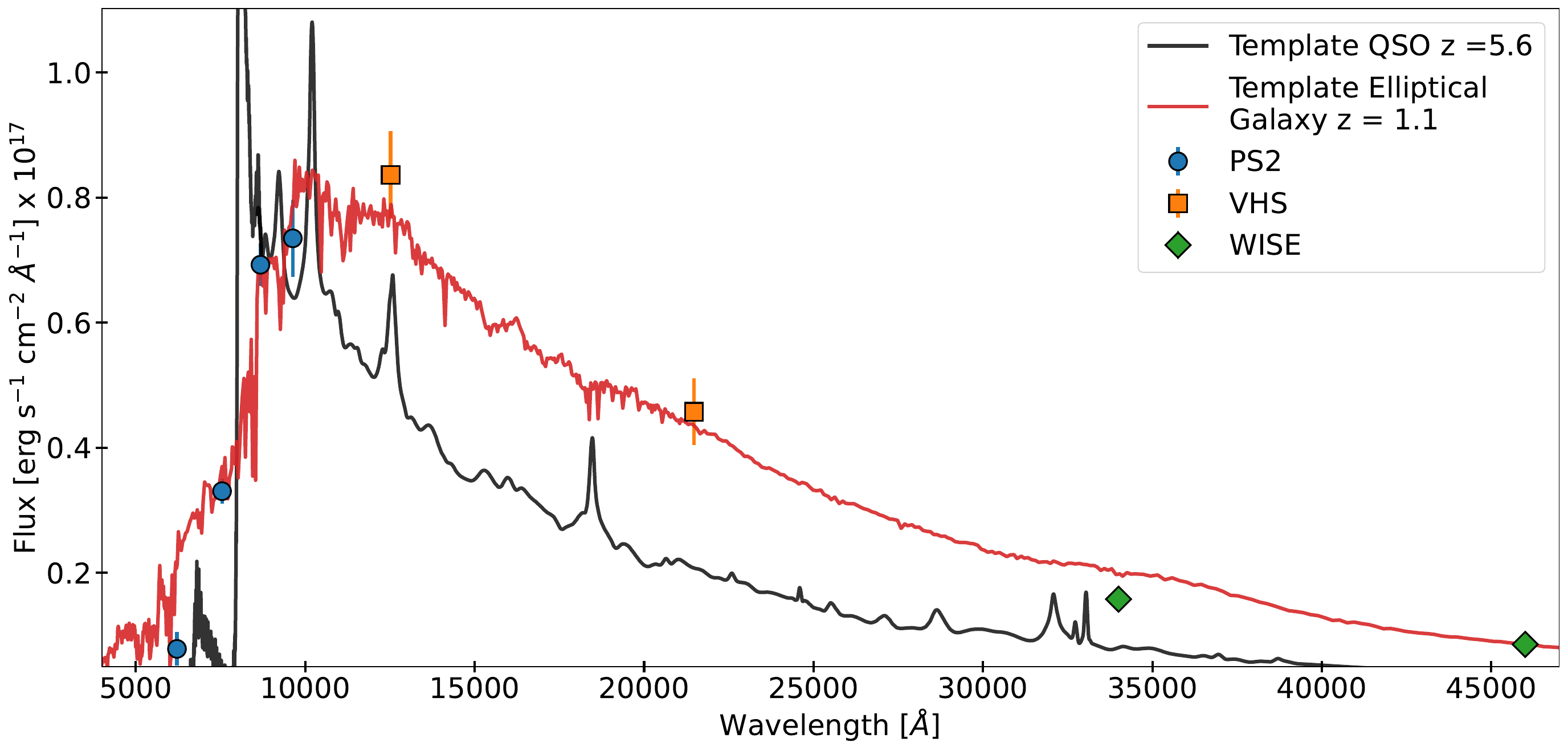}\\
   	\includegraphics[width=0.59\hsize]{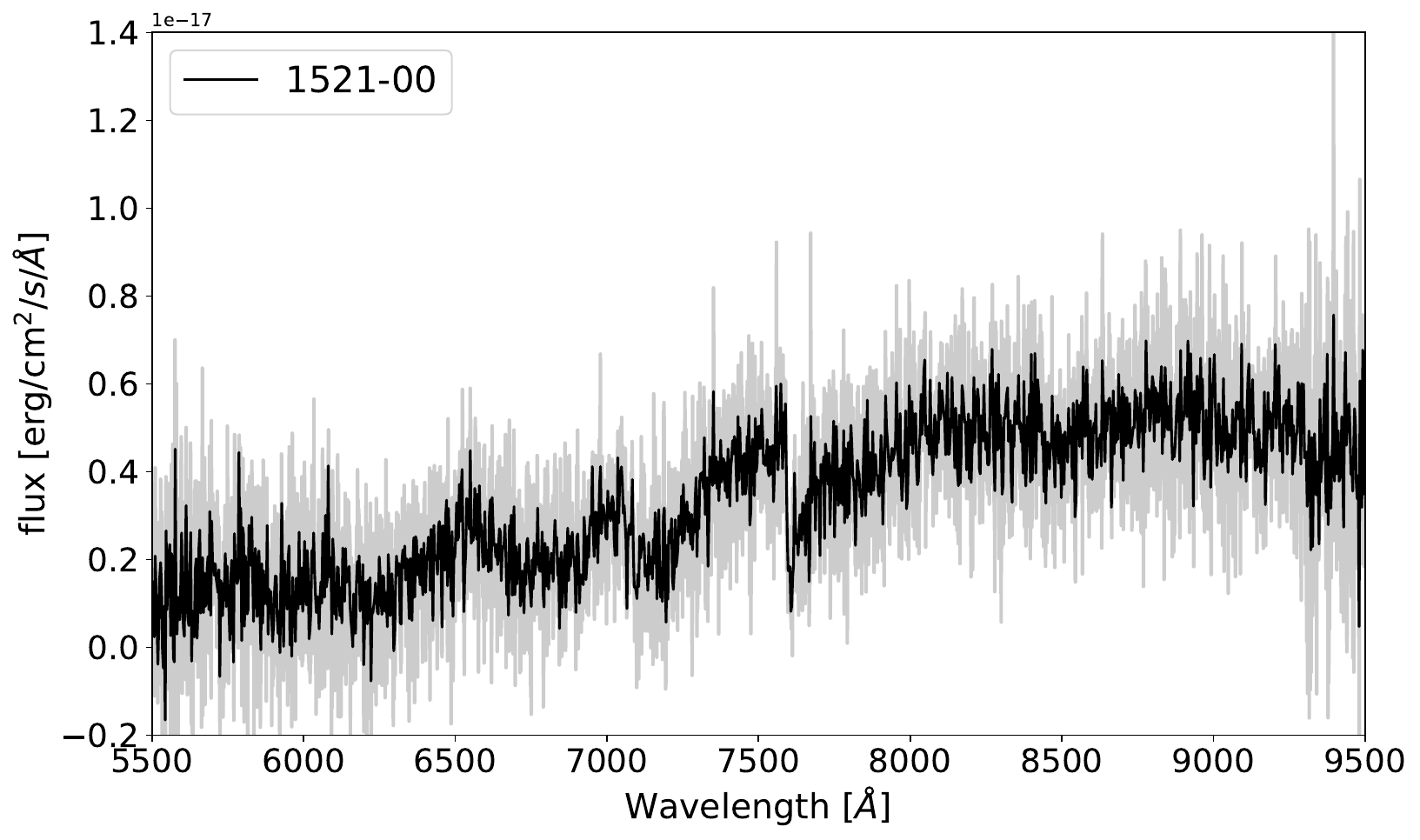}
    \includegraphics[width=0.4\hsize]{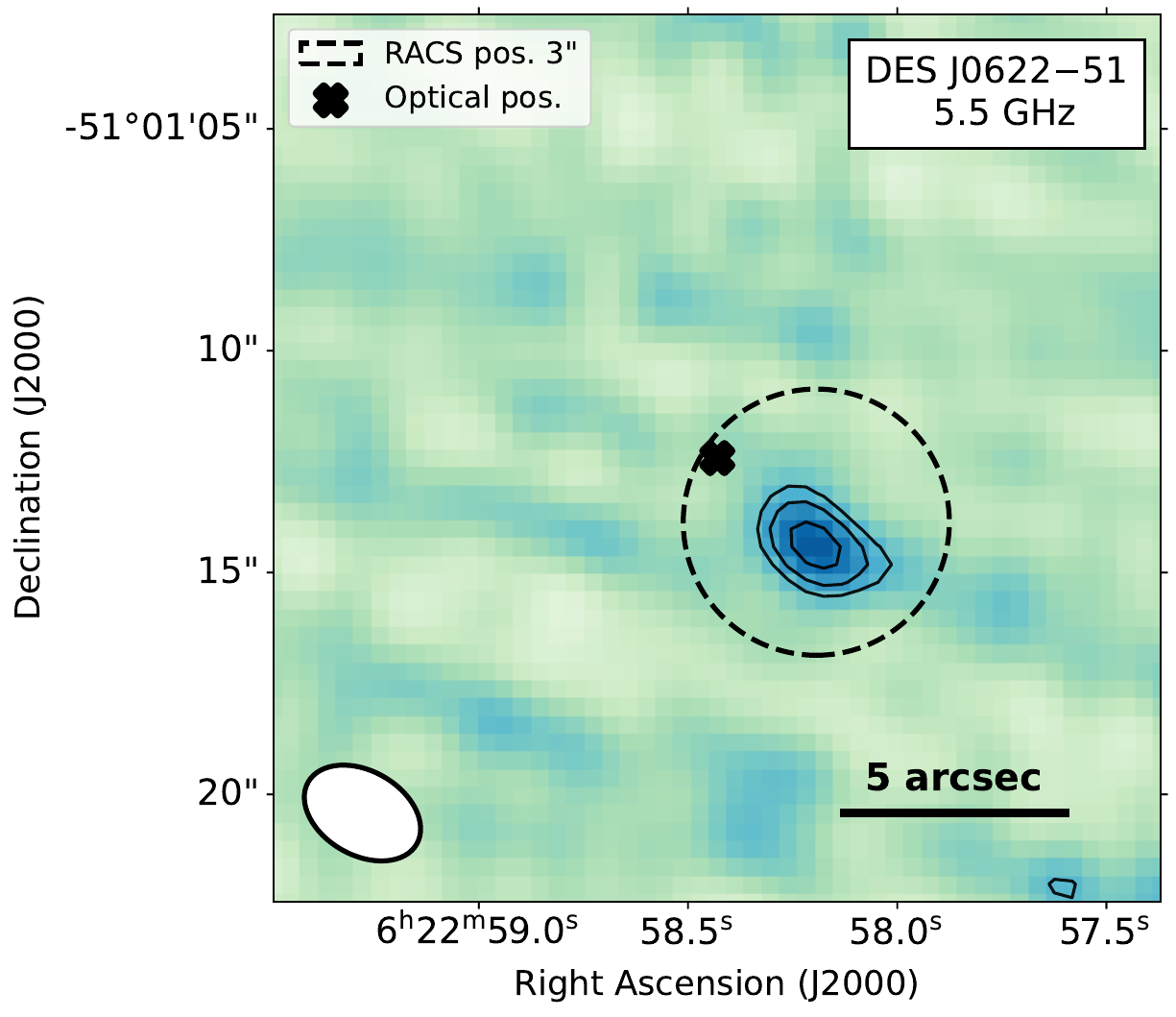}

      \caption{
      {\bf Top-left panel}: Example of high-$z$ candidate selected based on the initial optical/IR selections from DES and catWISE, but then discarded after a visual inspection of its $g$-band image. Note that the image has been smoothed with a Gaussian filter in order to make the faint signals at the position of the optical source more evident.
      {\bf Top-right panel}: Example of a selected candidate which was then discarded based on the NIR photometric points (from VHS). The black solid line is a template for high-$z$ quasars and the red solid line is a template for an elliptical galaxy. In this specific case, the source is likely an elliptical galaxy at $z\sim1.1$.
      {\bf Bottom-left panel:} Example of the LBT/MODS spectrum of a candidate (RACS~J1521$-$00) not confirmed to be at high redshift. In this specific case, the target is likely a type M brown dwarf. The grey spectrum is at a resolution of $\sim$0.9~\AA, whereas the black one is binned at $\sim$3.5~\AA. {\bf Bottom-right panel:} ATCA image at 5.5~GHz of RACS~J0622$-$51, a candidate selected based on its optical/NIR colours and a radio association in the RACS-low survey  within 3$''$ (dashed circle). The black cross shows the optical position of the high-$z$ candidate. The radio source detected with the ATCA revealed the RACS-low association to be spurious, with the radio source being $>10\sigma$ away from the optical counterpart, with $\sigma=0.24''$ the uncertainty on the ATCA position derived from the fit of the radio source. The black contours in the image are drawn at -3, 3, 3$\sqrt{2}$, 6 times the RMS of the image (RMS~=~27~\textmu Jy~beam$^{-1}$).}
    \label{fig:discard_cand}
\end{figure}

In this section we report the sources selected as high-$z$ candidates which turned out to be at low redshift or spurious radio-optical associations. 
In particular, in the top-left panel of Fig. \ref{fig:discard_cand} we report the example of a candidate detected in the g-band even after the requirement of a magnitude $> 24$ (in DES). In the top-right panel of Fig. \ref{fig:discard_cand}, we show the example of a candidate discarded after comparing its NIR colours from the VHS survey against an elliptical galaxy template (from \citealt{Polletta2007}). From the plot, it is clear that the selected candidate is more likely an elliptical galaxy at $z \sim 1.1$ which satisfied the point-like criteria detailed in Sec. \ref{sec:selection}.

In Tab. \ref{tab:list_contaminants} we list the objects selected as good high-$z$ radio quasars, but which turned out to be either stars or radio galaxies at low redshift (see e.g. Fig. A.2, right).
For the majority of these sources the high-$z$ nature was discarded based on spectroscopic observations (see Fig. \ref{fig:discard_cand}, bottom-left panel, for a brown dwarf selected as high-$z$ quasar candidate). In the case of RACS~J0622$-$51 follow-up radio observations with the ATCA telescope (see Sec. \ref{sec:radio_observations}) revealed that the association of the optical candidate with the RACS-low source was likely spurious (see bottom-right panel in Fig. \ref{fig:discard_cand}). Indeed, the more accurate radio position derived from ATCA is 3.1\arcsec away from the optical source, corresponding to $>10$ times the positional uncertainty derived from the fit ($\sigma=0.24$\arcsec).

\begin{landscape}
\begin{table}
	\centering
\renewcommand{\arraystretch}{1.1}
	\begin{tabular}{lcccccccccc}
	Target & R.A. & Dec. & drop & $S_{\rm 888MHz}$ & mag$z$ & drop & mag$z$--mag$W2$ & d$_{\rm ro}$ & Telescope\\
	 & (deg) & (deg) & (Survey) & (mJy~beam$^{-1}$) & (mag AB) & & (AB -- Vega) & (arcsec) &\\
	\hline
	\hline
	RACS~J0010$-$53 &    2.526012 &  $-$53.589526 & $r$-drop  (D)  & 1.2$\pm$0.4  & 20.84$\pm$0.04 & 1.64 & 3.35 & 1.47 & VLT \\
    RACS~J0249$-$28 &   42.311270 &  $-$28.855045 & $i$-drop  (D)  & 5.1$\pm$0.5  & 20.83$\pm$0.03 & 1.23 & 3.09 & 2.01 & VLT \\
    RACS~J0420$-$47 &   65.043381 &  $-$47.723408 & $r$-drop  (D)  & 10.3$\pm$0.7 & 21.22$\pm$0.04 & 1.62 & 3.63 & 2.13 & VLT \\
    RACS~J0427$-$53 &   66.930894 &  $-$53.393840  & $i$-drop (D)  & 3.6$\pm$0.6  & 20.04$\pm$0.01 & 1.08 & 3.20 & 1.47 & VLT \\
    RACS~J0616$-$48 &   94.199469 &  $-$48.347094 & $i$-drop  (D)  & 2.1$\pm$0.8  & 21.24$\pm$0.03 & 1.39 & 3.37 & 2.76 & VLT &  \\
    RACS~J0622$-$51 &   95.743477 &  $-$51.020087 & $r$-drop  (D)  & 1.9$\pm$0.7  & 20.74$\pm$0.03 & 1.82 & 3.72 & 2.71 & ATCA$^*$ \\
    RACS~J0905$-$10  & 136.262847 & $-$10.426427 & $r$-drop  (P)   & 6.2$\pm$1.4  & 20.15$\pm$0.07 & 1.58 & 2.41 & 1.14 & LBT &  \\
    RACS~J1521$-$00  & 230.377454 &  $-$0.300307  & $r$-drop (P)   & 12.6$\pm$0.5 & 20.98$\pm$0.16 & 1.27 & 3.51 & 0.37 & LBT &  \\
    RACS~J1552$-$29  & 238.041957 & $-$29.255812 & $r$-drop  (P)   & 6.4$\pm$0.3  & 19.92$\pm$0.05 & 1.28 & 3.72 & 1.20 & AAT \\
    RACS~J1557$-$18  & 239.321655 & $-$18.041279 & $r$-drop  (P)   & 1.2$\pm$0.4  & 20.36$\pm$0.03 & 1.91 & 4.24 & 1.12 & LBT &  \\
    RACS~J1614+14    & 243.735217 &   +14.315941  & $r$-drop  (P)  & 1.6$\pm$0.4  & 20.27$\pm$0.03 & 1.53 & 3.80 & 0.91 & LBT &  \\
    RACS~J1722+29    & 260.728554 &   +29.108219  & $i$-drop  (P)  & 1.5$\pm$0.4  & 21.29$\pm$0.13 & 1.64 & $<2.3$ & 1.30 & LBT &  \\
    RACS~J1802+10    & 270.599597 &   +10.922134  & $r$-drop  (P)  & 6.1$\pm$0.7  & 20.27$\pm$0.06 & 1.58 & 3.89 & 1.02 & LBT &  \\
    RACS~J2015$-$17  & 303.930440 & $-$17.577126 & $r$-drop   (P)  & 1.6$\pm$0.3  & 20.96$\pm$0.09 & 1.82 & 3.90 & 0.27 & TNG \\
    RACS~J2021$-$20  & 305.411544 & $-$20.655044 & $r$-drop   (P)  & 2.5$\pm$0.5  & 20.67$\pm$0.04 & 1.52 & 3.61 & 1.23 & AAT \\
    RACS~J2044$-$16  & 311.087282 & $-$16.383996 & $r$-drop   (P)  & 1.5$\pm$0.4  & 20.70$\pm$0.06 & 1.74 & $<1.7$& 1.11 & AAT \\
    RACS~J2103$-$40 &  315.773997 &  $-$40.697616 & $r$-drop  (D)  & 3.0$\pm$0.5  & 20.97$\pm$0.04 & 1.96 & 3.97 & 2.91 & VLT &  \\
    \hline
    Not observed yet \\
    \\
    RACS~J0632$-$59 & 98.245905  & $-$59.567634 & $r$-drop    (D)     & 1.2$\pm$0.4  & 21.01$\pm$0.04 & 1.81 & 2.93 & 1.14 & -- \\
    RACS~J0721+34    & 110.480455 & +34.5776441  & $i$-drop   (P)     & 2.1$\pm$0.5  & 21.29$\pm$0.15 & 3.74 & $<2.3$ & 1.42 & -- \\ 
    RACS~J0937$-$21  & 144.310777 & $-$21.019468 & $i$-drop   (P)     & 1.0$\pm$0.3  & 20.82$\pm$0.03 & 1.22 & $<1.8$ & 2.85 & -- \\ 
    RACS~J1536$-$28  & 234.204898 & $-$28.4014139 & $i$-drop   (P)    & 1.8$\pm$0.3  & 21.04$\pm$0.09 & 1.53 & 4.62 & 2.05 & -- \\

	\hline
		
	\end{tabular}
\caption{Properties, used for the selection, of the non-confirmed candidates. {\bf Col. (1, 2, 3):} name and coordinates of the source; {\bf Col. (4):} type of dropout and survey the source was selected from, in brackets; {\bf Col. (5):} peak flux density at 888~MHz from the RACS-low source lists; {\bf Col. (6):} $z$-band magnitude from DES or Pan-STARRS; {\bf Col. (7):} dropout $r-i$ or $i-z$ depending on the type of dropout in Col. 4; {\bf Col. (8):}  $z-W2$ colour, where $z$ is in the AB system and $W2$ in the Vega one. In case of a non detection in the $W2$ filter, we assumed an upper limit of mag$W2$~=~19; {\bf Col. (9):} radio-to-optical distance; {\bf Col. (10):} telescope used for the identification. The `*' indicates radio telescopes used to get more accurate position of the radio counterpart and discard the radio-optical association.}  
\label{tab:list_contaminants}

\end{table}
\end{landscape}

\section{Dedicated radio observations}
\label{sec:radio_observations}

In this section we list the dedicated radio observations available for the high-$z$ RACS sources that we used in order to build their radio spectra (see Fig. \ref{fig:radio_spectra_pt1}, \ref{fig:radio_spectra_pt2} and \ref{fig:radio_spectra_zless5}).
In particular, for $z>5$ sources already discussed in the literature, we considered the dedicated radio observations discussed in: \cite{Shao2020,Shao2022} for RACS~J0131$-$03, RACS~J0741+25, RACS~J0836+00, RACS~J1026+25, RACS~J1034+20, RACS~J1146+40 and RACS~J1427+33; \cite{Gloudemans2023} for RACS~J0309+27 and RACS~J2201+23; \cite{Keller2024} for RACS~J1427+33; \cite{An2023} for RACS~J1702+13.

In addition, we also performed observations for a sub-set of the newly identified $z>4.5$ radio quasars with the following radio telescopes radio telescopes (PI: Ighina): upgraded Giant Metrewave Radio Telescope (uGMRT) for RACS~J0320$-$35 and RACS~J0322$-$18 (project 42\_001), MeerKAT for RACS~J2252$-$51 (project DDT-20231128-LI-01) and ATCA (for 10 targets; projects C3477, C3535, C3601). 
The data reduction for the uGMRT and ATCA data-sets was performed similarly to \cite{Ighina2024,Ighina2024b}. Here we provide a short summary of the main steps.

uGMRT observations of RACS~J0320$-$35 and RACS~J0320$-$18 were performed using the GMRT wideband backend (GWB; \citealt{Reddy2017} with a bandwidth of 200~MHz) in band-3 (centred at 400~MHz) and in band-4 (centred at 650~MHz). The data reduction was performed using the CAsa Pipeline-cum-Toolkit for Upgraded GMRT data REduction (\texttt{CAPTURE}; \citealt{Kale2021}) code by applying further flagging depending on the specific observation and by applying four cycles of phase self-calibration. For the imaging analysis, we adopted a robust parameter of 0.5 (i.e. a compromise between resolution and sensitivity). 

ATCA observations were carried out at three different central frequencies: 2.1, 5.5 and 9~GHz (all with a bandwidth of 2~GHz). To process the data (calibration and imaging), we used the \texttt{MIRIAD} data-reduction package \citep{Sault1995} following a standard reduction. We also applied 1-3 cycles of phase self-calibration to these data-sets, depending on the brightness of the objects in the filed of each target. For imaging, we adopted a robust parameter of 0.5. 

Dedicated MeerKAT observations for RACS~J2252$-$51 were performed on 2023 December 01-02-03 in three different bands: L (centred at 1.3~GHz), S0 (centred at 2.1~GHz) and S4 (centred at 3.1~GHz). We reduced the data using {\sc Oxkat}\footnote{\url{https://github.com/IanHeywood/oxkat}} \citep[v0.4;][]{Heywood2020}, where the Common Astronomy Software Applications \citep[CASA;][]{2007ASPC..376..127M} package and {\sc Tricolour}\footnote{\url{https://github.com/ska-sa/tricolour}} \citep{2022ASPC..532..541H} were used for measurement sets splitting, cross-calibration, and flagging. {\sc CubiCal}\footnote{\url{https://github.com/ratt-ru/CubiCal}} \citep{2018MNRAS.478.2399K} was used for self-calibration, and {\sc Wsclean} \citep{2014MNRAS.444..606O} was used for continuum imaging. All processes were executed with {\sc Oxkat} default settings. 

For all these different observations (uGMRT, ATCA and MeerKAT) we performed a 2D Gaussian fit to the radio emission obtained in the final images using the \texttt{CASA} \citep{Mcmullin2007} software. We report in Tab. \ref{tab:new_radio_data} the results of the fit for these new radio observations.
We consider a further 5\% error in the flux density estimates obtained, added in quadrature, to account for uncertainties related to the absolute flux scale and the reduction process.

\begin{table}
    \centering
    \caption{Best-fit flux estimated from the dedicated radio observations of a sub-set of sources within the RACS sample.}  

\renewcommand{\arraystretch}{1.1}
\begin{tabular}{l c c c c c c c }
Source & Central Frequency & Flux & Flux error & Telescope \\
 & (GHz) & (mJy) & mJy &  \\
    \hline
    \hline

RACS J0036$-$37 &	2.1 	&	0.67 	&	0.13 	&	ATCA\\
 	&	5.5 	&	0.73 	&	0.07 	&	ATCA\\
 	&	9.0 	&	0.39 	&	0.07 	&	ATCA\\
    \hline

RACS J0110$-$39 &	2.1 	&	10.1 &	0.3 &	ATCA \\
 	&	5.5 	&	5.3 	&	0.4 	&	ATCA \\
 	&	9.0 	&	3.7 	&	0.2 	&	ATCA \\
    \hline

RACS J0127$-$44 &	2.1 	&	13.7 	&	0.2 	&	ATCA \\
 	&	5.5 	&	13.7 	&	0.2 	&	ATCA \\
 	&	9.0 	&	12.1 	&	0.3 	&	ATCA\\
    \hline

RACS J0202$-$17 &	2.1 &	16.1 &	0.1 &	ATCA \\
 &	5.5 &	19.1 &	0.1 &	ATCA \\
 &	9.0 &	13.8 &	0.1 &	ATCA \\
    \hline

RACS J0209$-$56 &	2.1 &	13.0 &	0.2 &	ATCA \\
 &	5.5 &	10.4 &	0.1 &	ATCA \\
 &	9.0 &	9.5 &	0.1 &	ATCA \\
    \hline

RACS J0320$-$35 &	0.4 &	5.6 &	0.9 &	uGMRT \\
 &	0.65 &	4.0 &	0.2 &	uGMRT \\
 &	2.1 &	2.09 &	0.09 &	ATCA \\
 &	5.5 &	0.84 &	0.03 &	ATCA \\
 &	9.0 &	0.63 &	0.03 &	ATCA \\
    \hline

RACS J0322$-$18 &	0.4 &	4.0 &	0.3 &	uGMRT \\
 &	0.65 &	2.5 &	0.1 &	uGMRT \\
 &	2.1 &	1.28 &	0.06 &	ATCA \\
 &	5.5 &	0.61 &	0.03 &	ATCA \\
 &	9.0 &	0.44 &	0.03 &	ATCA \\
    \hline

RACS J1011$-$01 &	5.5 &	4.0 &	0.2 &	ATCA \\
 &	9.0 &	2.7 &	0.2 &	ATCA \\
    \hline

RACS J1322$-$13 &	5.5 &	53.27 &	0.2 &	ATCA \\
 &	9.0 &	46.60 &	0.1 &	ATCA \\
     \hline

RACS J1426$-$11 &	5.5 &	2.54 &	0.2 &	ATCA \\
 &	9.0 &	1.6 &	0.1 &	ATCA \\
    \hline

RACS J2020$-$62 &	2.1 &	0.82 &	0.20 &	ATCA \\
 &	5.5 &	0.80 &	0.08 &	ATCA \\
 &	9.0 &	0.43 &	0.05 &	ATCA \\
    \hline

RACS J2252$-$51 &	1.3 &	16.91 &	0.07 &	MeerKAT \\
 &	2.2 &	15.43 &	0.05 &	MeerKAT \\
 &	3.1 &	12.30 &	0.04 &	MeerKAT \\
 &	2.1 &	16.1 &	0.2 &	ATCA \\
 &	5.5 &	6.7 &	0.1 &	ATCA \\
 &	9.0 &	4.0 &	0.1 &	ATCA \\
 &	5.5 &	6.51 &	0.05 &	ATCA \\
 &	9.0 &	3.65 &	0.03 &	ATCA \BBstrut \\
\hline
\hline 
    \end{tabular}
    \caption{Flux densities derived from dedicated observations of a sub-sample of the RACS high-$z$ quasars. The errors reported here are the ones derived from the 2D Gaussian fit of the source in the given image and do not take the uncertainty in the flux scale calibration (assumed to be 5\%) into account.}  

	\label{tab:new_radio_data}
\end{table}

\section{Properties of the $z<5$ RACS quasars}
\label{sec:radio_zless5}
\begin{figure*}[ht]
   \centering
	\includegraphics[width=\hsize]{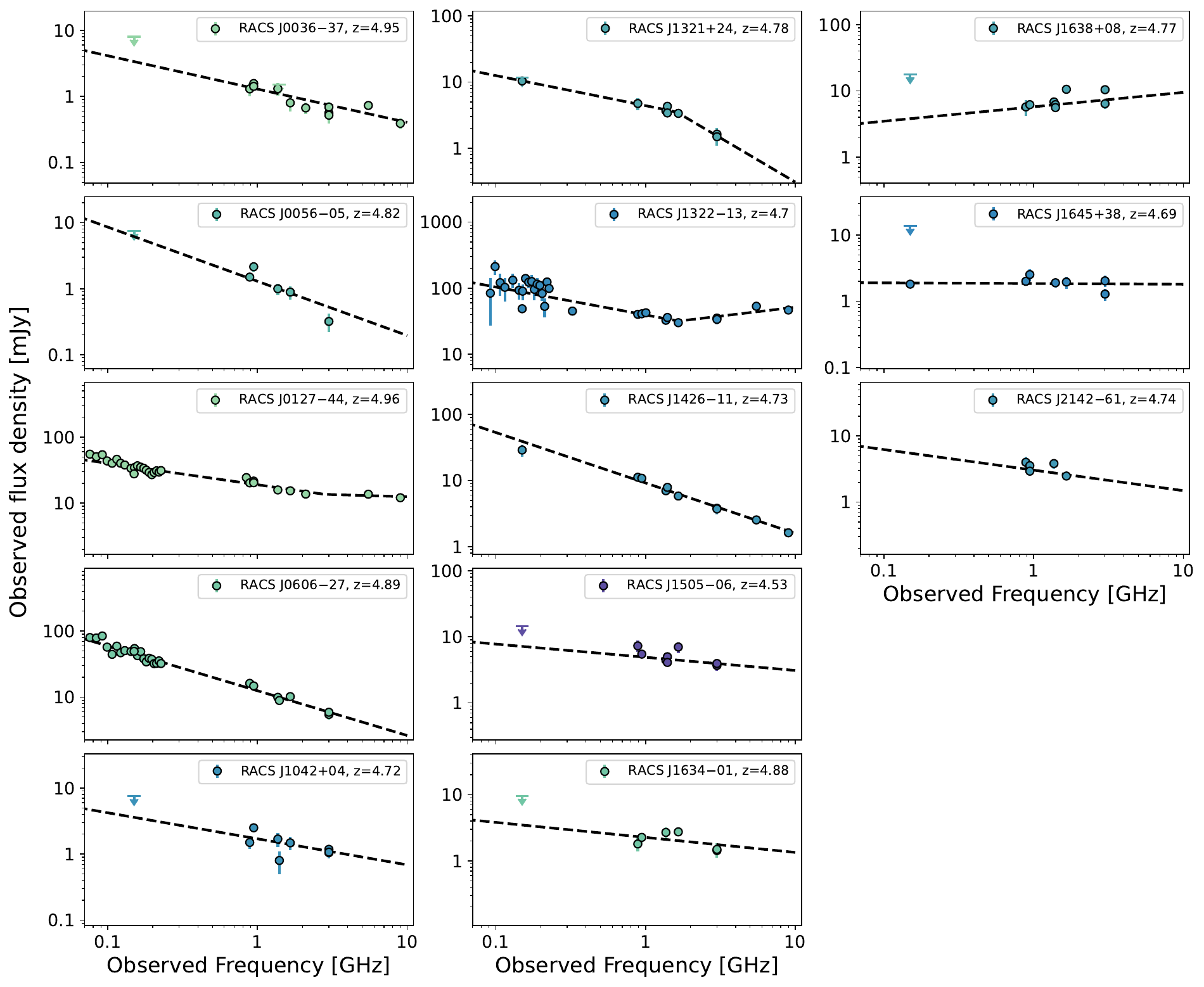}
\caption{Radio spectra of the $4.5<z<5$ radio quasars discovered from RACS-low. The data-points are colour-coded based on their redshift (same scale of Fig. \ref{fig:opt_spec_sample}) and are mainly from public surveys with some sources having dedicated observations (see Tab. \ref{tab:new_radio_data}).}
    \label{fig:radio_spectra_zless5}
\end{figure*}

In this section we report the multi-wavelength properties of the $z<5$ radio quasars identified from our selection, similarly to the ones reported in the main text for the well-defined sample at $z>5$. In particular, in Fig. \ref{fig:radio_spectra_zless5} we show their radio spectra built from publicly available surveys (see Sec. \ref{sec:radio_properties}) and dedicated observations (see Sec. \ref{sec:radio_observations}). Moreover, we also report their optical and radio luminosities, as well as radio spectral indices and R and $\tilde{\alpha}_{\rm ox}$ parameters in Tab. \ref{tab:lum_RACS_zless5}.

 \begin{landscape}

\begin{table*}[ht]
    \centering
    \caption{Multi-wavelength properties of the $z<5$ RACS quasars.}
    \renewcommand{\arraystretch}{1.25}
    \begin{tabular}{l c c c c c c c c c c}
Name & $z$ & log(L$_{2500\text{\normalfont\AA}}$) & log(L$_{4400\text{\normalfont\AA}}$) & log(L$_{\rm 5GHz}$) & log($\nu$L$_{\rm 1.4GHz}$) & log(R) & ${\alpha}_{\rm low}^{\rm r}$ & ${\alpha}_{\rm high}^{\rm r}$ & ${\nu}_{\rm break}$ & $\tilde{\alpha}_{\rm ox}$\Bstrut\\
 & & (erg~s$^{-1}$~Hz$^{-1}$) & (erg~s$^{-1}$~Hz$^{-1}$) & (erg~s$^{-1}$~Hz$^{-1}$) & (erg~s$^{-1}$) & & & GHz & & \\
    \hline
    \hline

        {\bf RACS~J0036$-$37}  & 4.95 & 31.20$\pm$0.03 & 31.46$\pm$0.03 & 32.78$\pm$0.07 & 42.21$\pm$0.07 & 1.32$\pm$0.08 	&	 0.50 $_{- 0.04 }^{+ 0.04 }$ 	 & 	 -- 	 & 	 -- 	& $>1.03$ \\
        {\bf RACS~J0056$-$05}  & 4.82 & 30.80$\pm$0.04 & 31.03$\pm$0.04 & 32.77$\pm$0.09 & 42.38$\pm$0.09 & 1.74$\pm$0.10 	&	 0.82 $_{- 0.01 }^{+ 0.01 }$ 	 & 	 -- 	 & 	 -- 	& --\\
        {\bf RACS~J0127$-$44}  & 4.96 & 30.95$\pm$0.03 & 30.99$\pm$0.04 & 33.94$\pm$0.07 & 43.26$\pm$0.07 & 2.95$\pm$0.08 	&	 0.32 $_{- 0.14 }^{+ 0.14 }$ 	 & 	 0.07 $_{- 0.06 }^{+ 0.06 }$ 	 & 	 2.82 $_{- 0.49 }^{+ 0.55 }$ 	& 1.12$\pm$0.08\\ 
        {\bf RACS~J0606$-$27}  & 4.89 & 31.16$\pm$0.03 & 31.41$\pm$0.03 & 33.77$\pm$0.07 & 43.29$\pm$0.07 & 2.36$\pm$0.08 	&	 0.68 $_{- 0.11 }^{+ 0.10 }$ 	 & 	 -- 	 & 	 -- 	& $>1.05$ \\ 
        {\bf RACS~J1042+04}    & 4.74 & 30.81$\pm$0.04 & 30.90$\pm$0.05 & 32.85$\pm$0.09 & 42.22$\pm$0.09 & 1.95$\pm$0.10 	&	 0.39 $_{- 0.17 }^{+ 0.14 }$ 	 & 	 -- 	 & 	 -- 	& $>0.94$ \\
        {\bf RACS~J1321+24}    & 4.78 & 30.81$\pm$0.04 & 30.90$\pm$0.05 & 33.28$\pm$0.07 & 42.68$\pm$0.07 & 2.38$\pm$0.09 	&	 0.46 $_{- 0.01 }^{+ 0.01 }$ 	 & 	 1.34 $_{- 0.39 }^{+ 0.39 }$ 	 & 	 1.60 $_{- 0.33 }^{+ 0.33 }$ 	& --\\
        {\bf RACS~J1322$-$13}  & 4.70 & 31.05$\pm$0.03 & 31.22$\pm$0.04 & 34.22$\pm$0.07 & 43.59$\pm$0.07 & 3.01$\pm$0.08 	&	 0.42 $_{- 0.02 }^{+ 0.03 }$ 	 & 	 $-$0.27 $_{- 0.03 }^{+ 0.03 }$ 	 & 	 1.66 $_{- 0.08 }^{+ 0.10 }$ 	& 1.02$\pm$0.03 \\ 
        {\bf RACS~J1426$-$11}  & 4.73 & 30.80$\pm$0.04 & 30.84$\pm$0.07 & 33.61$\pm$0.07 & 43.59$\pm$0.07 & 2.77$\pm$0.10 	&	 0.76 $_{- 0.04 }^{+ 0.04 }$ 	 & 	 -- 	 & 	 -- 	& $>0.97$ \\
        {\bf RACS~J1505$-$06}  & 4.53 & 30.93$\pm$0.04 & 31.07$\pm$0.04 & 33.27$\pm$0.09 & 42.53$\pm$0.09 & 2.20$\pm$0.10 	&	 0.20 $_{- 0.04 }^{+ 0.03 }$ 	 & 	 -- 	 & 	 -- 	& $>1.05$ \\ 
        {\bf RACS~J1634$-$01}  & 4.88 & 31.10$\pm$0.03 & 31.13$\pm$0.04 & 32.99$\pm$0.09 & 42.26$\pm$0.09 & 1.86$\pm$0.10 	&	 0.23 $_{- 0.04 }^{+ 0.04 }$ 	 & 	 -- 	 & 	 -- 	& -- \\ 
        {\bf RACS~J1638+08}    & 4.77 & 31.10$\pm$0.03 & 31.31$\pm$0.03 & 33.35$\pm$0.10 & 42.38$\pm$0.10 & 2.04$\pm$0.11 	&	 $-$0.22 $_{- 0.09 }^{+ 0.08 }$ 	 & 	 -- 	 & 	 -- 	& --\\ 
        {\bf RACS~J1645+38}    & 4.69 & 30.74$\pm$0.04 & 30.79$\pm$0.05 & 32.86$\pm$0.07 & 42.01$\pm$0.07 & 2.07$\pm$0.09 	&	 0.01 $_{- 0.01 }^{+ 0.01 }$ 	 & 	 -- 	 & 	 --    & --\\ 
        {\bf RACS~J2142$-$61}  & 4.74 & 31.31$\pm$0.03 & 31.43$\pm$0.03 & 33.10$\pm$0.12 & 42.42$\pm$0.12 & 1.67$\pm$0.12 	&	 0.31 $_{- 0.02 }^{+ 0.02 }$ 	 & 	 -- 	 & 	 -- 	& $>1.06$ \BBstrut\\ 
    \hline
    \hline

    \end{tabular}
    \tablefoot{{\bf (Col. 1)} Name of the source; {\bf (Col. 2)} spectroscopic redshift; {\bf (Col. 3} and {\bf 4)} UV (25500\AA) and optical (4400\AA) rest-frame luminosity computed from the observed magnitudes in the $z$ and $W1$ filters;  {\bf (Col. 5 and 6)} Radio luminosity at 5~GHz and integrated radio luminosity at 1.4~GHz computed from the best-fit radio spectrum of each source; {\bf (Col. 7)} radio-loudness parameter, defined as R~$=S_{\rm 5GHz}/S_{4400\AA}$; {\bf (Col. 8 and 9)} Best-fit spectral indices before and after the break frequency. Only value is shown for objects where a single power law was preferred; {\bf (Col. 10)} best-fit break frequency, in GHz and in the observed frame, for systems where a broken power law was preferred; {\bf (Col. 11)} $\tilde{\alpha}_{\rm ox}$ parameter, defined as $\tilde{\alpha}_{\rm ox}=0.303\times{\rm log(L_{\rm 10keV}/L_{2500\AA})}$. The X-ray information is from \cite{Merloni2024}.}  

\label{tab:lum_RACS_zless5}
\end{table*}
 \end{landscape}

\end{appendix}

\end{document}